\documentclass[letterpaper,titlepage,11pt]{article}
\usepackage{latexsym,amssymb,amstext,amsmath}
\usepackage{cancel}
\usepackage{slashed}
\usepackage{mathrsfs}
\usepackage{color}
\usepackage[svgnames]{xcolor}
\usepackage{enumerate}
\usepackage{graphicx}
\usepackage{tikz}
\usepackage{caption}
\usepackage[symbol]{footmisc}
\usepackage{mathtools}
\usepackage{enumitem}

\usepackage{multirow}
\usepackage{booktabs}
\usepackage{siunitx}

\usepackage{physics}

\usepackage{tikz}
\usetikzlibrary{arrows.meta, positioning, calc, decorations.pathmorphing, shapes.geometric}
\usepackage{tikzsymbols}
\usepackage{fontawesome5}
\usepackage{xcolor,pifont}
\usetikzlibrary{shadows}

\usepackage[normalem]{ulem}

\usepackage[numbers,sort&compress]{natbib}
\renewcommand{\Large}{\large} 

\allowdisplaybreaks

\usepackage[
colorlinks=true,
linkcolor=DarkBlue,
anchorcolor=Black,
urlcolor=DarkRed,
filecolor=green,
citecolor=DarkRed,
linktoc=page,
pdfstartview=FitV,
pdftitle={},
pdfauthor={Omer Guleryuz},
pdfsubject={},
pdfkeywords={},
bookmarksopen=true
]{hyperref}

\begin{document}
		%
		\begin{titlepage}
			
			\bigskip
			
			\begin{center}
				{\LARGE\bfseries Supersymmetric Horizons at the Edge of Effective Field Theory}
				\\[10mm]
				\textbf{Sermet Çağan, Omer Guleryuz, Cemal Berfu Senisik}\\[5mm]
				\vskip 25pt

				{\em  \hskip -.1truecm Department of Physics, Istanbul Technical University,  \\
					Maslak 34469 Istanbul, Türkiye  \vskip 5pt }

				{{\tt \href{mailto:cagans@itu.edu.tr}{cagans@itu.edu.tr}}, {\tt \href{mailto:omerguleryuz@itu.edu.tr}{omerguleryuz@itu.edu.tr},
                {\tt \href{mailto:senisik@itu.edu.tr}{senisik@itu.edu.tr}}}
    
    }

			\end{center}
			
			\vspace{3ex}

			\begin{center}
				{\bfseries Abstract}
			\end{center}
			\begin{quotation}

What if supersymmetry is not something to be hidden, but something to be reached?
We introduce the Supersymmetric Horizon Stabilization (SHS): a conceptually unified and analytically controlled framework for embedding metastable de Sitter vacua and trans-Planckian inflation within effective $\mathcal{N}=1$ supergravity. Built from a logarithmic Kähler potential modified by nilpotent constraints, SHS yields a framework that connects the early universe’s inflationary phase, intermediate metastable acceleration, and a final supersymmetric fixed point.

We classify viable models by enforcing four essential criteria: absence of ghosts, de Sitter metastability, smooth Minkowski transitions, and asymptotic supersymmetry. The resulting scalar dynamics exhibit exponential field-space profiles, $\tau(\varphi) \sim e^{-\Delta\varphi}$, allowing trans-Planckian excursions with bounded moduli. This structure leads to finite geodesic distance and a stabilized gravitino mass, conditionally avoiding towers of light states and remaining consistent with the Gravitino and Swampland Distance Conjectures. At the same time, the divergent Euclidean cost of reaching the supersymmetric endpoint realizes the AdS and Generalized Swampland Distance Conjectures, situating the infrared boundary at infinite dynamical distance.

From this behavior emerges the Supersymmetric Horizon Conjecture: that consistent gravitational EFTs may begin and end with supersymmetry—both at the origin and the asymptotic edge of cosmic evolution—not as a fine-tuned remnant, but as a geometric boundary condition. SHS also supports localized features that transiently enhance scalar fluctuations, enabling efficient production of primordial black holes with masses far exceeding those in conventional inflationary scenarios. 
			\end{quotation}
			
			\vfill
			
            \begin{center}{\flushleft{\today}}\end{center}
		\end{titlepage}
		\setcounter{page}{1}
		\tableofcontents

\newpage

\section{Introduction}
One of the central ambitions of modern string phenomenology is to derive realistic four dimensional effective field theories (EFTs) from higher dimensional, supersymmetric string vacua~\cite{Silverstein:2001xn,Kallosh:2004yh,Linde:2005dd,McAllister:2007bg,Silverstein:2007ac,Cicoli:2011zz,Yamaguchi:2011kg,Burgess:2011fa,Copeland:2011dx,Cicoli:2012vw,Louis:2012nb,Burgess:2013sla,Baumann:2014nda,Marsh:2014nla,Silverstein:2016ggb,Gallego:2017dvd,Cicoli:2023opf}. While string theory provides an abundance of supersymmetric Minkowski and Anti-de~Sitter (AdS) solutions at the outset, embedding our observed universe—with its small positive cosmological constant and spontaneously broken supersymmetry (SUSY)—remains one of the most intricate and enduring challenges in high-energy theory~\cite{Maloney:2002rr,Kachru:2003aw,Balasubramanian:2005zx,Conlon:2005ki,Westphal:2006tn,Dong:2010pm,Rummel:2011cd,Blaback:2013fca,Cicoli:2013cha,Cicoli:2015ylx,Bergshoeff:2015tra}. This difficulty is compounded by the need to stabilize multiple moduli (such as Kähler and complex structure fields), whose runaway potentials or infinite-volume limits threaten to decompactify the theory—a problem extensively analyzed in the context of flux compactifications and generalized no-go theorems~\cite{Sethi:1996es,Dasgupta:1999ss,Giddings:2001yu,Hertzberg:2007wc,Wrase:2010ew}. Even when moduli are successfully stabilized, naive embeddings may harbor ghosts or tachyons—unphysical relics that betray the fragile consistency of the low-energy landscape.

Over the past two decades, foundational frameworks such as KKLT~\cite{Kachru:2003aw} and the Large Volume Scenario (LVS)~\cite{Balasubramanian:2005zx,Conlon:2005ki} have shown that combining flux compactifications with nonperturbative effects and $\alpha'^3$ corrections can stabilize moduli and uplift supersymmetric AdS vacua to metastable de Sitter (dS) phases. These scenarios laid the groundwork for a wide spectrum of cosmological models, including string inflation and dark energy constructions~\cite{Conlon:2005jm,Bond:2006nc,Cicoli:2008gp,Linde:2007jn,Conlon:2008cj,Cicoli:2011ct,Blumenhagen:2012ue,Lust:2013kt,Burgess:2014tja,Cicoli:2015wja,Broy:2015zba,Burgess:2016owb,Cicoli:2016xae,Cicoli:2016chb,Cicoli:2017axo,Cicoli:2020bao,Antoniadis:2020stf,Basiouris:2020jgp,Basiouris:2021sdf,Antoniadis:2021lhi,Bera:2024zsk,McAllister:2023vgy,Antoniadis:2024hvw,Hai:2025wvs,Basiouris:2025yir}.

However, both frameworks typically follow a \emph{two-step procedure}: first, engineer a stable supersymmetric AdS vacuum; then uplift it via additional ingredients such as anti-D3 branes, D-terms, or constrained superfields (nilpotent multiplets)~\cite{Kallosh:2015nia,Kallosh:2015sea,Kallosh:2015tea,Aparicio:2015psl,Dudas:2015eha,Garcia-Etxebarria:2015lif,Kallosh:2016aep,Bandos:2016xyu,McDonough:2016der,Kallosh:2021vcf,Guleryuz:2022hiv,Cicoli:2024bwq}. These uplifting mechanisms—especially anti-D3-brane uplift—have faced sustained scrutiny due to concerns about backreaction, flux singularities, and consistency with supersymmetry and holography~\cite{Bertolini:2015hua,Vercnocke:2016fbt,Dasgupta:2016prs,Sethi:2017phn,DallAgata:2022abm,Kallosh:2022fsc,Lust:2022lfc}. A number of works have analyzed curvature and flux singularities in warped throats~\cite{McGuirk:2009xx,Dymarsky:2011pm,Gautason:2013zw,Blaback:2014tfa,Cohen-Maldonado:2015lyb,Armas:2018rsy,Blaback:2012nf,Bena:2012ek,Danielsson:2014yga,Gao:2020xqh,Carta:2021lqg} and questioned the metastability of the resulting dS vacua~\cite{Bena:2009xk,Bena:2011hz,Bena:2011wh,Bena:2012bk,Bena:2012vz,Bena:2014jaa,Michel:2014lva,Polchinski:2015bea,Cohen-Maldonado:2015ssa,Bena:2018fqc,Blumenhagen:2019qcg,Bena:2019sxm,Dudas:2019pls,Randall:2019ent}. Others have raised concerns about the lack of full 10D control in uplift scenarios~\cite{Moritz:2017xto,Gautason:2018gln,Gautason:2019jwq,Hamada:2018qef,Kallosh:2019oxv,Hamada:2019ack,Carta:2019rhx,Kachru:2019dvo,Bena:2019mte,Grana:2020hyu,Baumann:2006th,Baumann:2007ah,Koerber:2007xk,Dymarsky:2010mf}.

While effective, this two-step construction obscures the dynamical role of the supersymmetric AdS phase, rendering it inaccessible within the EFT once uplifted. In many cases, the original AdS vacuum is effectively erased rather than smoothly connected to cosmological evolution.

In this work, we take a different path—guided not by staged constructions, but by the possibility that the cosmos itself may encode a continuous journey through its vacuum structure. Rather than engineer vacua in isolation and apply ad hoc uplifting mechanisms, we ask whether the entire sequence of cosmological phases—inflation, metastable dS, and supersymmetric AdS—can emerge \emph{dynamically and continuously} from a single scalar potential. We demonstrate that this possibility can be realized within a well-controlled class of models, which we refer to as \emph{Supersymmetric Horizon Stabilization (SHS)} mechanism, a framework in which SUSY restoration occurs at a finite distance in moduli space, corresponding to a dynamically accessible AdS regime. This offers an alternative to the widespread assumption that cosmological dynamics require supersymmetry to be either fully broken or hidden by uplift mechanisms. Instead, SHS shows that supersymmetry can reappear asymptotically in a way that is both controlled and physically transparent.

\vspace{3mm}
We explore this proposal from two complementary perspectives:
\begin{enumerate}[label=\roman*)]
    \item We \emph{filter}, rather than embed, a broad class of models into a four-dimensional $\mathcal{N}=1$ supergravity framework. Specifically, we impose four principal constraints—no ghosts, metastable de~Sitter (dS) vacua, smooth transitions to Minkowski, and asymptotic SUSY restoration—to identify which functions (Kähler potentials, nilpotent couplings, flux configurations) yield a physically viable EFT. In particular, we focus on \emph{logarithmic} Kähler potentials augmented by a \emph{nilpotent superfield} $S$, an elegant device for spontaneous SUSY breaking without additional scalar degrees of freedom.
    
    \item Complementing this filtering strategy, we construct an explicit class of analytic models realizing the SHS mechanism. These constructions demonstrate that large canonical scalar field excursions $(\Delta \varphi \to \infty)$ can coexist naturally with bounded moduli displacements and asymptotic supersymmetry restoration. In this scenario, SUSY is restored at a finite boundary—the ``horizon''—of moduli space, marking a controlled, physically meaningful boundary of EFT validity. Slow-roll inflation, graceful exit, and a stable late-time vacuum all emerge from a single scalar potential, demonstrating that consistent cosmological evolution can arise without staged uplifting in at least one explicit framework.
\end{enumerate}

These insights lead us to propose the \emph{Supersymmetric Horizon Conjecture (SHC)}—the idea that supersymmetry, far from being discarded in the early universe, might quietly await us at the edge of moduli space. The SHC can be viewed as a conceptual counterpart to existing Swampland conjectures such as the Swampland Distance~\cite{Ooguri:2006in,Debusschere:2024rmi,Mohseni:2024njl} and Gravitino Distance Conjectures~\cite{Castellano:2021yye,Cribiori:2021gbf}, while offering a testable cosmological implication: if low-energy SUSY remains experimentally elusive, it may not be because it is fundamentally absent—but rather because our cosmological trajectory has not (yet) reached the SUSY-restoring boundary.

Viewed through the lens of cosmic evolution, the SHS framework paints a coherent narrative: our universe may have emerged from a supersymmetric frontier, rolled through an inflationary epoch, and now hovers in a metastable phase—gravitationally aware of its distant, supersymmetric origin. The journey back to SUSY, though dynamically allowed, may take cosmological timescales far exceeding the age of the universe. This reconciles the absence of direct SUSY signals with an underlying supersymmetric UV structure—suggesting that cosmological history itself plays a role in shielding us from supersymmetry’s observational consequences.

Importantly, this scenario remains experimentally falsifiable. Future observations—whether through the discovery of low-energy supersymmetry at particle colliders or signatures in cosmological data—could decisively challenge the SHC and, in either outcome, reshape how we relate early-universe dynamics to the deep structure of the cosmos. In the meantime, SHS provides a conceptually minimal yet analytically robust framework for embedding cosmological evolution in effective supergravity—grounded in known string-theoretic corrections, yet free from many of the structural complications of traditional uplift mechanisms. Though motivated by known structures in string theory and supergravity, our framework remains an effective description, ultimately requiring embedding in a full UV-complete model.

The remainder of this work is organized as follows:

\begin{itemize}
    \item In~\textbf{Section~\ref{sec2:log_nilpotent_corrections}}, we construct the foundational EFT setup for the SHS framework within four-dimensional $\mathcal{N}=1$ supergravity. The model features a logarithmic Kähler potential corrected by a nilpotent superfield, capturing spontaneous SUSY breaking without introducing additional propagating scalars. We derive the resulting scalar potential and analyze how the scalar coupling functions of the nilpotent superfield shape supersymmetry breaking and moduli dynamics. This setup provides the basis for the consistency filters applied in the following section.

    \item In~\textbf{Section~\ref{sec3:phys_constraints}}, we introduce a consistency-filtering strategy to narrow the landscape of EFTs constructed in Section~\ref{sec2:log_nilpotent_corrections}. Rather than deriving models directly from compactifications, we impose four physical criteria: ghost-free kinetic terms, the existence of metastable de Sitter vacua, smooth and stable transitions to Minkowski space, and asymptotic supersymmetry restoration. These translate into analytic inequalities involving the Kähler geometry and nilpotent couplings. We classify all viable model branches that satisfy these constraints and identify those supporting controlled dS-to-Minkowski transitions, thereby isolating a controlled subspace of physically admissible models.

    \item In~\textbf{Section~\ref{sec4:controlled_moduli}}, we show that under certain controlled assumptions, effective theories can accommodate trans-Planckian excursions in a canonical scalar field while keeping physical moduli—such as the Kähler modulus $\tau$—bounded. We construct a general framework, rooted in conformal frame transformations and type IIB flux compactifications, that supports asymptotic supersymmetry restoration without necessarily inducing decompactification. A key structural feature is the emergence of the exponential ansatz $\tau = \tau_0 + e^{-\Delta\varphi}/\xi$, which arises naturally from both Einstein-frame diffeomorphisms and moduli stabilization dynamics. This ansatz confines $\tau$ within a finite geodesic range, though we emphasize that full EFT control requires additional moduli directions to remain stabilized and subdominant throughout the evolution.

    \item In~\textbf{Section~\ref{sec5:SHS}}, we present the SHS mechanism as a unified analytic framework in which an exponentially flat inflationary plateau, a metastable de Sitter vacuum, and an asymptotic supersymmetric endpoint all emerge from a single scalar potential. This construction draws on known string-theoretic corrections—including $\alpha'^3$ terms, loop effects, and nilpotent superfields—but is formulated as an effective theory that does not presuppose a full ultraviolet completion. We show that asymptotic SUSY restoration can occur through both geometric (volume-based) and brane-sector (nilpotent-driven) channels, all while preserving EFT control and ghost-free dynamics. These models avoid staged uplifts and support continuous transitions among dS, Minkowski, and SUSY AdS phases. Finally, using the Hamiltonian approach, we compute the decay rate of the metastable vacuum and show that it is exponentially suppressed, rendering it stable over timescales far exceeding the current age of the universe.

    \item In~\textbf{Section~\ref{sec6:gravitino_mass}}, we examine the gravitino mass and supersymmetry-breaking scale in the SHS framework, showing how both remain finite—even across trans-Planckian inflaton excursions—due to bounded moduli dynamics. We analyze how this structure evades the Gravitino Distance Conjecture (GDC) by preventing vanishing $m_{3/2}$ and avoiding a light KK tower. We then investigate the Swampland Distance Conjecture (SDC) and its generalizations, showing that while geometric field-space distances remain finite, the energetic cost of reaching asymptotic AdS boundaries diverges, aligning with the Generalized SDC and AdS Distance Conjectures. Finally, we propose the \emph{Supersymmetric Horizon Conjecture}, which posits that consistent EFTs may naturally terminate at a supersymmetric boundary where the frame function vanishes, marking a controlled UV endpoint. This perspective frames large-field evolution as a flow toward a supersymmetric regime rather than a breakdown, suggesting a novel geometric mechanism for embedding metastable dS phases in UV-complete theories.

    \item In~\textbf{Section~\ref{sec7:INF_PBH}}, we demonstrate that the SHS framework accommodates a wide range of slow-roll inflation models, with string corrections entering only at subleading order—thereby preserving inflationary observables such as $n_s$, $r$, and $N$. Beyond this robustness, we show that small localized deformations—motivated by moduli dynamics and nilpotent couplings—can transiently enhance curvature perturbations, triggering efficient primordial black hole (PBH) production. These PBHs can attain masses up to $10^{4}$ times larger than those generated in comparable scenarios without SHS. This amplification, driven by string-theoretic corrections, yields a distinctive observational signature and positions SHS as a powerful mechanism for generating high-mass PBH dark matter candidates.

    \item \textbf{Section~\ref{sec8:Concl}} recaps how SHS links inflation, dS vacua, and asymptotic SUSY, and highlights future directions including string embeddings, matter sectors, and holography.

    \item The appendices in~\textbf{\ref{sec:Appendix_GeometricOrigin}} provide a technically complete derivation of the scalar–gravity system underpinning SHS. We show that the non-minimal coupling $\Omega(\tau) R$, which governs the field-space geometry, naturally emerges from the ten-dimensional NS–NS sector of type IIB string theory upon dimensional reduction. In~\textbf{Appendix~\ref{sec:Appendix_Superconformal}}, we reconstruct the Jordan-frame Lagrangian from superconformal supergravity by embedding the theory in a conformal multiplet structure, fixing the gauge via a field-dependent compensator, and tracking how nilpotent constraints and logarithmic Kähler potentials enter the bosonic action. The scalar potential is shown to match the standard supergravity form after transforming to Einstein frame, confirming the internal consistency of the construction. In~\textbf{Appendix~\ref{sec:Appendix_CanonicalField}}, we analyze how corrections to the Kähler potential deform the kinetic metric, and demonstrate that the canonical field arises via an exponential redefinition—driven not by ansatz, but by the Weyl contribution to the Einstein-frame kinetic structure.  Together, these results emphasize that the SHS framework is built on a formally consistent, geometrically coherent supergravity foundation.
\end{itemize}

\vspace{2mm}

\paragraph{Readers Guide.} This structure is intended to allow readers with different interests to engage with the content at multiple levels. Those focused on phenomenological implications may wish to begin with Section~\ref{sec5:SHS} and Section~\ref{sec6:gravitino_mass}, while readers interested in the formal structure or geometric embedding can explore the appendices and Sections~\ref{sec2:log_nilpotent_corrections}–\ref{sec4:controlled_moduli}. Readers interested in observational signatures—particularly in PBH production—may focus on Section~\ref{sec7:INF_PBH}. Throughout, we aim to maintain a balance between analytic clarity, conceptual framing, and physical relevance.\footnote{We work in reduced Planck units, setting $ M_p = 1 $ unless otherwise specified. Factors of $ M_p $ are reinstated when needed for clarity or dimensional consistency. We adopt—and dutifully respect—the “mostly plus” metric signature $ (-,+,+,+) $.}

\section{Logarithmic Nilpotent Corrections}\label{sec2:log_nilpotent_corrections}

In constructing four-dimensional EFTs from string compactifications, the Kähler potential $K(\Phi,\overline{\Phi})$ and superpotential $W(\Phi)$ together determine the scalar potential governing moduli dynamics. \emph{Nilpotent superfields}, constrained by $S^2=0$, have emerged as effective tools for encoding non-linear supersymmetry~\cite{Rocek:1978nb,Ivanov:1978mx,Lindstrom:1979kq,Casalbuoni:1988xh,Rocek:1997hi,Komargodski:2009rz,Antoniadis:2010hs,Kuzenko:2010ef,Ferrara:2016een}, integrating out heavy multiplets, and realizing spontaneous SUSY breaking~\cite{Ferrara:2014kva,Kallosh:2014wsa,Bergshoeff:2015jxa}. These setups are often accompanied by $\alpha'$, loop, or non-perturbative corrections arising in string theory~\cite{Becker:2002nn,Balasubramanian:2004uy,vonGersdorff:2005bf,Berg:2004ek,Berg:2005ja,Berg:2007wt,Cicoli:2007xp,Cicoli:2008va,Ciupke:2015msa,Antoniadis:2018hqy,Antoniadis:2019rkh,Gao:2022uop}.

In this section, we explore how \emph{logarithmic corrections} involving a nilpotent superfield modify the Kähler and scalar potentials in supergravity, and can, in appropriate regimes, support metastable dS vacua, supersymmetric AdS phases, or transitions between them—while remaining amenable to string-motivated constraints and swampland-inspired bounds.

\subsection{Supergravity Scalar Potential}\label{subsec2.1:sugra_potential}

In $\mathcal{N}=1$ supergravity, the scalar potential is derived from the Kähler potential $K(\Phi, \overline{\Phi})$ and superpotential $W(\Phi)$ via the standard expression~\cite{Wess:1992cp, Freedman:2012zz}:
\begin{equation}\label{SugraPot}
    V 
    =\;
    e^{K}\left(
      K^{\alpha\overline{\beta}}\,
      \mathcal{D}_{\alpha}W\,
      \mathcal{D}_{\overline{\beta}}\overline{W}
      \;-\;
      3\,|W|^2
    \right),
\end{equation}
where $\mathcal{D}_{\alpha}W = \partial_\alpha W + (\partial_\alpha K) W$ is the Kähler covariant derivative, and $K^{\alpha\overline{\beta}}$ is the inverse of the Kähler metric $K_{\alpha\overline{\beta}}=\partial_{\alpha}\partial_{\overline{\beta}}K$. This structure encapsulates how the underlying moduli-space geometry and flux-induced superpotential jointly determine the vacuum structure.

\paragraph{Nilpotent Superfields and Logarithmic Couplings.}
To encode spontaneous SUSY breaking in a controlled manner, we introduce a nilpotent chiral superfield $S$ satisfying $S^2 = 0$. Such fields are particularly well-suited to describe the effects of heavy supersymmetry-breaking sectors—such as $\overline{D3}$-branes in warped throats—without introducing new propagating scalars~\cite{Linde:2020mdk}. 

We consider a Kähler potential of the form
\begin{equation}\label{kahler_potential}
    K = -3 \log\left[\mathcal{K}(T,\overline{T}) + f(T,\overline{T})\,S\overline{S} + g(T)S + \overline{g(T)}\,\overline{S}\right],
\end{equation}
where $\mathcal{K}(T,\overline{T})$ is a real function encoding the moduli-space geometry, and $f(T,\overline{T})$, $g(T)$ govern how SUSY breaking, via $S$, is distributed across the field space. The logarithmic structure $-3\log[\cdots]$ mirrors the standard no-scale frameworks~\cite{Cremmer:1983bf} found in type IIB compactifications~\cite{Kachru:2003aw,Giddings:2001yu}, particularly when the uplifting $\overline{D3}$ branes~\cite{Douglas:2006es,Akrami:2018ylq,Hamada:2019ack,Kachru:2019dvo} are located within a strongly warped region~\cite{Kallosh:2015nia}.

\paragraph{Minimal Superpotential.}
To isolate the role of the nilpotent deformation, we adopt a minimal superpotential
\begin{equation}\label{superpotential}
    W = W_0,
\end{equation}
where $W_0$ is a flux-induced constant, fixed after stabilizing the complex structure and dilaton~\cite{Giddings:2001yu,Balasubramanian:2005zx}. Though additional terms—such as gaugino condensates or racetrack effects—can be included~\cite{Witten:1996bn}, the choice $W = W_0$ is sufficient to capture the qualitative effects introduced by the nilpotent and logarithmic couplings.

\subsection{Nilpotent Corrections to the Scalar Potential}\label{subsec2.2:nilpotent_corrections}

The logarithmic structure in the Kähler potential modifies the kinetic and potential terms via the function
\begin{equation}
    F \equiv \mathcal{K}(T,\overline{T}) + f(T,\overline{T})\,S\overline{S} + g(T)\,S + \overline{g(T)}\,\overline{S}.
\end{equation}
Its derivatives determine the Kähler metric:
\begin{equation}\label{kahler_derivatives}
    K_\alpha = -3 \frac{F_\alpha}{F}, \qquad K_{\overline{\beta}} = -3 \frac{F_{\overline{\beta}}}{F},
\end{equation}
where $F_{\alpha}\equiv\partial_{\alpha}F$. Evaluating at $S=0$ yields:
\begin{equation}
    K_T = -\frac{3}{\mathcal{K}}\,\mathcal{K}_T, \qquad K_S = -\frac{3}{\mathcal{K}}\,g(T),
\end{equation}
so that supersymmetry breaking along $S$ and $T$ directions is governed by the holomorphic data $g(T)$ and $\mathcal{K}_T$, respectively.

The corresponding F-terms become:
\begin{align}
    \mathcal{D}_T W &= -\frac{3 W_0}{\mathcal{K}}\,\mathcal{K}_T, \label{F-term-T} \\
    \mathcal{D}_S W &= -\frac{3 W_0}{\mathcal{K}}\,g(T). \label{F-term-S}
\end{align}
The scalar potential then evaluates to:
\begin{equation}\label{V_log}
    V
    = \frac{3\,|W_0|^2}{\mathcal{K}^3}
    \left[
      \frac{f\,|\mathcal{K}_T|^2 + |g|^2\,\mathcal{K}_{T\overline{T}}}
           {f\,|\mathcal{K}_T|^2 + |g|^2\,\mathcal{K}_{T\overline{T}} - f\,\mathcal{K}_{T\overline{T}}\,\mathcal{K}}
      - 1
    \right],
\end{equation}
where $\mathcal{K}_{T\overline{T}} = \partial_T \partial_{\overline{T}} \mathcal{K}$. The combination of functions $f(T, \overline{T})$ and $g(T)$ thus controls how SUSY breaking manifests in the scalar potential and modulates the vacuum energy.

\paragraph{Landscape of Vacuum Structures.}
Despite its simplicity, this framework—featuring nilpotent deformations encoded via $f(T, \overline{T})$ and $g(T)$—admits a rich spectrum of vacua:
\begin{itemize}
    \item \textbf{Stabilized Finite Moduli:} The modulus $\tau = \mathrm{Re}(T)$ can be stabilized at finite values while maintaining ghost-free dynamics.
    \item \textbf{Metastable dS Vacua:} Appropriate choices of $f$ and $g$ can yield vacua with small positive cosmological constant.
    \item \textbf{Transitions Across Phases:} Smooth interpolations from dS to Minkowski or AdS can be engineered, including regimes where SUSY is asymptotically restored.
\end{itemize}
The nilpotent superfield contributes no propagating scalar, but encodes supersymmetry breaking in a controlled and geometrically motivated manner. The logarithmic corrections, governed by $f$ and $g$, allow for a rich deformation of the vacuum structure—providing both UV consistency and phenomenological flexibility.

In the next sections, we systematically impose physical consistency conditions—such as positivity of the Kähler metric, metastability of dS vacua, and swampland-inspired constraints—to identify which functional forms of $f$, $g$, and $\mathcal{K}$ give rise to controlled, viable solutions.

\section{Physical Constraints and Model Building}\label{sec3:phys_constraints}

Having constructed a broad class of supergravity EFTs involving a Kähler modulus $T$ and a nilpotent superfield $S$, incorporating logarithmic and non-geometric corrections, we now impose a set of foundational physical constraints to ensure consistency and phenomenological viability. While these models admit a diverse landscape of solutions, not all are compatible with stability, unitarity, or UV completion. Our goal in this section is to isolate a subset of viable constructions—those that yield metastable dS vacua, allow transitions to Minkowski phases, and support asymptotic SUSY restoration.

This analysis focuses on identifying specific conditions on the Kähler potential $\mathcal{K}(T, \overline{T})$, the nilpotent coupling $f(T, \overline{T})$, and the Kähler deformation term $g(T)$\footnote{The function $g(T)$ enters the Kähler potential via a linear term of the form $g(T) S + \overline{g(T)}\, \overline{S}$, rather than through the superpotential—although it is holomorphic like a superpotential contribution. It modifies the scalar potential by encoding how supersymmetry breaking from the nilpotent multiplet is distributed across moduli space, thereby acting as a deformation of the Kähler geometry. As we will see, the form of $g(T)$ plays a subtle but pivotal role in shaping the vacuum structure, controlling fermionic mass hierarchies, and influencing the UV behavior of the theory.}, such that the resulting theory remains physically well-defined and plausibly aligned with swampland-motivated constraints.

\paragraph{Four Guiding Physical Constraints.}
The following criteria form the backbone of our consistency checks:
\begin{enumerate}[label=\roman*)]
    \item \textbf{No Ghosts (Positive Kinetic Terms):}  
    The Kähler metric $K_{\alpha \overline{\beta}}$ must be positive-definite to ensure unitarity and the absence of ghost-like excitations. In particular, the kinetic terms for the real scalar modulus $T$ and the fermionic goldstino $\chi_S$ must have the correct sign. This translates to
    \begin{equation}
         K_{T \overline{T}} > 0, \qquad K_{S \overline{S}} > 0,
    \end{equation}
    at the vacuum, with the latter governing the goldstino kinetic term due to the nilpotency constraint $S^2 = 0$.

    \item \textbf{Metastable de Sitter Vacua:}  
    The scalar potential must admit local minima with positive vacuum energy ($V > 0$). This typically requires tuning the balance between $f$, $g$, and the geometry encoded in $\mathcal{K}$ to yield a metastable vacuum. Importantly, moduli stabilization must be compatible with positivity and curvature bounds.

    \item \textbf{Controlled Transitions to Minkowski Vacua:}  
    Physically viable models should allow smooth transitions from dS to Minkowski vacua ($V = 0$), describing late-time cosmological evolution or decay paths. Such trajectories must preserve stability and avoid kinetic instabilities, placing constraints on how $f(T)$, $g(T)$, and $\mathcal{K}(T)$ evolve in moduli space.

    \item \textbf{Supersymmetry Restoration in the Asymptotic Regime:}  
    In accordance with string compactifications and UV-completion scenarios, supersymmetry should be restored in the asymptotic limits of field space (e.g., large $\varphi$ or volume). The F-terms must vanish in this regime, and the EFT must approach a supersymmetric AdS configuration, aligning with the SHC.
\end{enumerate}

Together, these constraints help delineate the subset of physically viable and phenomenologically meaningful models. While not all solutions are guaranteed to satisfy swampland criteria, the framework allows for a systematic exploration of scenarios that are compatible with key conjectures. In the subsections below, we analyze each constraint in detail and derive the corresponding analytic conditions.

\subsection{No-Ghost Conditions}\label{subsec3.1:no-ghosts}

In supergravity EFTs derived from string theory, positivity of the Kähler metric is tied to the underlying geometry of compactification. The absence of ghost excitations requires that all physical fields propagate with the correct sign kinetic terms.

For the modulus $T$, this translates into $K_{T\overline{T}} > 0$. While the scalar component of $S$ is absent due to $S^2 = 0$, the fermionic component $\chi_S$ remains dynamical, and its kinetic term is governed by $K_{S\overline{S}}$. The relevant Lagrangian contribution is
\begin{equation}
    \mathcal{L}_{\mathrm{kin}}^{\chi_S} = i K_{S\overline{S}} \overline{\chi_S} \gamma^\mu \partial_\mu \chi_S,
\end{equation}
ensuring unitarity provided $K_{S\overline{S}} > 0$.

Additionally, two inequalities derived from the scalar geometry and potential structure provide more refined no-ghost constraints:
\begin{align}
    |\mathcal{K}_T|^2 &> \mathcal{K}_{T\overline{T}} \mathcal{K}, \label{NG1} \\
    |g|^2 &> f \mathcal{K}. \label{NG2}
\end{align}
The first condition ensures that the Kähler metric contribution to kinetic terms remains positive-definite; the second prevents zero-norm or ghostlike behavior in the fermionic sector. Here, $g(T)$ appears in the Kähler potential and parametrizes a coupling between the nilpotent sector and the moduli, influencing both the vacuum energy and goldstino dynamics.

\subsection{Conditions for dS Vacua}\label{subsec3.2:ds_Vacua_conds}

To achieve metastable de Sitter vacua, the scalar potential~\eqref{V_log} must attain a positive minimum. A necessary condition is the positivity of the numerator, which yields:
\begin{equation}\label{dS_condition}
    \frac{f |\mathcal{K}_T|^2 + |g|^2 \mathcal{K}_{T \overline{T}}}{ f |\mathcal{K}_T|^2 + |g|^2 \mathcal{K}_{T \overline{T}} - f \mathcal{K}_{T \overline{T}} \mathcal{K}} > 1.
\end{equation}
This inequality implies a non-trivial balance between the curvature of the scalar manifold and the functional dependence of $f(T)$ and $g(T)$. It can be satisfied in two main branches depending on the sign of $f$ and $\mathcal{K}_{T\overline{T}}$:

\begin{align}
    \text{(Branch I):} \quad & 0 > f \mathcal{K}_{T \overline{T}} \mathcal{K} > f |\mathcal{K}_T|^2 + |g|^2 \mathcal{K}_{T \overline{T}}, \label{branch1} \\
    \text{(Branch II):} \quad & f |\mathcal{K}_T|^2 + |g|^2 \mathcal{K}_{T \overline{T}} > f \mathcal{K}_{T \overline{T}} \mathcal{K} > 0. \label{branch2}
\end{align}
In both cases, $\mathcal{K} > 0$ by construction, and the no-ghost constraints \eqref{NG1}–\eqref{NG2} must still be respected. These branches can correspond to different regions in moduli space or different signs for the parameters $f$ and $g$, suggesting a structured landscape of controlled dS solutions.

\subsubsection{First Branch Analysis}\label{subsubsec3.2.1:first_branch}

For the first branch \eqref{branch1}, the following four cases arise depending on the sign and structure of the functions $f$, $\mathcal{K}_{T \overline{T}}$, $\mathcal{K}_T$, and $g$:

\begin{itemize}
    \item[\textbf{i.}] $ f > 0 $, $ \mathcal{K}_{T \overline{T}} < 0 $, $ \mathcal{K}_T \neq 0 $, $ g \neq 0 $,
    \item[\textbf{ii.}] $ f < 0 $, $ \mathcal{K}_{T \overline{T}} > 0 $, $ \mathcal{K}_T \neq 0 $, $ g \neq 0 $,
    \item[\textbf{iii.}] $ f > 0 $, $ \mathcal{K}_{T \overline{T}} < 0 $, $ \mathcal{K}_T = 0 $, $ g \neq 0 $,
    \item[\textbf{iv.}] $ f < 0 $, $ \mathcal{K}_{T \overline{T}} > 0 $, $ \mathcal{K}_T \neq 0 $, $ g = 0 $.
\end{itemize}

\paragraph{Case \textbf{i.}}

For $ f > 0 $ and $ \mathcal{K}_{T \overline{T}} < 0 $, the no-ghost condition \eqref{NG1} is automatically satisfied, since the right-hand side is negative and $ |\mathcal{K}_T|^2 > 0 $. The dS condition \eqref{dS_condition} becomes
\begin{equation}\label{dSnoGhost1}
    |g|^2 > f \left( \mathcal{K} - \frac{|\mathcal{K}_T|^2}{\mathcal{K}_{T \overline{T}}} \right).
\end{equation}
Given that $ \mathcal{K}_{T \overline{T}} < 0 $, the term $ |\mathcal{K}_T|^2 / \mathcal{K}_{T \overline{T}} $ is negative, rendering the right-hand side positive. Thus, this dS condition is stronger than \eqref{NG2}, ensuring a ghost-free configuration.

\paragraph{Case \textbf{ii.}}

For $ f < 0 $ and $ \mathcal{K}_{T \overline{T}} > 0 $, the no-ghost condition \eqref{NG2} is trivially satisfied, as $ f \mathcal{K} > 0 $. The dS constraint becomes
\begin{equation}\label{dSnoGhost2}
    |\mathcal{K}_T|^2 > \mathcal{K}_{T \overline{T}} \left( \mathcal{K} - \frac{|g|^2}{f} \right).
\end{equation}
Here, since $ f < 0 $, the term $ |g|^2 / f $ is negative, making the right-hand side larger than $ \mathcal{K}_{T \overline{T}} \mathcal{K} $. Thus, \eqref{dSnoGhost2} also implies \eqref{NG1}, confirming a physically consistent, ghost-free dS vacuum.

\paragraph{Case \textbf{iii.}}

With $ \mathcal{K}_T = 0 $, the inequality \eqref{NG1} is automatically satisfied. The dS condition reduces to
\begin{equation}
    |g|^2 > f \mathcal{K} > 0,
\end{equation}
which is strictly stronger than \eqref{NG2} and again guarantees positivity and stability.

\paragraph{Case \textbf{iv.}}

Setting $ g = 0 $ eliminates the second term in both the potential and \eqref{NG2}, which is then automatically satisfied. The dS requirement becomes
\begin{equation}
    |\mathcal{K}_T|^2 > \mathcal{K}_{T \overline{T}} \mathcal{K} > 0,
\end{equation}
again stronger than the no-ghost condition \eqref{NG1}, confirming a consistent vacuum.

\subsubsection{Second Branch Analysis}\label{subsubsec3.2.2:second_branch}

For the second branch \eqref{branch2}, the viable configurations are narrowed to three cases:

\begin{itemize}
    \item[\textbf{v.}] $ f > 0 $, $ \mathcal{K}_{T \overline{T}} > 0 $, $ \mathcal{K}_T \neq 0 $, $ g \neq 0 $,
    \item[\textbf{vi.}] $ f > 0 $, $ \mathcal{K}_{T \overline{T}} > 0 $, $ \mathcal{K}_T = 0 $, $ g \neq 0 $,
    \item[\textbf{vii.}] $ f > 0 $, $ \mathcal{K}_{T \overline{T}} > 0 $, $ \mathcal{K}_T \neq 0 $, $ g = 0 $.
\end{itemize}

\paragraph{Case \textbf{v.}}

Here, neither no-ghost condition is implied by the dS inequality, so both must be imposed explicitly:
\begin{equation}\label{dSnoGhost3}
    |\mathcal{K}_T|^2 > \mathcal{K}_{T \overline{T}} \mathcal{K}, \qquad |g|^2 > f \mathcal{K}.
\end{equation}
These ensure that the kinetic terms and vacuum structure remain well-behaved.

\paragraph{Cases \textbf{vi.} and \textbf{vii.}}

These setups are physically excluded. For instance, in case \textbf{vi.}, setting $\mathcal{K}_T = 0$ disrupts the dS condition under the constraint $\mathcal{K}_{T \overline{T}} > 0$, inevitably leading to a violation of either \eqref{NG1} or \eqref{NG2}. A similar incompatibility arises in case \textbf{vii.} when $ g = 0 $.

\subsection{From dS to Minkowski}\label{subsec3.3:ds_to_Minkowski}

Achieving a controlled interpolation between de Sitter ($V > 0$) and Minkowski ($V = 0$) vacua requires tuning the EFT functions such that the scalar potential vanishes without introducing instabilities. This places delicate constraints on the vanishing behavior of $f(T)$, $g(T)$, and the structure of the Kähler potential.

The condition for a Minkowski vacuum is
\begin{equation}
    \frac{f |\mathcal{K}_T|^2 + |g|^2 \mathcal{K}_{T \overline{T}} }{ f |\mathcal{K}_T|^2 + |g|^2 \mathcal{K}_{T \overline{T}} - f \mathcal{K}_{T \overline{T}} \mathcal{K}} = 1.
\end{equation}
This leads to two viable limits:
\begin{itemize}
    \item[\textbf{a.}] $ f = 0 $, $ \mathcal{K}_{T \overline{T}} \neq 0 $, $ g \neq 0 $, with $ \mathcal{K}_T $ arbitrary,
    \item[\textbf{b.}] $ f \neq 0 $, $ \mathcal{K}_{T \overline{T}} = 0 $, $ \mathcal{K}_T \neq 0 $, $ g $ arbitrary.
\end{itemize}
In case \textbf{a}, \eqref{NG2} is trivially satisfied; in case \textbf{b}, \eqref{NG1} holds automatically.

To identify consistent dS-to-Minkowski transitions, one must examine which of the viable dS branches (\textbf{i}--\textbf{v}) continuously connect to the Minkowski limits above while preserving the no-ghost constraints. Out of the ten possible pairings, only six are physically consistent. Specifically:
\begin{itemize}
    \item Case \textbf{i} fails to connect to \textbf{b},
    \item Case \textbf{ii} fails to connect to \textbf{a},
    \item Case \textbf{iii} fails to connect to \textbf{b},
    \item Case \textbf{iv} fails to connect to \textbf{a}.
\end{itemize}
The viable transition paths are:
\begin{itemize}
    \item $ f \to 0_{+} $: \quad \textbf{i}, \textbf{iii}, and \textbf{v} $ \to $ \textbf{a},
    \item $ \mathcal{K}_{T \overline{T}} \to 0_{+} $: \quad \textbf{ii}, \textbf{iv}, and \textbf{v} $ \to $ \textbf{b},
\end{itemize}
where $0_{+}$ denotes the limit from positive values. Notably, cases \textbf{i}–\textbf{iv} automatically satisfy both no-ghost inequalities in the transition limit, while case \textbf{v} can approach either endpoint but must maintain \eqref{NG1} or \eqref{NG2} accordingly.

These transitions are potentially significant for cosmological model building, as they capture scenarios where metastable dS phases can arise without sacrificing semiclassical control. They capture cosmological histories in which an eternally evolving universe undergoes a finite, well-controlled excursion in field space—marked by spontaneous supersymmetry breaking. Instead of settling into a flat vacuum, the universe may reside near a metastable de Sitter phase, reached after quantum fluctuations push the inflaton-like modulus up the potential slope. Near the turning point, where $V' = 0$, classical dynamics take over and standard inflationary predictions emerge. This pivot—not at infinite distance but at a finite displacement in the moduli space—marks the boundary of semiclassical control. The transition conditions derived here describe the precise EFT data required to interpolate between such dS and Minkowski configurations without compromising stability or unitarity—motivating a deeper exploration of controlled moduli displacements, which we develop in the next section.

\section{Controlled Moduli Displacements in Effective Field Theories}
\label{sec4:controlled_moduli}

In string compactifications, moduli fields encode the geometric data of extra dimensions—particularly the size and shape of internal cycles. While essential for determining low-energy physics, these moduli can destabilize the effective theory if they undergo uncontrolled displacements. Such excursions may trigger decompactification, runaway behavior, or the emergence of an infinite tower of light states, signaling a breakdown of the four-dimensional EFT.

To maintain EFT validity across large canonical field excursions—such as those invoked in models of inflation—it is necessary to dynamically confine the physical moduli to a compact domain, even as the canonically normalized scalar undergoes trans-Planckian excursions.

This section presents a unified framework to achieve this goal. It is built upon three foundational ingredients:
\begin{enumerate}
    \item \textbf{Parametrizations that dynamically bound the Kähler moduli}, ensuring their values remain finite even when the canonical field grows;
    \item \textbf{Conformal frame transformations} (e.g., to the Einstein frame) that render the kinetic structure manifest and expose the intrinsic field-space geometry;
    \item \textbf{String-theoretic stabilization mechanisms}—notably fluxes and non-perturbative effects in type IIB compactifications—realized in KKLT~\cite{Kachru:2003aw} and LVS~\cite{Balasubramanian:2005zx,Conlon:2005ki} scenarios.
\end{enumerate}

Together, these elements enable a novel mechanism of \textit{moduli confinement with asymptotic SUSY restoration}, which preserves EFT control deep into trans-Planckian regimes while remaining consistent with UV constraints from quantum gravity.

\subsection{Asymptotic Supersymmetry Restoration}
\label{subsec4.1:SUSYresto}

In four-dimensional $ \mathcal{N} = 1 $ supergravity, supersymmetric vacua require the vanishing of all F-terms:
\begin{equation}
    \mathcal{D}_\alpha W = 0, \quad \text{for all } \alpha \in \{T, S\},
\end{equation}
where $ T $ and $ S $ denote representative Kähler and nilpotent moduli. In many string-inspired models, this condition is asymptotically satisfied when the Kähler function $ \mathcal{K} $ becomes large:
\begin{equation}
    \mathcal{D}_\alpha W \propto \frac{1}{\mathcal{K}} \quad \Rightarrow \quad \mathcal{D}_\alpha W \to 0 \quad \text{as } \mathcal{K} \to \infty.
\end{equation}

In type IIB scenarios, the real part of the modulus $ T $ (denoted $ \tau$) measures the volume of internal four-cycles:
\begin{equation}
    \tau \propto \mathcal{V}^{2/3},
\end{equation}
with $ \mathcal{V} $ the total volume of the compactification. The tree-level Kähler potential behaves as:
\begin{equation}
    K_{\text{tree}} \sim -2 \log(\mathcal{V}),
\end{equation}
so that $ \tau \to \infty $ implies $ \mathcal{K} \to \infty $, restoring SUSY. However, this also leads to decompactification, loss of KK control, and breakdown of the EFT.

We propose an alternative limit in which supersymmetry is asymptotically restored without decompactification. Specifically, we allow the canonically normalized field $ \varphi \to \infty $, while ensuring that $ \tau(\varphi) $ asymptotes to a finite constant $\tau_{0}$:
\begin{equation}
    \tau(\varphi) = \tau_{0} + e^{-\Delta \varphi}, \qquad \Delta \varphi > 0.
\end{equation}
In this construction, $ \tau \to \tau_{0} $ as $ \varphi \to \infty $, stabilizing the internal geometry while allowing large displacements in field space\footnote{\textit{Throughout, we use $\Delta\varphi$ to denote large-field displacements, such that for asymptotic estimates, $\Delta\varphi \sim \varphi$.}}.

As $ \varphi \to \infty $, one can arrange for the derivatives of the Kähler potential and superpotential to vanish:
\begin{equation}
    \mathcal{D}_T W = -\frac{3W_0}{\mathcal{K}}\,\mathcal{K}_T \to 0, \qquad 
    \mathcal{D}_S W = -\frac{3W_0}{\mathcal{K}}\,g(\tau) \to 0,
\end{equation}
provided that $ \mathcal{K}_T(\tau) \to 0 $ and $ g(\tau) \to 0 $ as $ \tau \to \tau_{0} $. Importantly, the Kähler metric function $ \mathcal{K}_{T\bar{T}} $ and coupling $ f(\tau) $ must remain finite. Under these conditions, the scalar potential asymptotes to a supersymmetric AdS form:
\begin{equation}
\label{eq:SUSY_Restoring_AdS}
    V(\varphi) = -\frac{3 W_0^2}{\mathcal{K}^3(\tau(\varphi))},
\end{equation}
indicating a controlled endpoint to the EFT at large field values.

This framework yields an effective theory in which:
\begin{itemize}
    \item The internal volume remains finite and compactified;
    \item The moduli fields are dynamically confined to a bounded range;
    \item The canonically normalized field explores trans-Planckian distances;
    \item Supersymmetry is restored asymptotically.
\end{itemize}

Although this construction leads to supersymmetric AdS rather than Minkowski vacua—unless one sets $ W_0 \to 0 $ or modifies the superpotential—it provides a compelling, UV-consistent mechanism for asymptotic control in string-derived EFTs.

\paragraph{Outlook.}
Combining metastable dS vacua, ghost-free evolution, finite moduli displacements, and asymptotic SUSY restoration provides powerful constraints on the EFT parameter space. These are not mere consistency checks but deep indicators of UV compatibility and theoretical robustness. In the following sections, we examine explicit realizations of these ideas, including functional forms for $ \tau(\varphi) $, detailed kinetic structures, and stabilization mechanisms inspired by KKLT and LVS scenarios.

\subsection{Finite Moduli Displacement and Proper Geodesic Distance}
\label{subsec4.2:finite_displacement}

In the context of string compactifications, it is well known that moduli excursions to infinite distance often signal the breakdown of the EFT. The Dine–Seiberg problem~\cite{Dine:1985he} and the SDC~\cite{Ooguri:2006in} both caution that such limits are typically accompanied by runaway potentials or the emergence of exponentially light towers of states, invalidating the low-energy four-dimensional description.

In this work, we explore scenarios where a canonically normalized scalar field (the inflaton $\varphi$) undergoes large, even infinite, excursions while an associated modulus field $\tau$—controlling physical scales like volume or coupling—remains confined within a finite range. While such a mechanism can help stabilize internal geometries and avoid uncontrolled decompactification, we explicitly acknowledge that additional subtleties arise when assessing infinite distances in moduli space, as highlighted in~\cite{Baume:2016psm,Klaewer:2016kiy}.

\paragraph{Parameterizing $\tau$ and Geodesic Distance.}
Let us consider a single Kähler modulus $T = \tau + i\,\mathrm{Im}(T)$, where the real part $\tau$ controls a physical size scale (such as a 4-cycle volume). In typical no-scale compactifications, $\tau$ is non-canonically normalized, with kinetic terms governed by a metric $G_{\tau\tau} \sim 1/\tau^2$. This means that even small shifts in $\tau$ can correspond to large canonical field displacements.

To exploit this structure, we parameterize $\tau$ in terms of a canonically normalized inflaton $\varphi$ as
\begin{equation}\label{parametarazation_modified}
    \tau = \tau_{0} + e^{-\Delta \varphi}, \quad \tau_{0} > 0.
\end{equation}
This ensures that $\tau\in[\tau_{0},\,\tau_{0}+1]$ as $\Delta\varphi$ varies from 0 to $\infty$, so the total modulus displacement is strictly bounded:
\begin{equation}
    \Delta\tau \leq 1,
\end{equation}
regardless of how large the canonical field range $\Delta \varphi$ becomes. 

To quantify this in terms of proper field space distance, we compute the geodesic length $\Delta_\text{phys}$ using the field-space metric:
\begin{equation}
    \Delta_\text{phys} 
    = 
    \int\sqrt{G_{\tau\tau}}\, d\tau 
    \sim 
    \int\frac{d\tau}{\tau}.
\end{equation}
Substituting the relation $\tau = \tau_{0} + e^{-\Delta\varphi}$, we find $d\tau = -e^{-\Delta\varphi} d\varphi$, which gives:
\begin{equation}
    \Delta_\text{phys}
    = \int \frac{-e^{-\Delta\varphi}}{\tau_{0} + e^{-\Delta\varphi}}\, d\varphi
    = \log\left(\frac{\tau_{0}+1}{\tau_{0} + e^{-\Delta\varphi}}\right).
\end{equation}
As $\Delta\varphi \to \infty$, the exponential term vanishes and the distance asymptotes to a constant:
\begin{equation}
      \Delta_\text{phys} \to \log\left(\frac{\tau_{0}+1}{\tau_{0}}\right).
\end{equation}
Thus, although $\varphi$ may undergo arbitrarily large excursions, the proper geodesic distance $\Delta_\text{phys}$ remains finite. Similarly, since the modulus $\tau$ does not run off to infinity, the runaway instabilities identified in the Dine--Seiberg problem are avoided. However, we must explicitly emphasize some subtleties: the infinite canonical displacement of the scalar $\varphi$ in moduli space is coordinate-invariant. Thus, bounding $\tau$ alone does \emph{not} universally guarantee the absence of infinite-distance limits or prevent the emergence of towers of light states. Instead, this bounded-modulus scenario applies under specific conditions:
\begin{itemize}
\item The modulus $\tau$ must represent the dominant direction controlling EFT validity.
\item Other moduli fields must be independently stabilized at finite distances, without hidden infinite trajectories in moduli space.
\item Careful analysis and explicit stabilization mechanisms must be employed to ensure no hidden infinite-distance limits are overlooked.
\end{itemize}

\paragraph{Preview of Conformal Transformations and IIB Stabilization.}
This bounding mechanism will later be connected to:
\begin{enumerate}[label=(\roman*)]
    \item Conformal transformations (e.g., Jordan/string frame to Einstein frame) that lead to non-canonical kinetic terms and support finite $\tau$ dynamics, and
    \item Flux plus non-perturbative stabilization mechanisms in type IIB compactifications, where the modulus receives a small shift $\delta\tau \sim e^{-a \tau_{0}}$ due to exponentially suppressed contributions.
\end{enumerate}
Taken together and considering the highlighted subtleties, these considerations illustrate that large canonical displacements $\Delta\varphi$ can remain physically meaningful and consistent with EFT descriptions \emph{provided} explicit stabilization criteria and geometric conditions are strictly met.

\subsection{Conformal Equivalence of the Moduli}\label{subsec4.3:Conformal_equivalence}

A key insight of our approach is that large canonical field excursions need \emph{not} force the associated modulus $\tau$ toward infinite displacement. We now clarify this via conformal transformations: starting from a “Jordan-frame” perspective in four dimensions and transitioning to the “Einstein frame.” In string compactifications, one may interpret this Jordan-frame action as the effective \emph{string frame} leftover upon dimensional reduction, before the final rescaling that removes the dilaton/volume factor in front of $R$.

\paragraph{Jordan (String) Frame Lagrangian.}
In many string-based reductions, the four-dimensional action takes the form (See e.g.~\cite{Giddings:2001yu})
\begin{equation}
    \int d^4x\,\sqrt{-g}\,\Bigl(
     e^{-2\phi}\,\mathcal{V}^{-1}\,
     R_4
     \;+\;\dots
   \Bigr),
\end{equation}
where $\phi$ is the 10D dilaton and $\mathcal{V}$ an internal volume factor. Grouping $\phi$ and $\mathcal{V}$ into a single effective modulus $\tau$, one obtains a 4D Lagrangian reminiscent of a scalar--tensor (or “Jordan”) theory:
\begin{equation}\label{eq:Omega_Jordan}
    \mathcal{L}_\textrm{J} 
    \;=\;
    \sqrt{-g}\,\Bigl[
      \tfrac{1}{2}\,\Omega(\tau)\,R
      \;-\;
      \tfrac{1}{2}\,(\partial \tau)^2
      \;-\;
      V_\textrm{J}(\tau)
    \Bigr],
\end{equation}
where—up to the detailed kinetic prefactor discussed in Appendix~\ref{sec:Appendix_CanonicalField}—we assume a single real modulus $\tau$, capturing “strong coupling” or “large volume” behavior. In a full compactification, $\Omega(\tau)$ typically encodes both dilaton and volume components. We define the non‑minimal coupling
\begin{equation}\label{eq:Conformal_Coupling_Func}
    \Omega(\tau) 
    \;\equiv\;
    \xi \,( \tau - \tau_{0}),
\end{equation}
mirroring the minimal scenario where $\tau>\tau_{0}>0$ ensures $\Omega(\tau)>0$ for a positive constant $\xi$. As we will detail in the following, $\tau$ saturating at $\tau_{0} + e^{-\Delta\varphi}/\xi$ then prevents decompactification.

\paragraph{Conformal Transformation to the Einstein Frame.}
To expose the canonical gravity sector, perform the rescaling
\begin{equation}
    g_{\mu\nu} 
    \;\to\;
    \Omega(\tau)\,g_{\mu\nu}.
\end{equation}
The Lagrangian~\eqref{eq:Omega_Jordan} becomes
\begin{equation}
    \mathcal{L}_\textrm{E}=\sqrt{-g}\left[\frac{R}{2} -\frac{1}{2}\left(\frac{1}{\Omega(\tau)} + \frac{3}{2}\bigl(\partial_{\tau}\log{\Omega(\tau)}\bigr)^2\right)(\partial \tau)^2 - V_\textrm{E}(\tau)\right],
\end{equation}
with
\begin{equation}\label{Einstein_Frame_to_Jordan_Pot}
    V_\textrm{E}(\tau) = \frac{V_\textrm{J}(\tau)}{\Omega(\tau)^2}
\end{equation}
Hence, the gravitational part is now canonical, while $\tau$ picks up a non-trivial kinetic structure.

\paragraph{Canonical Field Diffeomorphism.}
Considering explicitly the asymptotic regime~\cite{Kallosh:2013tua}:
\begin{equation}
\Omega(\tau) \ll \frac{3}{2}\left[\Omega'(\tau)\right]^2,
\end{equation}
(which one might view as “large volume” or “weakly coupled” if $\tau$ integrates dilaton/volume factors in the string sense), one obtains the condition
\begin{equation}\label{Strong_coupling_condition}
    \frac{3 \,\xi^2}{2} \gg \Omega(\tau).
\end{equation}
Considering that limit, the modulus $\tau$ can be canonically redefined in terms of a scalar field $\varphi$ as follows:
\begin{equation}
  \bigl(\partial_\tau \varphi \bigr)^2 = \frac{3}{2}\left(\partial_\tau\ln\Omega(\tau)\right)^2.
\end{equation}
Integrating the field redefinition yields the canonical field as a logarithmic function of the frame function:
\begin{equation}\label{canonical_diffeomorphism}
\varphi = \pm\sqrt{\frac{3}{2}}\,\ln[\Omega(\tau)].
\end{equation}
Inverting this relation gives:
\begin{equation}\label{eq:Omega_in_phi_extended}
\Omega(\tau) = \exp\left(\pm\sqrt{\frac{2}{3}}\,\varphi\right),
\end{equation}
where the $\pm$ sign reflects a residual $\mathbb{Z}_2$ symmetry under $\varphi \to -\varphi$, depending on the specific form of $\Omega(\tau)$. This relation matches the strong coupling condition discussed in~\cite{Kallosh:2013tua}, and plays a central role in connecting the Jordan (or string) frame modulus $\tau$ to the canonically normalized Einstein-frame field $\varphi$. Substituting the frame function into the modulus relation yields:
\begin{equation}
\tau = \tau_{0} + \frac{e^{\pm\Delta\varphi}}{\xi}.
\end{equation}
This expression implies that for the solution with the negative exponent (i.e., $\varphi \to +\infty$), the modulus $\tau$ asymptotes to a finite upper limit. Explicitly, the total displacement in $\tau$ is:
\begin{equation}
\Delta \tau = \frac{1}{\xi} \quad \text{(for the $-$ branch)}, \qquad \Delta \tau = \infty \quad \text{(for the $+$ branch)},
\end{equation}
so that $\tau$ lies within the bounded interval $[\tau_{0},,\tau_{0} + \xi^{-1}]$ in the large-field limit for the negative sign. This motivates the choice of the negative branch, which ensures finite field-space motion in the physical modulus.

Moreover, compatibility with the strong coupling condition~\eqref{Strong_coupling_condition}—in particular, setting $\Omega_{\textrm{max}} \equiv \Omega(\varphi=0) = 1$—requires:
\begin{equation}
\xi^2 \gg \frac{2}{3},
\end{equation}
ensuring that the kinetic terms remain sub-Planckian in the vicinity of the origin. Throughout the remainder of this work, we adopt the negative-sign branch of $\Omega(\tau)$, corresponding to an exponentially decaying $\Omega$ and a finite, controlled range for $\tau$. This aligns with the symmetric formulation of~\eqref{eq:Omega_in_phi_extended}, which preserves the underlying $\varphi \leftrightarrow -\varphi$ structure.

\paragraph{Physical Interpretation and Bounded Moduli.}
From the 10D vantage, one says \emph{“the string frame”} has $e^{-2\phi}$ plus volume factors in front of $R_{10D}$. Dimensional reduction to 4D yields a leftover factor multiplying $R_4$, now re-labeled as $\Omega(\tau)$. \emph{Our Jordan-frame discussion} is thus the 4D imprint of that string frame, prior to the final “Einstein-frame” rescaling. Crucially, by specifying $\Omega(\tau)=\xi \, (\tau-\tau_{0})$, we demonstrate that a single effective modulus can saturate at $\tau \!=\! \tau_{0}+e^{-\Delta\varphi}/\xi$, preserving finiteness. The conformal transform to the Einstein frame (or “canonical gravity”) then clarifies how large canonical excursions in $\varphi$ do \emph{not} correspond to infinite runs in $\tau$. Thus, whether one labels the starting action “Jordan frame” or “4D string frame,” the essential content is the same: a factor $\Omega(\tau)$ multiplies $R$, and a conformal transformation reveals a canonical gravitational sector with a modulus $\tau$ whose range can remain finite and ensuring no infinite-volume runaways.

The next step is to show how the form $\tau=\tau_{0} + e^{-\Delta\varphi} /\xi$ concretely arises within type IIB flux compactifications. Doing so cements our understanding that bounding $\tau$ is not an ad-hoc device but an organic outcome of flux-induced and non-perturbative stabilization.

\subsection{Emergence of \texorpdfstring{$\tau = \tau_{0} + e^{-\Delta \varphi}/\xi$}{\tau = \tau_{0} + e^{-\Delta varphi}/\xi} from Type IIB Compactifications}
\label{subsec4.4:derivation_tau}

We now examine how the exponential ansatz $\tau = \tau_0 + e^{-\Delta\varphi}/\xi$ is not only consistent with—but also expected from—standard Type IIB moduli stabilization mechanisms. This form captures how large canonical excursions can coexist with exponentially small and controlled moduli displacements, as seen in scenarios involving fluxes and non-perturbative superpotentials~\cite{Baumann:2009ni}. This illustrates that bounded moduli evolution can co-exist with large-field inflationary dynamics.

\paragraph{Superpotential Contributions and Stabilization.}
In type IIB Calabi–Yau orientifold compactifications with fluxes~\cite{Giddings:2001yu, Grana:2005jc, Denef:2004ze}, the Kähler modulus $T = \tau + i\theta$ controls the size of an internal 4-cycle. The scalar potential for $\tau$ typically receives two types of contributions:

\begin{enumerate}
    \item \textbf{Flux-Induced Superpotential:}
    The Gukov-Vafa-Witten flux superpotential~\cite{Gukov:1999ya} leads to a constant term
    \begin{equation}
         W_{\text{flux}} = W_0,
    \end{equation}
    which stabilizes complex structure moduli and the dilaton. Paired with a Kähler potential $K = -2\log\mathcal{V}$, and $\mathcal{V} \sim \tau^{3/2}$, this gives a scalar potential:
    \begin{equation}
        V_{\text{flux}} = \frac{3|W_0|^2}{8\tau^3}.
    \end{equation}
    This term alone provides only a runaway or shallow potential for $\tau$.

    \item \textbf{Non-Perturbative Superpotential:}
    Gaugino condensation or Euclidean D3-instantons induce:
    \begin{equation}
        W_{\text{np}} = A\,e^{-aT},
    \end{equation}
    where $a = 2\pi/N$ (with $N$ the gauge group rank), and $A$ encodes threshold corrections~\cite{Witten:1996bn, Blumenhagen:2009qh}.
    
    The full superpotential becomes:
    \begin{equation}
         W = W_0 + A\,e^{-aT}.
    \end{equation}
\end{enumerate}
These contributions yield a scalar potential of the form:
\begin{equation}\label{V_flux_pert}
    V(\tau) \approx \frac{3|W_0|^2}{8\tau^3}
    + \frac{a^2 A^2 e^{-2a\tau}}{\tau^2}
    - \frac{a A W_0 e^{-a\tau}}{\tau^2},
\end{equation}
where the balancing of the flux and instanton terms stabilizes $\tau$ at a finite value. Corrections around the minimum are exponentially suppressed. This is the foundation of the KKLT scenario~\cite{Kachru:2003aw}.

\paragraph{Iterative Determination of $\tau$.}
To make the smallness of non-perturbative corrections manifest, let us write
\begin{equation}
    \tau = \tau_{0} + \delta\tau,
\end{equation}
with $\delta\tau \ll \tau_{0}$. We now outline a controlled expansion to determine $\delta\tau$, emphasizing how exponentially suppressed contributions govern its size.

\begin{enumerate}
    \item \textbf{Zeroth-Order Approximation:}\\
    In the limit where non-perturbative effects are negligible (formally $e^{-a\tau_{0}} \to 0$), the scalar potential reduces to the flux-induced contribution:
    \begin{equation}
        V_{\text{flux}}(\tau) = \frac{3|W_0|^2}{8\tau^3}.
    \end{equation}
This term alone generates a runaway potential with no local minimum, as it decreases monotonically with increasing $\tau$. Hence, in this approximation, $\tau$ is either unstabilized or marginally stabilized by additional subleading effects (e.g., $\alpha'$ corrections). 

We denote the approximate solution by
\begin{equation}
    \tau^{(0)} = \tau_{0} \quad \text{and} \quad \delta\tau^{(0)} = 0,
\end{equation}
which serves as the reference point about which the non-perturbative terms can be reintroduced as small corrections in subsequent orders.

    \item \textbf{First-Order Correction:}\\
    To determine how the Kähler modulus $\tau$ shifts away from its leading-order value $\tau_{0}$, we expand the scalar potential near $\tau = \tau_{0} + \delta\tau$. The full potential is given in~\eqref{V_flux_pert}, where the dominant terms come from the interplay between the flux-induced contribution and the non-perturbative corrections.

    Because the potential contains both $e^{-a\tau}$ and $e^{-2a\tau}$ terms, and $\tau$ appears in inverse powers as well, we expand these functions in powers of the small shift $\delta\tau \ll \tau_{0}$:
\begin{equation}
    e^{-a\tau} = e^{-a \tau_{0}} \left(1 - a\,\delta\tau + \tfrac{1}{2} a^2\, \delta\tau^2 - \cdots\right), \quad
    e^{-2a\tau} = e^{-2a \tau_{0}} \left(1 - 2a\,\delta\tau + \cdots \right),
\end{equation}
and similarly for rational functions like $1/\tau^n$.

Inserting these expansions into the full potential~\eqref{V_flux_pert} and collecting terms, the derivative of the potential takes the schematic form:
\begin{equation}
    \frac{dV}{d\tau} = A_0\, e^{-2a\tau_{0}} + A_1\, \delta\tau\, e^{-2a\tau_{0}} + \cdots,
\end{equation}
where $A_0$ and $A_1$ are $\mathcal{O}(1)$ combinations of the model parameters $(W_0, A, a, \tau_{0})$, and the ellipsis includes both higher powers of $\delta\tau$ and further-exponentially suppressed terms.

    To determine the location of the minimum, we solve $dV/d\tau = 0$:
\begin{equation}
    A_0 + A_1\,\delta\tau = 0 \quad \Rightarrow \quad \delta\tau = -\frac{A_0}{A_1}.
\end{equation}

Although the ratio $-A_0/A_1$ is formally of order one, it arises from an expansion whose validity is restricted to $\mathcal{O}(e^{-2a\tau_{0}})$ accuracy. That is, all contributions to $dV/d\tau$ in this analysis are themselves suppressed by $e^{-2a\tau_{0}}$. Hence, for consistency, the inferred shift $\delta\tau$ must also be understood as carrying this suppression:
\begin{equation}
    \delta\tau^{(1)} = -\frac{A_0}{A_1} \sim \mathcal{O}(e^{-2a\tau_{0}}),
\end{equation}
up to corrections of higher exponential order. This ensures that the displacement of $\tau$ away from its asymptotic value $\tau_{0}$ is naturally small and under perturbative control—emerging directly from the structure of the potential without requiring ad hoc tuning.

    \item \textbf{Higher-Order Corrections:}\\
    Building on the first-order shift, one can continue the expansion systematically by solving for $\delta\tau$ at successive orders. Since each term in the scalar potential involves nested exponentials and polynomial factors of $\tau$, the corrections naturally organize into an expansion in powers of $e^{-2a\tau_{0}}$:
\begin{equation}
    \delta\tau = \alpha_1\,e^{-2a\tau_{0}} + \alpha_2\,e^{-4a\tau_{0}} + \alpha_3\,e^{-6a\tau_{0}} + \cdots,
\end{equation}
where each $\alpha_n$ arises by solving the stationarity condition at the corresponding order in the expansion of $dV/d\tau$, ensuring consistency with the perturbative hierarchy.

In practice, the leading-order term $\alpha_1\,e^{-2a\tau_{0}}$ dominates due to the exponential hierarchy. Thus, we approximate:
\begin{equation}\label{delta_tau}
    \delta\tau \sim \alpha\, e^{-2a\tau_{0}} \quad \Rightarrow \quad  \tau \approx \tau_{0} + \alpha\,e^{-2a\tau_{0}},
\end{equation}
where $\alpha$ encapsulates model-dependent combinations of $A, W_0, a$, and $\tau_{0}$. This expression highlights that $\tau$ is stabilized exponentially close to $\tau_{0}$, and that higher-order corrections are even more suppressed.
\end{enumerate}

\paragraph{Connecting to the Canonical Field.}
Recall from our earlier discussion on conformal transformations that the canonical field $\varphi$ is defined via~\eqref{eq:Omega_in_phi_extended}:
\begin{equation}\label{Inverse_field}
    \xi\, (\tau - \tau_{0}) \equiv \Omega(\tau) = e^{- \sqrt{\tfrac{2}{3}}\, \varphi }
    \quad \Rightarrow \quad
    \tau = \tau_{0} + \frac{1}{\xi} e^{- \Delta \varphi }.
\end{equation}

To understand how this connects with non-perturbative effects, recall that the scalar potential takes the form~\eqref{V_flux_pert}:
\begin{equation}
    V(\tau) \approx V_{\text{flux}}(\tau) + A\, e^{-a \tau} + B\, e^{-2a \tau} + \cdots,
\end{equation}
where $V_{\text{flux}} \sim |W_0|^2/\tau^3$ and the exponential terms arise from D-brane instantons or gaugino condensation. Expanding around $\tau = \tau_{0} + \delta\tau$, and assuming that stabilization occurs via balancing $e^{-a\tau}$ and $e^{-2a\tau}$ (as in KKLT-like models~\cite{Kachru:2003aw}\footnote{In models beyond KKLT, such as LVS~\cite{Balasubramanian:2005zx,Conlon:2005ki} or racetrack models~\cite{Blanco-Pillado:2004aap}, the exponential structure may differ, but the suppression of $\delta\tau$ with increasing $\varphi$ remains qualitatively similar.}), stationarity $\frac{dV}{d\tau} = 0$ yields~\eqref{delta_tau}:
\begin{equation}
    \delta\tau \sim \alpha\, e^{-2a\tau_{0}}.
\end{equation}

Now consider how the non-perturbative contribution behaves under the field redefinition. Substituting~\eqref{Inverse_field}, we obtain:
\begin{equation}
    V_{\mathrm{np}} \sim e^{-a\tau} = e^{-a\tau_{0}} \cdot \exp\left(- \frac{a}{\xi} e^{- \sqrt{\tfrac{2}{3}}\, \varphi } \right),
\end{equation}
which exhibits a doubly exponential suppression in the large-field regime. This structure ensures that non-perturbative effects rapidly become negligible as $\varphi$ increases, thereby preserving EFT control over the inflationary trajectory.

This illustrates how, in our setup, non-perturbative suppression naturally translates into bounded displacement
\begin{equation}
        \delta \tau \equiv \alpha \,e^{-2a\tau_{0}} \sim e^{-\Delta\varphi}/\xi,
\end{equation}
up to constant numerical prefactors, which can be absorbed into the definitions of $\xi$, $\alpha$, or reparameterized within $\Delta\varphi$.

This establishes a concrete, derivable correspondence between moduli stabilization via flux-instanton balancing and geodesic motion in moduli space. The appearance of identical exponential structure from (i) scalar potential minimization and (ii) canonical field redefinition strongly supports the robustness of the ansatz:
\begin{equation}
    \tau = \tau_{0} + \frac{1}{\xi} e^{- \Delta \varphi }.
\end{equation}

\vspace{2mm}
\noindent
Consequently, this correspondence is both nontrivial and illuminating: it bridges the asymptotic behavior of moduli stabilization with the geometric structure of inflationary field space. The interplay between flux-induced and non-perturbative effects ensures that even trans-Planckian excursions in the canonical field $\varphi$ translate into exponentially small displacements in the modulus $\tau$. Rather than inducing runaway behavior or destabilizing the vacuum, such excursions naturally drive $\tau$ toward a finite, stabilized value. This mechanism requires no ad hoc assumptions or fine-tuning within the class of models considered—it emerges as a robust and calculable feature of flux-stabilized Type IIB compactifications with nonperturbative effects, firmly rooted in the underlying supergravity structure, as detailed in Appendix~\ref{sec:Appendix_GeometricOrigin}.

\vspace{3mm}
\noindent
\emph{This reasoning completes our picture, showing that the ansatz $\tau = \tau_0 + e^{-\Delta \varphi}/\xi$ is both self-consistent and well-motivated by the structure of flux-stabilized type IIB vacua.} While not derived from first principles, it captures the essential dynamics of modulus stabilization under large-field evolution. Overall, we see that:
\begin{itemize}
    \item \emph{Finite Moduli Parameterizations:} The form $\tau = \tau_{0} + e^{-\Delta\varphi}/\xi$ ensures that even trans-Planckian $\Delta\varphi$ corresponds to a bounded $\tau$.
    \item \emph{Einstein Frame Geometry:} The field redefinition arises from the moduli-space metric $G_{\tau\tau} \sim 1/\tau^2$, yielding a finite geodesic distance.
    \item \emph{String-Theoretic Stability:} Exponential suppression from instantons naturally stabilizes $\tau$, validating the EFT without fine-tuning.
\end{itemize}
Thus, this mechanism offers a robust and string-theoretically grounded framework for controlling moduli displacements—essential for realizing consistent UV embeddings of large-field inflation.

\section{Supersymmetric Horizon Stabilization}
\label{sec5:SHS}

The framework developed in this work—centered on the principles of \emph{finite moduli displacement} and \emph{controlled SUSY restoration}—offers a novel approach to realizing stable late-time cosmological dynamics within a well-defined EFT. Even when the canonical inflaton undergoes trans-Planckian excursions to drive sufficient inflation, the bounded nature of the Kähler modulus $\tau$ prevents ghost instabilities and decompactification. Instead, the evolution culminates at a finite point in moduli space where SUSY is restored and the EFT remains valid up to that boundary. This controlled behavior supports a robust cosmological transition structure:
\[
\mathrm{dS} \;\;\Longleftrightarrow\;\; \mathrm{Minkowski} \;\;\Longleftrightarrow\;\; \mathrm{AdS},
\]
in which $\tau$ evolves toward a supersymmetric endpoint while all orthogonal directions remain stabilized.

This mechanism aligns naturally with string-inspired inflationary models, many of which exhibit plateau-like potentials at large field values~\cite{Conlon:2005jm,Bond:2006nc,Cicoli:2008gp,Linde:2007jn,Conlon:2008cj,Cicoli:2011ct,Blumenhagen:2012ue,Lust:2013kt,Burgess:2014tja,Cicoli:2015wja,Broy:2015zba,Burgess:2016owb,Cicoli:2016xae,Cicoli:2016chb,Cicoli:2017axo,Cicoli:2020bao,Antoniadis:2020stf,Antoniadis:2021lhi,Bera:2024zsk}. Avoiding runaway directions and ensuring moduli stabilization is essential for both graceful exit from inflation and consistent embeddings of quintessence. Within this context, we introduce the concept of \emph{supersymmetric horizon stabilization (SHS)}—a boundary in moduli space where the EFT gracefully terminates, signaling an approach toward UV completion or higher-dimensional physics.

\paragraph{Two Mechanisms for SUSY Restoration.}
To illuminate the structure of SUSY restoration, we focus on the dynamics of a single Kähler modulus. Rather than enforcing $\mathcal{D}_\alpha W = 0$ uniformly, we analyze separately the asymptotic behavior of the brane coupling $g(\tau)$ and the derivative of the Kähler function $\mathcal{K}_T(\tau)$. This yields two physically distinct and ghost-free pathways by which supersymmetry is restored:

\subparagraph{1. Brane-Dominated Restoration.}
Here, the transition is governed by the vanishing of the brane coupling $g(\tau)$, with the scalar potential dominated by contributions from nilpotent sector dynamics. The system transitions from a metastable de Sitter phase (\textbf{iii.}) to a supersymmetric AdS vacuum (\textbf{a.}) via a Minkowski saddle, following the flow:
\begin{equation}
    f > 0 \to f = 0 \to f < 0, \quad \text{with} \quad \mathcal{K}_T = 0, \quad \mathcal{K}_{T \bar{T}} < 0, \quad g \to 0.
\end{equation}
The scalar potential reduces to:
\begin{equation}\label{eq:Brane_Dominated_Pot}
    V = \frac{3W_0^2}{\mathcal{K}^3} \left( \frac{|g|^2}{|g|^2 - f\mathcal{K}} - 1 \right),
\end{equation}
yielding the following phase classification:
\begin{itemize}
    \centering
    \item[] \textbf{dS:} $ |g|^2 > f\,\mathcal{K} > 0 $,
    \item[] \textbf{Minkowski:} $ f = 0,\; |g|^2 \neq 0 $,
    \item[] \textbf{AdS:} $ |g|^2 > 0 > f\,\mathcal{K} $, with \textbf{SUSY AdS} when $g = 0,\; f < 0$.
\end{itemize}

Direct exponential scalings, such as $f\mathcal{K} \sim e^{-\Delta\varphi}$, $g^2 \sim e^{-2\Delta\varphi}$, typically violate ghost-free conditions such as~\eqref{NG2}. A more robust and physically motivated approach is to introduce a sign-changing functional structure:
\begin{equation}\label{sc_func}
    f\mathcal{K} \sim |g|^2 - h, \quad g \to 0, \quad h > 0,
\end{equation}
leading to a simplified and transparent expression:
\begin{equation}
    V = \frac{3W_0^2}{\mathcal{K}^3} \left( \frac{|g|^2}{h} - 1 \right),
\end{equation}
where $g$ and $h$ capture brane and loop-corrected contributions.

\subparagraph{2. Volume-Dominated Restoration.}
In contrast, this pathway relies on the geometric volume dynamics encoded in the Kähler function. SUSY is restored through the vanishing of $\mathcal{K}_T$, rather than the brane coupling. The transition occurs via \textbf{iv.} $\to$ \textbf{b.}, following:
\begin{equation}
    \mathcal{K}_{T \bar{T}} > 0 \to 0 \to \mathcal{K}_{T \bar{T}} < 0, \quad \text{with} \quad \mathcal{K}_T \to 0,\; g = 0,\; f < 0.
\end{equation}
The potential becomes:
\begin{equation}
    V = \frac{3W_0^2}{\mathcal{K}^3} \left( \frac{|\mathcal{K}_T|^2}{|\mathcal{K}_T|^2 - \mathcal{K}\mathcal{K}_{T\bar{T}}} - 1 \right),
\end{equation}
and the associated phase diagram is:
\begin{itemize}
    \centering
    \item[] \textbf{dS:} $ |\mathcal{K}_T|^2 > \mathcal{K} \mathcal{K}_{T \bar{T}} > 0 $,
    \item[] \textbf{Minkowski:} $ \mathcal{K}_{T \bar{T}} = 0,\; \mathcal{K}_T \neq 0 $,
    \item[] \textbf{AdS:} $ |\mathcal{K}_T|^2 > 0 > \mathcal{K} \mathcal{K}_{T \bar{T}} $, with \textbf{SUSY AdS} at $ \mathcal{K}_T = 0,\; \mathcal{K}_{T \bar{T}} < 0 $.
\end{itemize}

To ensure ghost-freedom in this scenario, we propose:
\begin{equation}\label{sc_func2}
    \mathcal{K} \mathcal{K}_{T \bar{T}} \sim |\mathcal{K}_T|^2 - h, \quad \mathcal{K}_T \to 0, \quad h > 0,
\end{equation}
leading to the reduced potential:
\begin{equation}
    V = \frac{3W_0^2}{\mathcal{K}^3} \left( \frac{|\mathcal{K}_T|^2}{h} - 1 \right),
\end{equation}
with dynamics governed by the effective volume modulus $\mathcal{K} \sim \mathcal{V}^{2/3}$.

\vspace{0.5em}
Together, these mechanisms illustrate how SHS can be realized through either brane-sector suppression or geometric stabilization. Both paths admit smooth transitions across cosmological phases and are fully compatible with string-theoretic expectations for volume control and supersymmetry restoration.

\subsection{A String-Motivated Framework for Symmetric SUSY Restoration}\label{subsec5.1:String_Moti_SUSY_Resto}
Building upon earlier examples, we now present a pedagogically transparent and technically complete construction that realizes SUSY restoration along a controlled EFT trajectory. Our objective is to design a scalar potential that transitions smoothly through three cosmological phases: a metastable dS vacuum, an intermediate Minkowski saddle, and a supersymmetric AdS minimum, while maintaining EFT validity and compactification control.

This is achieved by employing a sign-changing nilpotent coupling structure of the form~\eqref{sc_func}, which ensures a ghost-free and stable transition. Imposing the simultaneous conditions $g(\tau = \tau_{0}) \to 0$ and $\mathcal{K}_T(\tau = \tau_{0}) \to 0$ simplifies the scalar potential and guarantees symmetric SUSY restoration on both sides of the vacuum boundary:
\begin{equation}\label{eq:Scalar_Pot}
    V = \frac{3 W_{0}^{2}}{\mathcal{K}^3}
    \left(
        \frac{\mathcal{K}_{T \overline{T}}(|g|^2 - h)}{\frac{|\mathcal{K}_{T}|^2}{\mathcal{K}}(|g|^2 - h) + \mathcal{K}_{T \overline{T}} h}
    \right).
\end{equation}

Among the viable phase trajectories introduced in Section~\S\ref{subsec3.3:ds_to_Minkowski}, we focus on the transition path characterized by $f > 0 \to f = 0 \to f < 0$ with $\mathcal{K}_{T \bar{T}} < 0$, describing a flow from dS to AdS through a Minkowski saddle. While both $g(\tau)$ and $\mathcal{K}_T(\tau)$ vanish at $\tau = \tau_{0}$, their interplay ensures that Eq.~\eqref{eq:Scalar_Pot} remains regular and non-singular. This preserves control over the EFT even for large canonical field excursions $\varphi \to \infty$, as $\tau$ remains bounded and approaches $\tau_{0}$ exponentially.

\paragraph{Motivation from String Compactifications.}
In type IIB string theory, the tree-level K\"ahler potential for the overall volume modulus $T = \tau + i\theta$ is classically given by the no-scale form~\cite{Cremmer:1983bf}:
\begin{equation}
    K_{\text{tree}} = -3 \ln(T + \bar{T}), \qquad \mathcal{V} = (T + \bar{T})^{3/2}.
\end{equation}
However, realistic compactifications receive both $ \alpha'^3 $ and loop corrections:
\begin{itemize}
    \item \textbf{$\alpha'^3$ Corrections:} The leading $\alpha'^3$ correction to the Kähler potential, when combined with the tree-level contribution, enter as~\cite{Becker:2002nn,Balasubramanian:2005zx}:
    \begin{equation}
        K = -2 \ln\left(\mathcal{V} + \frac{\Xi}{2 g_s^{3/2}}\right),
    \end{equation}
    with $\Xi \propto -\chi(X_3) \zeta(3)$ encoding topological curvature corrections. Here, $\chi(X_3)$ is the Euler characteristic of the Calabi–Yau threefold $X_3$, $\zeta(3)\approx 1.202$ is the Riemann zeta function evaluated at 3, and $g_s$ denotes the string coupling constant.

    \item \textbf{Loop Corrections:} Perturbative string loop corrections at orders 
$\mathcal{O}(g_s^2 \alpha'^2)$ and $\mathcal{O}(g_s^2 \alpha'^4)$
also contribute to the Kähler potential by mediating exchanges of closed Kaluza--Klein and winding strings between parallel or intersecting stacks of branes and O-planes~\cite{Berg:2004ek,Berg:2005ja,Berg:2007wt}. In LVS, these loop corrections typically take the schematic form~\cite{Berg:2007wt,Cicoli:2007xp,Cicoli:2024bxw}:
\begin{equation}
    \delta K_{\text{loop}} = \frac{g_s C_{\text{loop}}}{\mathcal{V}^{2/3}},
\end{equation}
where $C_{\text{loop}}$ encapsulates geometric and topological information related to the underlying compactification.
\end{itemize}
Combining these effects, the effective Kähler potential becomes:
\begin{equation}
    K = -2 \ln\left(\mathcal{V} + \frac{\xi}{2g_s^{3/2}}\right) + \frac{g_s C_{\text{loop}}}{\mathcal{V}^{2/3}}.
\end{equation}

\paragraph{Constructing the Effective Model.} 
In the following, we construct this scenario systematically, considering a general corrected Kähler potential:
\begin{equation}
K = -3\ln[\mathcal{K}(T + \bar{T})],
\end{equation}
with $\mathcal{K}(T + \bar{T})$ encompassing all correction effects. Assuming stabilization at the real slice $ T = \bar{T} \to \tau_{0} \in \mathbb{R} $, and parameterizing fluctuations around this stabilized point as:
\begin{equation}
T = \tau_{0} + e^{-\sqrt{\frac{2}{3} }\varphi}/\xi,
\end{equation}
the function $\mathcal{K}$ near this point can be expanded as a Taylor series:
\begin{equation}
    \mathcal{K}(T + \bar{T}) = \mathcal{K}(2\tau_{0}) + \mathcal{K}^{\prime}(2\tau_{0})\bigl[(T + \bar{T}) - 2\tau_{0}\bigr] + \frac{1}{2}\mathcal{K}^{\prime\prime}(2\tau_{0})\bigl[(T + \bar{T}) - 2\tau_{0}\bigr]^2 + \dots
\end{equation}
Given the physical naturalness of having an extremum along the real direction in no-scale models, it follows that
\begin{equation}
    \mathcal{K}^{\prime}(2\tau_{0}) = 0.
\end{equation}
To maintain consistency with the no-scale framework and simplify the analysis, we choose the normalization condition:
\begin{equation}
    \mathcal{K}(2\tau_{0}) = \mathcal{V_\text{0}}^{2/3}.
\end{equation}
Thus, to leading order, the function $\mathcal{K}(T + \bar{T})$ around the stabilized point is approximated as~\footnote{The truncation of $\mathcal{K}$ at quadratic order is valid in the regime where the canonical field $\varphi$ induces only exponentially small deviations in $\tau$. This ensures that higher-order corrections in $(T + \bar{T} - 2\tau_{0})$ remain negligible even as $\varphi \to \infty$, consistent with large-volume control in type IIB flux compactifications.}
:
\begin{equation}
    \mathcal{K}(T + \bar{T}) \approx \mathcal{V_\text{0}}^{2/3} + \frac{1}{2}\mathcal{K}^{\prime\prime}(2\tau_{0})\bigl[(T + \bar{T}) - 2\tau_{0}\bigr]^2.
\end{equation}
Given the conformal coupling function (\ref{eq:Conformal_Coupling_Func}):
\begin{equation}\label{Coupling_Func_T}
    \Omega(T) = \xi(T - \tau_{0}), \qquad \Omega(\tau) = e^{-\sqrt{2/3}\varphi},
\end{equation}
the Kähler function can be written as
\begin{equation}\label{Kahler_Func_T}
     \mathcal{K} = \mathcal{V_\text{0}}^{2/3} + \varepsilon\bigl[\Omega(T) + \Omega(\overline{T})\bigr]^2, \qquad \mathcal{K} = \mathcal{V_\text{0}}^{2/3} + 4\varepsilon\Omega(\tau)^2,
\end{equation}
where we defined the relation
\begin{equation}
    \varepsilon = \frac{\mathcal{K}^{\prime\prime}(2\tau_{0})}{2\xi^2}.
\end{equation}
This implies the second derivative becomes:
\begin{equation}
    \mathcal{K}^{\prime\prime}(2\tau_{0}) = 2\varepsilon\xi^2
\end{equation}
and must be negative to ensure $\mathcal{K}_{T \bar{T}} = \mathcal{K}^{\prime\prime}(2\tau_{0}) < 0$, thus requiring $\varepsilon < 0$. Consequently, $\varepsilon$ parameter encodes the impact of string-derived perturbative effects.

This concise yet robust construction explicitly encapsulates the crucial features required for asymptotic SUSY restoration along the selected physical pathway—specifically, the vanishing of the first derivative and the negativity of the second derivative at the stabilization point. While our analysis focuses on the single Kähler modulus case, the structure readily generalizes to multiple moduli settings, where similar corrections play a central role in stabilizing geometric moduli consistent with EFT control. Consequently, this analysis reveals a profound connection between phenomenological moduli stabilization scenarios and fundamental corrections arising from type IIB string theory, highlighting how fundamental string-theoretic corrections can be coherently integrated into phenomenological models of supersymmetric horizon restoration.

\subsection{From Toy Models to String-Based Realizations of SUSY Restoration}\label{subsec5.2:Toy_to_String}

We now present explicit constructions based on the corrected Kähler potential incorporating perturbative $\alpha'^3$ and string loop effects, as introduced in Eq.~\eqref{Kahler_Func_T}. These constructions include both toy and exact models that robustly exhibit all three key cosmological phases: a metastable dS plateau, a Minkowski-like intermediary region, and a supersymmetric AdS minimum. The models achieve this within the framework of the parameterization given in~\eqref{sc_func}, and with a naturally small cosmological constant in the dS phase.

In the asymptotic large-field limit $\varphi \to \infty$, the modulus $\tau \to \tau_{0}$ and the brane coupling vanishes as $g \to 0$. This behavior enforces the condition $f \propto (|g|^2 - h)/\mathcal{K} < 0$, corresponding to the supersymmetric AdS vacuum. The scalar potential~\eqref{eq:Scalar_Pot} then stabilizes at
\begin{equation}
    V_{(\varphi \to \infty)} = -\frac{3 W_0^2}{\mathcal{V}_0^2},
\end{equation}
matching the energy scale of SUSY restoration as defined in Eq.~\eqref{eq:SUSY_Restoring_AdS}. It is convenient to denote this as
\begin{equation}\label{Inf_Energy_Scale}
    V_0 \equiv \frac{3 W_0^2}{\mathcal{V}_0^2}.
\end{equation}

In the opposite limit $\varphi \to 0$ (implying $\tau \to \tau_{0} + \xi^{-1}$), the potential simplifies to
\begin{equation}
    V_{(\varphi \to 0)} = \frac{3 W_0^2}{\mathcal{K}^2} \left(\frac{|g|^2 - h}{8 \varepsilon (|g|^2 - h) + h \mathcal{K}}\right).
\end{equation}
To obtain a positive and small vacuum energy (i.e., a metastable dS vacuum), the condition $|g|^2 \gtrsim h$ must be satisfied. Assuming a large internal volume ($\varepsilon \ll \mathcal{V}_0^{2/3}$), the expression further simplifies to
\begin{equation}
    V_{(\varphi \to 0)} \approx V_0 \left(\frac{|g|^2}{h} - 1\right).
\end{equation}

This simplified approximation remains valid across the entire field range when the following condition is met:
\begin{equation}\label{eq:General_Approximation}
    \mathcal{V}_0^{2/3} \gg 8 \varepsilon \left(\frac{|g|^2}{h} - \frac{3}{2}\right).
\end{equation}
Under this constraint, the scalar potential universally reduces to
\begin{equation}\label{eq:V_approx}
    V(\varphi) \approx V_0 \left(\frac{|g|^2}{h} - 1\right),
\end{equation}
matching the earlier brane-dominated SUSY restoration form given in Eq.~\eqref{eq:Brane_Dominated_Pot}, but now generalized to include nonzero $\mathcal{K}_T$, consistent with the SUSY horizon scenario.

\paragraph{Effective Potentials from Brane Couplings and Corrections.}

\begin{figure}[htb]
\includegraphics[width=\textwidth, height=\textwidth, keepaspectratio]{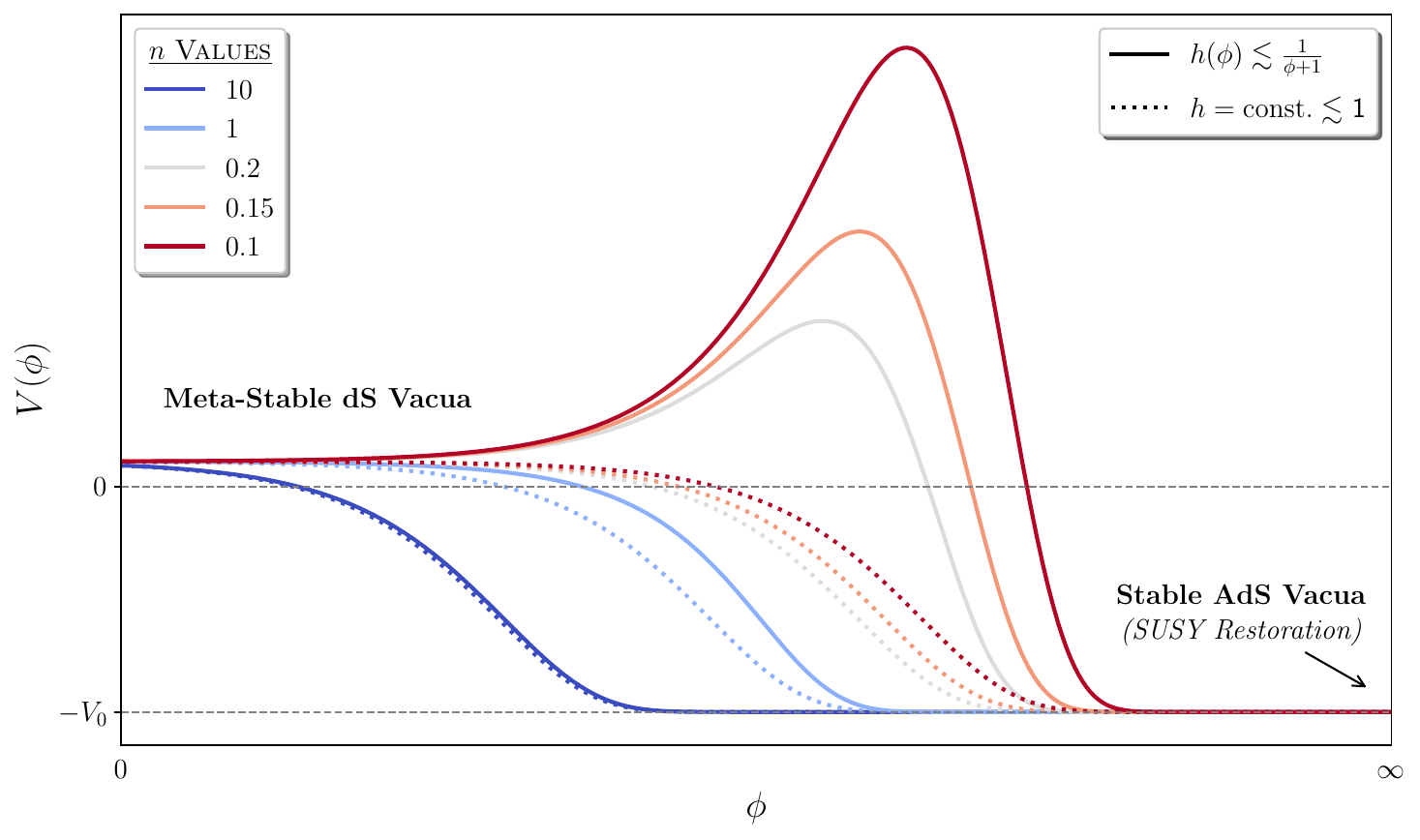}
\caption{\label{fig:1}%
Scalar potential $V(\varphi)$ for the toy model with various $n$ values. Solid curves use scalar-dependent $h(\varphi)$; dotted curves use constant $h$. The form of $g(\varphi)$ controls the sharpness and lifetime of the dS phase.}
\end{figure}

As a preliminary step toward string-theoretic embeddings, we examine a toy model using simple functional ansätze for the brane couplings. Consider the exponential form:
\begin{equation}\label{eq:Test_Function1}
    g(\tau) \propto \Omega(\tau)^n \equiv e^{-n\sqrt{\frac{2}{3}}\,\varphi},
\end{equation}
with $n > 0$. This yields $g(\varphi = 0) = 1$, and a viable dS vacuum can be realized by choosing:
\begin{equation}
    h(\varphi) \lesssim \frac{1}{\varphi + 1}, \quad \text{or simply} \quad h = \text{constant} \lesssim 1.
\end{equation}
Figure~\ref{fig:1} illustrates the scalar potential evolution under these assumptions.

Key features include:
\begin{itemize}
  \item[\textbf{i.}] Constant $h$ leads to extended dS phases without a potential barrier; scalar-dependent $h \sim (\varphi + 1)^{-1}$ introduces an explicit barrier.
  \item[\textbf{ii.}] Smaller $n$ values in $g(\varphi)$ enhance both the height of the barrier and the longevity of the dS phase.
\end{itemize}

These analytically tractable toy models offer more than illustrative simplicity—they reveal the essential ingredients needed to realize multi-phase cosmological dynamics within a controlled EFT. The interplay between exponential brane couplings and perturbative corrections not only enables the emergence of a metastable de Sitter phase and its graceful transition to a supersymmetric AdS vacuum, but also highlights how seemingly minimal assumptions—such as $g(\varphi) \sim e^{-n\varphi}$ and scalar-dependent $h(\varphi)$—can encode rich vacuum structures. Crucially, the shapes and lifetimes of dS phases are governed by tunable parameters like $n$ and $h$, making these constructions directly relevant for connecting UV-motivated inflationary mechanisms to phenomenological observables such as the spectral index and tensor-to-scalar ratio.

Moreover, the structural robustness of these models—particularly the exponential localization of SUSY restoration and the natural suppression of runaway behavior—strongly suggests their generalizability. They provide a natural stepping stone toward more realistic string-inspired cosmology scenarios that implement the SHS mechanism. The flexibility of the framework means that a broad family of cosmological models can be constructed by varying coupling functions or compactification data, while still maintaining analytic control and capturing essential features expected from string-motivated supergravity constructions.

\paragraph{Embedding Inflationary Dynamics in Type IIB Geometry.}
Motivated by the analytic insight gained from our toy model, we now present a concrete realization of the scalar potential within the framework of Type IIB flux compactifications. Our goal is to specify forms for the brane coupling functions $g(\tau)$ and $h(\varphi)$ that are consistent with both moduli stabilization and string-theoretic corrections.

These functions must satisfy two key criteria:
\begin{itemize}
  \item They must allow for a long-lived de Sitter phase and dynamically connect to a supersymmetric AdS vacuum at large $\varphi$,
  \item They must incorporate perturbative string corrections, especially from loops and $\alpha'^3$ terms, that deform the Kähler potential and scalar potential through moduli-dependent prefactors.
\end{itemize}

Accordingly, and consistent with the approximation condition~\eqref{eq:General_Approximation}, we adopt:
\begin{equation}\label{good_choice}
\begin{aligned}
  g(\tau) &= \mathcal{V}_0^{1/3}\,\Omega(\tau)^{-\varepsilon} = \mathcal{V}_0^{1/3} e^{\varepsilon\sqrt{\frac{2}{3}}\,\varphi}, \\[5pt]
  h(\varphi) &= \frac{\mathcal{K}}{1 + s(\varphi)},
\end{aligned}
\end{equation}
where $\varepsilon$ quantifies the magnitude of subleading corrections that break no-scale structure. The function $s(\varphi)$ can be chosen to reproduce known inflationary potentials, e.g., $V_{\text{inf}}(\varphi) = V_0\, s(\varphi)$, while $g(\tau)$ captures the modulus dependence of brane couplings consistent with geometric suppression at strong coupling.

Although these forms are not uniquely determined—particularly due to the nilpotency of the goldstino superfield, which imposes minimal constraints—they serve to analytically capture the suppression of brane couplings at strong-coupling regimes (i.e., large $\varphi$), as expected in realistic Type IIB flux compactifications. As a result, the precise value of $\varepsilon$ and the inflationary profile $s(\varphi)$ can be adapted to match top-down models. This flexibility provides a promising bridge between bottom-up EFT constructions and explicit string compactifications, where loop and $\alpha'$ corrections can be computed from first principles.

\paragraph{Moduli-Coupled Inflaton Potential.}
Inserting these into the scalar potential~\eqref{eq:V_approx} yields a remarkably transparent analytic form:
\begin{equation}\label{eq:modelPotential}
    V(\varphi) = \bigl(V_0 + V_{\text{inf}}(\varphi)\bigr)\,A(\varphi) - V_0,
\end{equation}
with
\begin{equation}
    V_0 \equiv \frac{3 W_0^2}{\mathcal{V}_0^2}, \qquad
    A(\varphi) = \frac{g(\tau)^2}{\mathcal{K}} = \frac{\mathcal{V}_0^{2/3} e^{\varepsilon \sqrt{\tfrac{8}{3}}\,\varphi}}{\mathcal{V}_0^{2/3} + 4\varepsilon\, e^{-\sqrt{\tfrac{8}{3}}\,\varphi}}.
\end{equation}
The prefactor $A(\varphi)$ acts as a warp-like factor, capturing how the volume modulus $\tau(\varphi)$ responds to displacements of the canonical inflaton. It controls the effective coupling between the inflaton and the compactification geometry, and shapes the potential in a physically meaningful way.

Note that, as expected in the analytically transparent limit $\varepsilon \to 0^{-}$ where all the string-theoretic corrections vanish with $A(\varphi) \to 1$, the scalar potential simplifies to the well-known form:
\begin{equation}
    V(\varphi)_{\varepsilon \to 0^{-}} \approx V_{\text{inf}}(\varphi),
\end{equation}
recovering standard inflationary dynamics across all field values.

\paragraph{Asymptotic Behavior.}
This potential exhibits a smooth and physically meaningful interpolation between two asymptotic regimes:
\begin{itemize}
  \item \textbf{Small-field regime} $(\varphi \ll 1$): Exponential terms in $A(\varphi)$ are negligible, so $A(\varphi) \gtrsim 1$. The potential simplifies to
  \begin{equation}
      V(\varphi) \approx V_{\text{inf}}(\varphi) + \Lambda,
  \end{equation}
  where the uplift term $\Lambda$ is a small positive constant determined by $\varepsilon, W_0,$ and $\mathcal{V}_0$. This regime supports slow-roll inflation in a nearly flat de Sitter phase.

  \item \textbf{Large-field regime} ($\varphi \to \infty$): The exponential suppression in $A(\varphi) \sim e^{\varepsilon \sqrt{8/3}\,\varphi} \to 0$ becomes dominant (for $\varepsilon < 0$). As a result, the potential asymptotes to
  \begin{equation}
       V(\varphi) \to -V_0,
  \end{equation}
  reflecting a supersymmetric AdS vacuum determined by the flux sector. This exponential falloff ensures that the modulus stabilizes dynamically as $\varphi$ increases, avoiding uncontrolled runaways.
\end{itemize}
This dual behavior positions~\eqref{eq:modelPotential} as the central effective potential of the SHS framework. It naturally accommodates both slow-roll inflation in a stabilized moduli background and a graceful transition to a vacuum determined by string-theoretic inputs. The simplicity and analytic accessibility of this form are instrumental for both qualitative insight and consistent model-building in the string landscape. 

\begin{figure}[htb]
\centering
\includegraphics[width=\textwidth, height=\textwidth, keepaspectratio]{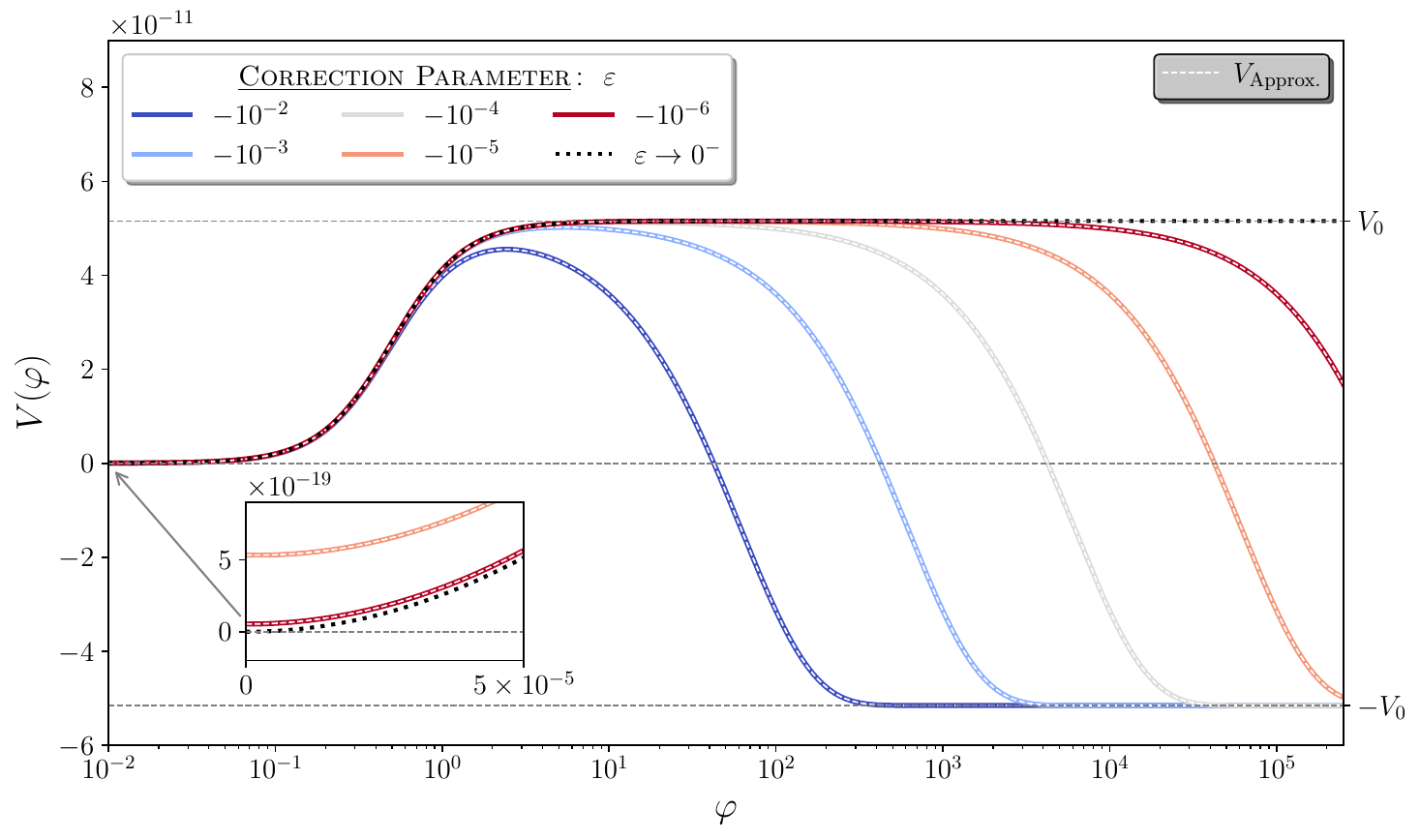}
\caption{\label{fig:2}%
Scalar potential $V(\varphi)$ plotted as a function of the canonical inflaton field. Solid curves represent the exact expression from Eq.~\eqref{eq:Scalar_Pot} for various values of the correction parameter $\varepsilon$, while the dashed (white) curves show the approximate potential from Eq.~\eqref{eq:V_approx}. The dotted line corresponds to the limiting inflationary potential $V(\varphi)_{\varepsilon \to 0^{-}} \approx V_{\text{inf}}(\varphi)$, chosen here as the KKLTI form $(V_{\text{inf}}(\varphi) = V_0\,\varphi^2 / (\varphi^2 + 1/4)$~\cite{Kachru:2003sx,Kallosh:2019eeu}. The inset zooms in on the region near $\varphi = 0$, illustrating how perturbative corrections shift the potential structure relative to the uncorrected case. For the representative parameter choices $W_0 = 1$, $\mathcal{V}_0 = 2.412 \times 10^5$, and small corrections $|\varepsilon| < 10^{-5}$, the model yields approximately 60 e-folds of inflation and remains in agreement with current CMB constraints~\cite{Planck:2018jri,Planck:2018vyg,BICEP:2021xfz} on the spectral index $n_s$ and the tensor-to-scalar ratio $r$.}
\end{figure}

Figure~\ref{fig:2} illustrates the full analytic and numerical behavior of the scalar potential across relevant regimes, including the uplifted dS phase near $\varphi = 0$ and the supersymmetric AdS vacuum at large $\varphi$.

\paragraph{dS Vacuum and Cosmological Constant.}
At $\varphi = 0$, and assuming $V_{\text{inf}}(0) = 0$, the scalar potential becomes:
\begin{equation}\label{Cosmological_consty}
V(0) \approx \frac{-12 \,\varepsilon \,W_0^2 }{\mathcal{V}_0^{8/3}} \equiv \Lambda,
\end{equation}
where we used $|\varepsilon| \ll \mathcal{V}_0^{2/3}$. This describes a metastable de Sitter vacuum, uplifted by subleading corrections to the Kähler potential~\cite{Berg:2005ja,Cicoli:2007xp,Becker:2002nn}.

Matching the observed vacuum energy $\Lambda_{\text{obs}} \sim 10^{-120}$ gives:
\begin{equation}
\varepsilon \approx -\frac{\Lambda_{\text{obs}}\,\mathcal{V}_0^{8/3}}{12\,W_0^2}.
\end{equation}
For benchmark values $W_0 = 1$, $\mathcal{V}_0 = 2.412 \times 10^5$, this yields:
\begin{equation}
\varepsilon \approx -2.18 \times 10^{-107}.
\end{equation}

Although small, such a value is technically natural: $\varepsilon$ arises only through perturbative corrections. It may originate from:
\begin{itemize}
  \item \emph{String loop effects}, scaling with $g_s^2 / \mathcal{V}$~\cite{Berg:2007wt},
  \item \emph{Higher-derivative curvature terms}, such as the $\alpha'^3 \mathcal{R}^4$ corrections~\cite{Becker:2002nn},
  \item \emph{Warped throats or localized sources}, which can exponentially suppress such terms~\cite{Giddings:2001yu}.
\end{itemize}

\paragraph{Full Supergravity Construction.}
The following Kähler potential defines the complete underlying supergravity model:
\begin{equation}\label{Full_Sugra_Kahler_SHS}
K(T,\bar{T},S,\bar{S}) =\;  -3 \log\Bigg[
\mathcal{K}(T,\bar{T}) +
\left(
\frac{g(T)\bar{g}(\bar{T})}{\mathcal{K}(T,\bar{T})} - \frac{1}{1 + s(\varphi)}
\right) S\bar{S}  + g(T) S + \bar{g}(\bar{T}) \bar{S}
\Bigg],
\end{equation}
which, in the real direction $T = \bar{T} = \tau$, reduces to:
\begin{equation}
K(\tau, S, \bar{S}) =\;  -3 \log\Bigg[
\mathcal{K}(\tau) +
\Bigg(
\underbrace{\frac{\mathcal{V}_0^{2/3} \, \Omega(\tau)^{-2\varepsilon} }{\mathcal{K}(\tau)}}_{\text{(meta)stable dS}}
-
\underbrace{\frac{1}{1 + s(\varphi)}}_{\text{Inflation}}
\Bigg) S\bar{S} +\,
\underbrace{\mathcal{V}_0^{1/3} \, \Omega(\tau)^{-\varepsilon}}_{\text{SUSY /}\cancel{\text{SUSY}}} (S + \bar{S})
\Bigg].
\end{equation}
We adopt the following Kähler function, as defined in~\eqref{Kahler_Func_T}:
\begin{equation}
     \mathcal{K}(T, \overline{T}) = \mathcal{V}_0^{2/3} + \varepsilon \left[ \Omega(T) + \Omega(\overline{T}) \right]^2, \quad \mathcal{V}_0 > 0, \quad \varepsilon < 0,
\end{equation}
in conjunction with the conformal factor introduced in~\eqref{Coupling_Func_T}:
\begin{equation}
    \Omega(T) = \xi (T - \tau_{0}), \quad \xi > 0,
\end{equation}
where $\Omega(\tau)$ can be re-expressed in terms of the canonically normalized scalar field $\varphi$ as:
\begin{equation}
    \Omega(\tau) = \exp\left(-\sqrt{\tfrac{2}{3}}\,\varphi\right).
\end{equation}
We further employ the parameterization~\eqref{sc_func}:
\begin{equation}
f \,S \bar{S}= \left(\frac{|g|^2 - h}{\mathcal{K}}\right)\,S \bar{S}, \qquad g \to 0, \qquad h > 0,
\end{equation}
together with the coupling choices in~\eqref{good_choice}, to support all three cosmological regimes—namely, inflation, a metastable dS vacuum, and a supersymmetric AdS horizon.~\footnote{The coupling functions specified in~\eqref{good_choice} are not unique; any functional forms consistent with the constraint~\eqref{sc_func} can be employed to construct alternative realizations of three-phase cosmological scenarios.}

The superpotential is constant:
\begin{equation}
W = W_0,
\end{equation}
and the nilpotent superfield $S$ satisfies $S^2 = 0$, removing the scalar component and focusing solely on its SUSY-breaking/uplifting auxiliary role.

\paragraph{Conceptual Structure and Distinctions.}
The SHS framework offers a structurally unified and parametrically flexible approach to embedding inflationary and vacuum dynamics within effective string theory. In contrast to conventional KKLT~\cite{Kachru:2003aw} and LVS~\cite{Balasubramanian:2005zx,Conlon:2005ki} constructions—which typically require fine-tuning $ W_0 \ll 1 $ to uplift an AdS vacuum—the SHS setup supports inflationary dynamics and a small cosmological constant without unnatural constraints on the flux superpotential $ W_0 $. 

The inflationary energy scale—governing both the dynamics of inflation and the asymptotic AdS vacuum—is set by the geometric combination
\begin{equation}
    V_0 \equiv \frac{3 W_0^2}{\mathcal{V}_0^2},
\end{equation}
allowing for a broad range of $W_0$ values, as long as the compactification volume $\mathcal{V}_0$ is sufficiently large to ensure $\mathcal{V}_0 \gg |\varepsilon|$, preserving perturbative control. This parametric freedom enables consistent tuning of the scalar potential without invoking hard hierarchies or UV-sensitive uplift ingredients.

Beyond this flexibility, SHS departs from the conventional \textit{two-step structure} of KKLT/LVS, wherein a supersymmetric AdS vacuum is first engineered and subsequently uplifted using additional brane or D-term ingredients~\cite{Kachru:2003aw,Balasubramanian:2005zx,Kallosh:2014wsa}. In many such constructions, the uplift effectively \textbf{masks the dynamical connection} to the original AdS vacuum, which becomes inaccessible in the post-uplift theory. This feature has motivated recent critiques from both swampland and holographic perspectives~\cite{Moritz:2017xto,Lust:2022lfc}.

By contrast, the SHS mechanism generates a \textbf{dynamically continuous scalar potential} that interpolates between three physically distinct regimes: a slow-roll inflationary phase, a metastable de Sitter vacuum, and a supersymmetric AdS minimum. These phases emerge organically from a single effective potential governed by string-theoretic corrections and moduli-coupled brane interactions. In particular, the AdS regime remains \textbf{dynamically accessible} in the large-field limit, suggesting an asymptotically supersymmetric regime where gravitational backreaction is under control.

While we do not claim a full UV completion—which would require explicit realization within a globally consistent Calabi–Yau compactification with stabilized complex-structure moduli and axio-dilaton—the SHS setup is nonetheless \textbf{UV-motivated}, \textbf{geometrically transparent}, and built from ingredients compatible with known string corrections such as $\alpha'^3 \mathcal{R}^4$ terms~\cite{Becker:2002nn} and string loop effects~\cite{Berg:2005ja,Cicoli:2007xp}.

Altogether, this framework offers a conceptually minimal and analytically tractable alternative to staged uplifting constructions. It supports realistic inflationary dynamics, avoids severe parameter tuning, and retains structural access to the supersymmetric sector—thereby providing a new toolset for embedding early-universe cosmology in a geometrically controlled and analytically transparent supergravity framework.

\subsection{Decay Rate of the Metastable Vacuum}\label{subsec5.4:Decay_Rate}
A key aspect of the analysis involves computing the decay rate of the metastable dS vacuum to the true AdS vacuum. From a physical standpoint, this decay must be sufficiently suppressed to ensure that the de Sitter phase persists over cosmological timescales. The theoretical framework for such quantum vacuum transitions was first developed in~\cite{Coleman:1977py, Callan:1977pt}, where the decay of a scalar field from a false (metastable) vacuum to a true vacuum was formulated in flat spacetime. This formalism was later extended to include gravitational effects in~\cite{Coleman:1980aw}, where the decay rate is computed semiclassically using a Euclidean (i.e., Wick-rotated) path integral centered on the so-called bounce solution.

In the late 1980s and early 1990s, an alternative formulation was developed by Fischler, Morgan, and Polchinski (FMP)~\cite{Fischler:1989se, Fischler:1990pk}, offering a Hamiltonian approach to vacuum decay. In contrast to the Euclidean treatment of~\cite{Coleman:1980aw}, the FMP framework operates directly in Lorentzian signature and avoids analytic continuation via Wick rotation. In what follows, we adopt this method to compute the decay rate in our model, as it is particularly well suited to scenarios with time-dependent backgrounds and real-time evolution.

\paragraph{CdL Mechanism.} We begin by reviewing the Euclidean approach to vacuum decay, often referred to as the Coleman–de Luccia (CdL) mechanism. Consider a single scalar field minimally coupled to gravity, described by the action
\begin{equation}
    S = \int \dd[4]{x} \sqrt{-g} \left[\frac{R}{2 \kappa}-\frac{1}{2} g^{\mu \nu}\partial_{\mu}{\varphi} \partial_{\nu}{\varphi} - V(\varphi) \right], \label{eq:mainTransitionAction}
\end{equation}
where $V(\varphi)$ denotes the scalar potential, $g_{\mu \nu}$ is the spacetime metric, $R$ is the Ricci scalar, and $\kappa = 8 \pi G$. Performing a Wick rotation to Euclidean signature
\begin{equation}
    S_E = \int \dd[4]{x} \sqrt{\tilde{g}}\left[\frac{R_E}{2 \kappa} -\frac{1}{2} \tilde{g}^{\mu \nu}\partial_{\mu}{\varphi} \partial_{\nu}{\varphi} + V(\varphi)\right]
\end{equation}
and assuming an $O(4)$-symmetric ansatz, we can obtain the equations of motion by varying the Euclidean action with respect to scalar field and the metric. These equations take the form:
\begin{equation}
    \begin{aligned}
        \varphi'' + 3 \frac{a'}{a}\varphi' -\dv{V(\varphi)}{\varphi} &= 0, \\
        (a')^2 - 1 -  \frac{\kappa}{3} \left[ \frac{1}{2}(\varphi')^2 -V(\varphi)\right] &= 0,
    \end{aligned}
\end{equation}
where $a(\xi)$ denotes the radial scale factor, and primes denote derivatives with respect to the Euclidean radial coordinate $\xi$. The bounce solution to this system corresponds to a non-trivial field configuration interpolating between the false vacuum $\varphi_{\text{fv}}$ and the true vacuum $\varphi_{\text{tv}}$. The semiclassical tunneling exponent is given by the difference in Euclidean action,
\begin{equation}
    B = S_E\left[\varphi_{\text{tv}}\right] - S_E\left[\varphi_{\text{fv}}\right],
\end{equation}
and the decay rate per unit volume is approximately
\begin{equation}
    \Gamma/V \sim e^{-B/\hbar}.
\end{equation}

This formalism was developed in~\cite{Coleman:1980aw} and has since been widely applied, including in studies of vacuum metastability in the Standard Model~\cite{Isidori:2007vm} and in analyzing the effects of gravitational backgrounds such as black holes on Higgs vacuum decay~\cite{Gregory:2013hja, Gregory:2016xix, Burda:2015yfa}. Although solving the coupled field and geometry equations can be numerically challenging, various efficient methods have been developed~\cite{Herodotou:2021ker}.

\paragraph{Hamiltonian (FMP) Approach.} The main difference of the FMP method lies in its real-time description of the decay process, formulated within the Hamiltonian framework, in contrast to the Euclidean approach employed in the CdL formalism \cite{Fischler:1989se,Fischler:1990pk}. Another important difference is that the CdL formalism requires a series of analytic continuations, often extending beyond the scope justified by WKB quantum mechanics, to reconstruct the Lorentzian geometry \cite{Cespedes:2020xpn}.

Now, let us consider the same action in \eqref{eq:mainTransitionAction}, a homogeneous and isotropic metric and a scalar field
\begin{equation}
    \begin{aligned}
        \dd{s^2} &= - \dd{t^2} + a^2(t) \left[\dd{\chi^2 +\sin^2{(\chi)}\left(\dd{\theta^2} +\sin^2{(\theta)}\dd{\phi^2}\right)}\right], \\
        \varphi &= \varphi(t).
    \end{aligned}
\end{equation}
Imposing these into \eqref{eq:mainTransitionAction}, we obtain
\begin{equation}
    S = \int \dd{t} 2 \pi^2 a^3 \left[\frac{R}{2\kappa} + \frac{1}{2}\dot{\varphi}^2 -V(\varphi)\right].
\end{equation}
Using this action, we can simply find the Hamiltonian as
\begin{equation}
    \begin{aligned}
        \mathcal{H} &= \pi_{\varphi} \dot{\varphi} + \pi_{a} \dot{a} - \mathcal{L}\\
        &= - \frac{6 \pi^2}{\kappa} a \left[\dot{a}^2 +1 -\frac{1}{3}\kappa a^2 \left(\frac{1}{2}\dot{\varphi}^2+V(\varphi)\right)\right],
    \end{aligned}
\end{equation}
where 
\begin{equation}
    \begin{aligned}
        \pi_{\varphi} &= -i \pdv{}{\varphi}, \\
        \pi_{a} &= -i \pdv{}{a}
    \end{aligned}
\end{equation}
in the operator forms. So the Wheeler-DeWitt equation take the form of
\begin{equation}
    \mathcal{H} \Psi = \left[-\frac{1}{4 \pi^2 a^3} \pdv[2]{}{\varphi} + \frac{\kappa}{24 \pi^2 a}\pdv[2]{}{a} -\frac{6 \pi^2}{\kappa}a + 2 \pi^2 a^3V(\varphi)\right] \Psi = 0, \label{eq:WheelerDeWittEqn}
\end{equation}
which is also a Hamiltonian constraint \cite{Kristiano:2018oyv}. Let us now define a general solution 
\begin{equation}
    \Psi(a, \varphi) = A(a, \varphi) e^{-i \Phi(a, \varphi)}.
\end{equation}
Imposing this into \eqref{eq:WheelerDeWittEqn}, and using the WKB-approximation, we can obtain the Hamilton-Jacobi (HJ) equation for this system \cite{Kristiano:2018oyv}. Solving the HJ equation gives us 
\begin{equation}
    \Phi = \int \left[-\frac{1}{6}\kappa a^2 \dd{\phi^2} + \dd{a^2}\right]^{1/2}\left[-2 U(\varphi, a)\right]^{1/2}, \label{eq:solHJEqn}
\end{equation}
where 
\begin{equation}
    U(\varphi, a) = \frac{1}{2}\left(\frac{12\pi^2}{\kappa}\right)^2 a^2 \left[1-\frac{1}{3}\kappa a^2 V(\varphi)\right].
\end{equation}

Now, let us define $\alpha \equiv a \sqrt{\kappa/3}$ and $\varphi = \varphi(\alpha)$. Thus we have
\begin{equation}
    \begin{aligned}
        \dd{a} &= \sqrt{\frac{3}{\kappa}} \dd{\alpha}, \\
        \dd{\varphi} &= \varphi^{\prime} \dd{\alpha},
    \end{aligned}
\end{equation}
where $\varphi^{\prime}$ denotes $\dv{\varphi}{\alpha}$. Imposing in \eqref{eq:solHJEqn}, we obtain,
\begin{equation}
    \Phi = \int \dd{\alpha} \left(\frac{12\pi^2}{\kappa}\right) \sqrt{\left(\varphi^{\prime}\right)^2 - \frac{\kappa}{6}} \sqrt{\frac{3 \alpha}{2 \kappa}} \sqrt{1-\alpha V(\varphi)}, \label{eq:exponentialParameter}
\end{equation}
where $\varphi$ and $\alpha$ correspond to stationary configurations that extremize the action, i.e $\var{\Phi} = 0$. Thus we obtain
\begin{equation}
\varphi^{\prime\prime} = \left[ \frac{3}{\kappa} \, \partial_\varphi V(\varphi) - \frac{\varphi^{\prime}}{\alpha} (1 - 2 \alpha^2 V(\varphi)) \right]
\left( \frac{1 - \frac{1}{6} \kappa \alpha^2 (\varphi^{\prime})^2}{1 - \alpha^2 V(\varphi)} \right)
- \frac{\varphi^{\prime}}{\alpha} \left( 2 - \frac{1}{6} \kappa \alpha^2 (\varphi^{\prime})^2 \right) \label{eq:phiDiffEq}
\end{equation}
\cite{Kristiano:2018oyv}.
Solving this equation with \eqref{eq:modelPotential}, we can find $\varphi$ and $\varphi^{\prime}$. Thus we can calculate \eqref{eq:exponentialParameter}, which is proportional to the tunneling probability $P \sim e^{-2\abs{\Phi}/\hbar}$ \cite{Kristiano:2018oyv}.

\paragraph{Numerical Calculations.}
Substituting Eq.\eqref{eq:modelPotential} into Eq.\eqref{eq:phiDiffEq} yields a differential equation governing the evolution of $\varphi$ as a function of $\alpha$. For illustrative purposes, we alter the correction parameter from $\epsilon = -10^{-2}$ to $\epsilon = -10^{-6}$. The differential equation was solved numerically using a Runge-Kutta method with boundary conditions $\alpha_{\text{false}} = 2.365\times10^{-5}$, $\alpha_{\text{true}} = 2.365\times10^{-105}$ and $\varphi(\alpha_{\text{false}}) = 0$, ensuring a well-posed initial value problem. The solution to this equation exhibits a monotonically increasing behavior in $\varphi$, in agreement with the expected behavior. The monotonic increase in $\varphi$ reflects a smooth transition from the false to the true vacuum, without oscillatory behavior, which is a hallmark of single-field tunneling dynamics under this potential \cite{Coleman:1977py, Coleman:1980aw}.
 
\begin{figure}[htb]
\centering
\includegraphics[width=\textwidth, height=\textwidth, keepaspectratio]{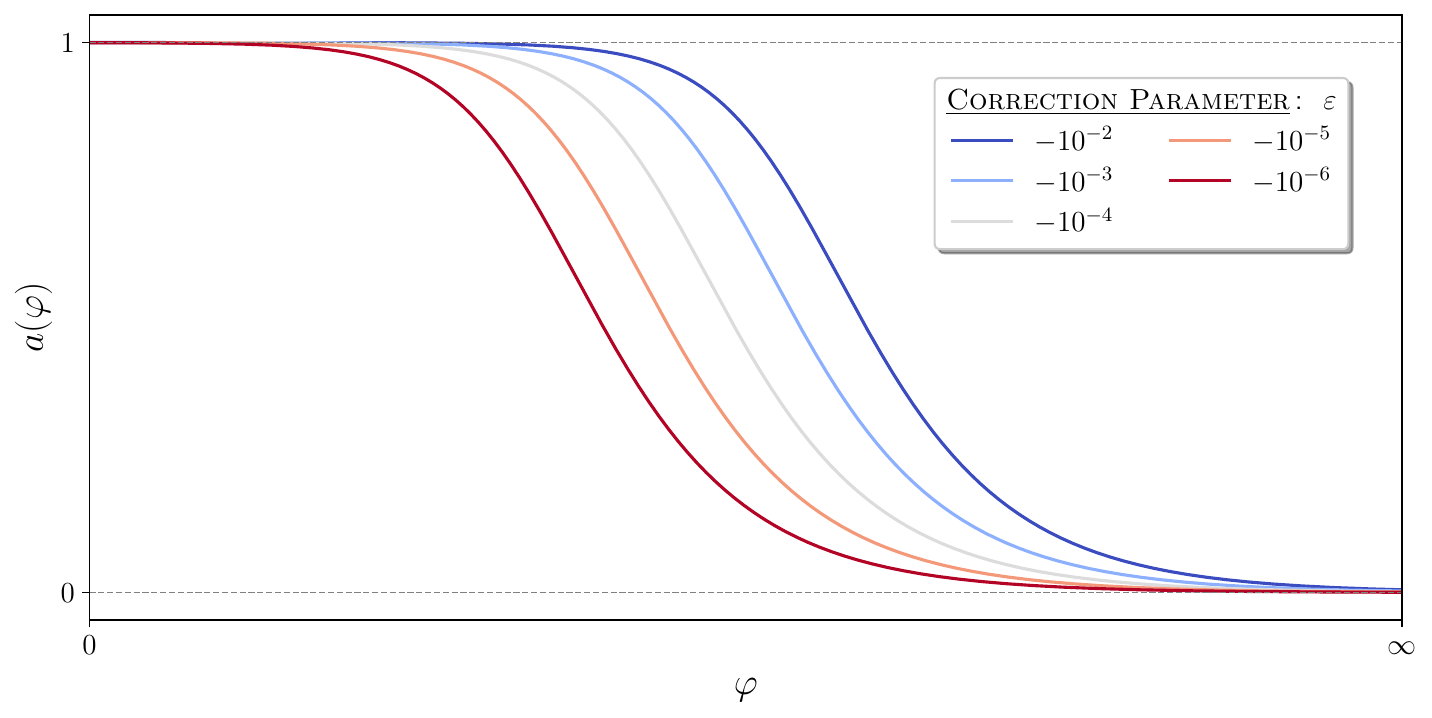}
\caption{\label{fig:3}%
Scale factor $a(\varphi)$ plotted as a function of the canonical inflaton field for different $\varepsilon$ values. As the field $\varphi$ exhibits a monotonically increasing behavior, $a$ evolves from $a_{\text{false}}=1$ to $a_{\text{true}}=0$, consistent with 
expectations.
}
\end{figure}

Substituting this result into Eq.~\eqref{eq:exponentialParameter} yields $\Phi \sim -3.643 \times 10^{7}$ with the correction parameter selection $\varepsilon = -10^{-5}$. A natural question to consider is whether this outcome is sensitive to variations in the parameter $\epsilon$. We find that altering $\epsilon$ within the range $\epsilon = -10^{-2}$ to $\epsilon = -10^{-6}$ leads to negligible changes in $\Phi$, and hence in the decaying rate. This insensitivity is a direct consequence of our potential’s functional form: upon insertion into \eqref{eq:exponentialParameter}, all $\varepsilon$-dependent terms are suppressed with respect to other terms and thus do not materially affect the solution. Consequently, the tunneling probability is estimated generally as $P \sim e^{-6.909 \times 10^{41}}$. Thus we found the lifetime of the metastable dS vacua at the order of magnitude $\tau \sim e^{6.909 \times 10^{41}} \text{s} \gg \tau_{\text{universe}}$\footnote{$\tau_{\text{universe}} \sim 10^{17} \text{s}$.}. This result suggests that the false vacuum is extraordinarily long-lived, rendering the decay process effectively negligible on cosmological timescales.

\section{Gravitino Mass, the Scale of SUSY Breaking and the Swampland}
\label{sec6:gravitino_mass}

A central prediction of any $\mathcal{N}=1$ supergravity model is the gravitino mass, given by
\begin{equation}\label{eq:gravitino_mass}
    m_{3/2} = e^{K/2} |W|.
\end{equation}
In our framework, with the constant superpotential 
\begin{equation}
    W = W_0,
\end{equation}
and the logarithmically corrected Kähler potential discussed in Section~\ref{sec2:log_nilpotent_corrections}, the gravitino mass becomes a sensitive probe of the underlying moduli stabilization mechanism. Notably, the SHS naturally permits $W_0 = \mathcal{O}(1)$, avoiding severe fine-tuning typically required in standard KKLT or LVS constructions.

For example, the approximate Kähler potential in SHS is given by
\begin{equation}
     K \simeq -3 \ln\left[\mathcal{V}_0^{2/3} + 4\varepsilon\,\Omega(\tau)^2\right],
\end{equation}
where $\Omega(\tau)=\xi\, (\tau-\tau_{0})$ with the finite displacement parameterization $\tau = \tau_{0} + e^{-\Delta\varphi}/\xi$. Thus, the gravitino mass takes the exact form:
\begin{equation}\label{eq:m32_exact}
    m_{3/2} = W_0 \left[\mathcal{V}_0^{2/3} + 4\varepsilon\,e^{-2\Delta\varphi}\right]^{-3/2}.
\end{equation}
Using this result, one can further identify the cosmological constant term \eqref{Cosmological_consty} in terms of the gravitino mass as:
\begin{equation}
  \Lambda \approx -12 \varepsilon \cdot \left( \frac{m_{3/2}}{W_{0}^{1/4}} \right)^{8/3}.
\end{equation}
Following this, we distinguish two regimes relevant to our analysis:
\paragraph{Volume-Dominated Regime.}  
When
\begin{equation}
    4\varepsilon\, e^{-2\Delta\varphi} \ll \mathcal{V}_0^{2/3},
\end{equation}
the correction term is negligible, and the gravitino mass simplifies to
\begin{equation}\label{eq:Gravitino_mass_volume_dominant}
    m_{3/2} \approx \frac{W_0}{\mathcal{V}_0} \quad \Longrightarrow \quad m_{3/2}^2 \approx \frac{V_0}{3},
\end{equation}
where the vacuum and inflation energy scale $V_0$ given in \eqref{Inf_Energy_Scale} is directly set by the amplitude of primordial fluctuations, as constrained by CMB normalization at the pivot scale. Thus, remarkably, the gravitino mass becomes directly linked to observable cosmological parameters. More generally, the volume dependence can be expressed as
\begin{equation}\label{eq:Gen_gravitino_mass}
    m_{3/2} \sim \frac{W_0}{\mathcal{V}_0^{p}},
\end{equation}
with the exponent $p$ encoding details of moduli stabilization dynamics (typically $p = 1$ for KKLT or $p = 3/2$ for LVS). For our chosen numerical values, $W_0 = 1$ and $\mathcal{V}_0 = 2.412 \times 10^5$, the resulting gravitino mass naturally aligns within a phenomenologically acceptable window.

\paragraph{Correction-Dominated Regime.}  
Conversely, when
\begin{equation}
    4\varepsilon\, e^{-2\Delta\varphi} \gg \mathcal{V}_0^{2/3},
\end{equation}
the correction term dominates and the gravitino mass becomes
\begin{equation}\label{eq:Gravitino_mass_correction_dominated}
    m_{3/2} \approx \frac{W_0}{(4\varepsilon)^{3/2}}\, e^{3\Delta\varphi},
\end{equation}
leading to an exponential increase with $\Delta\varphi$. Such scaling is generally inconsistent with a controlled EFT.

\subsection{Connection to the Gravitino Distance Conjecture}\label{subsec6.1:GDC_SHS}

The \textit{Gravitino Distance Conjecture} \cite{Castellano:2021yye,Cribiori:2021gbf} posits that the limit $m_{3/2} \to 0$ lies at infinite geodesic distance in moduli space and is necessarily accompanied by an infinite tower of light states. That is, as supersymmetry is restored and the gravitino mass vanishes, the EFT breaks down due to an emerging tower of modes---typically Kaluza--Klein (KK) states or string excitations---whose mass scale $M_{\text{tower}}$ is correlated with $m_{3/2}$ via a scaling law:
\begin{equation}
\frac{M_{\text{tower}}}{M_p} \sim \left(\frac{m_{3/2}}{M_p}\right)^{\delta}, \qquad 0 < \delta \leq 1.
\end{equation}
The exponent $\delta$ depends on the path in moduli space and compactification geometry, with $\delta = 2/3$ appearing frequently in large-volume limits, and $\delta = 1/2$ or $1$ in other settings~\cite{Castellano:2021yye}. Crucially, $\delta$ is bounded away from zero---the light tower cannot decouple more slowly than the gravitino itself.

\paragraph{Quantitative Illustration in SHS.} 
We explicitly acknowledge that although the modulus $\tau$ itself remains bounded, guaranteeing that $m_{3/2}$ does not vanish, other potentially neglected moduli directions could lead to infinite-distance trajectories and emerging towers of states.

Our claims of EFT validity thus explicitly depend on:\begin{itemize}
\item Single-modulus dominance or explicit stabilization of all other moduli directions.
\item Assurance that the infinite $\varphi$-trajectory does not correspond to hidden infinite moduli displacements.
\end{itemize}
Thus, the would-be dangerous modulus $\tau$ remains bounded within a finite range:
\begin{equation}
\tau = \tau_{0} + \frac{e^{-\Delta\varphi}}{\xi},
\end{equation}
where $\Delta\varphi \to \infty$ does \emph{not} imply $\tau \to \infty$. Consequently, the 4D gravitino mass  (restoring $M_p$ explicitly)
\begin{equation}
m_{3/2} = W_0\left[\mathcal{V}_0^{2/3} + 4\varepsilon\, e^{-2\Delta\varphi} \right]^{-3/2},
\end{equation}
stabilizes at a finite value as $e^{-2\Delta\varphi} \to 0$, yielding:
\begin{equation}
m_{3/2} \simeq \frac{M_p}{\mathcal{V}_0} \simeq 4.14 \times 10^{-6} M_p \quad \Rightarrow \quad m_{3/2} \sim 10^{13}\, \text{GeV}.
\end{equation}
Here, we took $W_0 = 1$ and $\mathcal{V}_0 = 2.412 \times 10^5$. This confirms that SUSY breaking remains high-scale in SHS, avoiding the GDC’s vanishing limit.

\paragraph{KK Tower and Scaling Relation.} The KK scale sets the mass threshold above which extra-dimensional modes become relevant, and is controlled by the compactification volume. In large-volume compactifications, the KK mass is related to the Planck scale as:
\begin{equation}
M_{\text{KK}} \sim \frac{M_p}{\mathcal{V}_0^{2/3}} \simeq \frac{2.4 \times 10^{18}}{3.9 \times 10^3} \simeq 6.15 \times 10^{14}\, \text{GeV},
\end{equation}
which corresponds to:
\begin{equation}
\frac{M_{\text{KK}}}{M_p} \simeq 2.56 \times 10^{-4}, \qquad \frac{m_{3/2}}{M_p} \simeq 4.14 \times 10^{-6}.
\end{equation}
Relating $M_{\text{KK}}$ and $m_{3/2}$ via the GDC scaling law:
\begin{equation}
\delta = \frac{\log(M_{\text{KK}} / M_p)}{\log(m_{3/2} / M_p)} \simeq \frac{\log(2.56 \times 10^{-4})}{\log(4.14 \times 10^{-6})} \simeq \frac{-8.27}{-12.39} \simeq 0.67,
\end{equation}
which sits precisely in the expected range, consistent with $\delta = 2/3$.

\paragraph{Avoiding the GDC Breakdown.} In scenarios aiming for low-scale SUSY (e.g., $m_{3/2} \sim \text{TeV}$), one must typically send some modulus---such as the overall volume $\mathcal{V}$---to infinity to suppress SUSY breaking. According to the GDC, this necessarily drags the KK tower down to $M_{\text{KK}} \sim 10^8$ GeV (for $\delta = 2/3$), eroding the scale separation between IR physics and quantum gravity. Such configurations risk falling into the swampland. 

In contrast, SHS \emph{geometrically forbids} such decompactification: moduli like $\tau$ remain trapped within a finite range. Hence, the GDC’s dangerous regime is never approached, the gravitino mass stays high ($\sim 10^{13}$ GeV), and the KK tower remains safely decoupled ($\sim 10^{14-15}$ GeV).

\paragraph{Key takeaway.} The SHS framework offers a clear resolution to the GDC constraint: by bounding moduli motion, it keeps $m_{3/2}$ finite and avoids any runaway to infinite distance. The KK tower remains heavy, and the EFT stays valid throughout inflationary and post-inflationary epochs. This provides a controlled and predictive example of how high-scale SUSY and large-field inflation can coexist with quantum gravity consistency conditions imposed by the GDC.

\subsection{Connection to the Generalized Swampland Distance Conjecture}
\label{subsec6.2:GSDC}

The traditional \emph{SDC}~\cite{Ooguri:2006in} asserts that EFTs coupled to quantum gravity lose validity when scalar fields traverse an infinite geodesic distance in field space. This breakdown is attributed to the emergence of an infinite tower of light states, signaling the EFT's entry into the so-called ``swampland.'' The proper distance is computed via the field-space metric:
\begin{equation}
\Delta_\text{phys} = \int \sqrt{G_{ij}(\phi) \, d\phi^i  d\phi^j}.
\end{equation}

Recent developments~\cite{Debusschere:2024rmi,Mohseni:2024njl} have refined this idea, noting that in dynamical settings—such as cosmological evolution, tunneling, or domain walls—the physical cost of field excursions should also account for potential energy. This leads to the \emph{Generalized Swampland Distance Conjecture (GSDC)}, where a dynamically weighted action-like cost function replaces distance:
\begin{equation}
\Delta_\text{eff} = \int dt\, \sqrt{\dot{\varphi}^2 + 2V(\varphi)}.
\end{equation}
This quantity coincides with the mechanical action in the Maupertuis principle and is directly related to the tension of domain walls interpolating between vacua.\footnote{This form follows by changing variables in the time integral using $ d\varphi = \dot{\varphi}\,dt $, yielding:
\begin{equation}
\Delta_\text{eff} = \int d\varphi\, \sqrt{1 + \frac{2V(\varphi)}{\dot{\varphi}^2}},
\end{equation}
which reduces to the geodesic distance in the limit $ V \to 0 $. } Alternatively, for static domain walls in flat spacetime, the tension is given by:
\begin{equation}\label{Tension}
T = \int_{\varphi_i}^{\varphi_f} d\varphi\, \sqrt{2\big(V(\varphi) + \rho_E\big)},
\end{equation}
where $ \rho_E $ is an integration constant fixed by boundary conditions.

\paragraph{SHS Framework and Finite Physical Distance.}
In our \emph{Supersymmetric Horizon Stabilization} framework, the inflaton $\varphi$ undergoes a trans-Planckian excursion $\Delta\varphi \to \infty$, while the underlying modulus $\tau$, controlling internal volume, remains finite:
\begin{equation}
\tau(\varphi) = \tau_{0} + \frac{1}{\xi} e^{-\sqrt{2/3}\,\varphi}, \quad \tau_{0}, \xi > 0.
\end{equation}
The field-space metric $G_{\tau\tau} \sim 1/\tau^2$ then yields a finite geodesic length:
\begin{equation}
\Delta_\text{phys} = \int_C^{\tau_{0} + \xi^{-1}} \frac{d\tau}{\tau} = \log\left(1 + \frac{1}{\tau_{0}\xi}\right).
\end{equation}
This finite displacement satisfies the original SDC condition from the viewpoint of $\tau$ alone. However, infinite-distance limits in moduli space are invariant under field redefinitions. Thus, the infinite canonical displacement of the field $\varphi$ must itself be viewed as a genuine infinite-distance limit in moduli space, independently of the finite displacement observed for $\tau$. This infinite canonical displacement cannot be neglected, as string theory constructions typically associate infinite towers of states with every infinite moduli-space trajectory.

Therefore, our analysis of the finite displacement of $\tau$ alone is insufficient to universally guarantee EFT validity. Indeed, even if one direction (e.g., volume modulus $\tau$) remains constant or nearly constant, combinations of moduli (such as complex-structure moduli in Calabi–Yau spaces) can still approach infinite limits, such as $\tau = \tau_1 \cdot \tau_2$ with $\tau_1 \propto 1/\tau_2 \to 0$,
producing light towers of Kaluza–Klein modes or winding modes.

Hence, our original claim—that bounded displacement of the modulus $\tau$ alone prevents EFT breakdown due to infinite towers—is overly restrictive and must be qualified explicitly. EFTs with infinite distances under the SHS construction are thus conditionally valid, considering these specific conditions:
\begin{itemize}
\item The modulus $\tau$ is genuinely the only modulus relevant for EFT control, or at least no other moduli are driven to infinite limits.
\item Other moduli fields, if present, must be stabilized by independent explicit mechanisms (such as fluxes or non-perturbative corrections), ensuring no hidden infinite-distance trajectories in moduli space emerge.
\end{itemize}

Consequently, our framework illustrates conditions under which EFT descriptions might remain valid despite infinite canonical displacements, provided explicit stabilization mechanisms prevent infinite moduli trajectories in any neglected directions. Without such mechanisms, infinite towers associated with hidden moduli directions will indeed emerge.

\paragraph{Asymptotic Structure and Energetic Cost.}
The scalar potential in the SHS model generally takes the analytical form given in eq.~\eqref{eq:modelPotential}:
\begin{equation}
V(\varphi) = \left(V_0 + V_{\text{inf}}(\varphi)\right) A(\varphi) - V_0,
\end{equation}
where
\begin{equation}
    A(\varphi) = \frac{\mathcal{V}_0^{2/3} e^{\varepsilon\sqrt{8/3}\,\varphi}}{\mathcal{V}_0^{2/3} + 4\varepsilon e^{-\sqrt{8/3}\,\varphi}}
\end{equation}
for the coupling choices given in eq.~\eqref{good_choice}. At the asymptotic limit, as $\varphi \to \infty$:
\begin{equation}
A(\varphi) \to 0 \quad \Rightarrow \quad V(\varphi) \to -V_0 \equiv V_\infty < 0.
\end{equation}
Hence, for large $\varphi$, we can make the approximation
\begin{equation}
    V(\varphi) \approx - V_0 + \delta V(\varphi), \quad  \delta V(\varphi) \sim e^{-\Delta \varphi}
\end{equation}
up to constant factors. Assuming slow-roll, with $3H\dot{\varphi} \approx -V'$ and $H^2 \approx |V|/3$, we estimate:
\begin{equation}
V'(\varphi) \sim - e^{-\Delta\varphi}, \qquad \dot{\varphi} \sim \frac{e^{-\Delta\varphi}}{\sqrt{V_0}}.
\end{equation}
Then:
\begin{equation}
\dot{\varphi}^2 \sim \frac{1}{V_0} e^{-2\Delta\varphi}, \qquad \frac{2V(\varphi)}{\dot{\varphi}^2} \sim -V_0^2 e^{2\Delta\varphi} + \ldots.
\end{equation}
Thus, the integrand in $\Delta_\text{eff}$ becomes imaginary at large $\varphi$. This signals the breakdown of the Lorentzian formulation near AdS minima with $V<0$.

\paragraph{Domain Walls and the Euclidean GSDC.}
To resolve the breakdown of the Lorentzian cost function in the presence of negative potentials, the GSDC can be reinterpreted via a Wick rotation $t_E = i t$, yielding a Euclideanized cost function:
\begin{equation}
\Delta_\text{eff}^{(E)} = \int dt_E\, \sqrt{\left(\frac{d\varphi}{dt_E}\right)^2 - 2V(\varphi)},
\label{eq:EuclideanCost}
\end{equation}
which remains real and positive-definite for $V(\varphi) < 0$. This coincides with the domain wall tension integral~\eqref{Tension} for kink-like scalar field profiles interpolating between vacua.

The scalar field profile in Euclidean time obeys the equation of motion:
\begin{equation}
\frac{d^2\varphi}{dt_E^2} = \frac{dV}{d\varphi}.
\end{equation}
Integrating once yields:
\begin{equation}
\left(\frac{d\varphi}{dt_E}\right)^2 = 2\left(V(\varphi) + \rho_E\right),
\end{equation}
where $ \rho_E $ is an integration constant determined by boundary conditions. Substituting into Eq.~\eqref{eq:EuclideanCost}, we obtain:
\begin{equation}
\Delta_\text{eff}^{(E)} = \int dt_E\, \sqrt{2\rho_E} = \sqrt{2\rho_E} \int dt_E.
\end{equation}
Changing variables using $ dt_E = \frac{d\varphi}{\sqrt{2(V(\varphi) + \rho_E)}} $, we find:
\begin{equation}
\Delta_\text{eff}^{(E)} = \int d\varphi\, \frac{\sqrt{2\rho_E}}{\sqrt{2(V(\varphi) + \rho_E)}}.
\end{equation}

In the asymptotic limit $ \varphi \to \infty $, we have:
\begin{equation}
V(\varphi) \to -V_0, \qquad V(\varphi) + \rho_E \to \rho_E - V_0 > 0,
\end{equation}
assuming $ \rho_E > V_0 $. Then the integrand becomes:
\begin{equation}
\frac{\sqrt{2\rho_E}}{\sqrt{2(V(\varphi) + \rho_E)}} 
= \left(1 + \frac{V(\varphi)}{\rho_E}\right)^{-1/2} 
\approx 1 + \frac{V_0}{2\rho_E} + \mathcal{O}(e^{-\lambda\varphi}).
\end{equation}
Thus, the cost diverges linearly:
\begin{equation}
\Delta_\text{eff}^{(E)} \sim \int d\varphi \left(1 + \frac{V_0}{2\rho_E} + \mathcal{O}(e^{-\Delta\varphi}) \right) \to \infty,
\end{equation}
confirming that the UV wall lies at infinite canonical distance, despite the moduli remaining confined.

\medskip
\noindent{\textbf{Key takeaway:}}
Finite displacement in a single modulus direction ($\tau$) alone does not universally avoid the breakdown of EFT. Instead, additional conditions on moduli stabilization must be explicitly satisfied to rigorously evade the emergence of infinite towers of states, ensuring validity under the GSDC. Without such conditions, the infinite canonical displacement of $\varphi$ must be recognized as an infinite-distance limit, leading generically to EFT breakdown. Moreover, the Euclidean effective distance diverges as the scalar field asymptotically approaches the supersymmetric AdS boundary. This confirms the GSDC: even when $\Delta_\text{phys}$ is finite, the dynamical cost $\Delta_\text{eff}^{(E)}$ diverges, indicating EFT breakdown.

\paragraph{Relation to the AdS Distance Conjecture.}
The divergence of the Euclidean effective cost $\Delta_\text{eff}^{(E)}$ in the asymptotic AdS region is also consistent with the \emph{AdS Distance Conjecture (AdSDC)}~\cite{Lust:2019zwm}, which posits that approaching the boundary of moduli space in an AdS vacuum results in an infinite tower of light states becoming exponentially light. The conjecture can be viewed as the AdS analogue of the SDC, governed by a breakdown of EFT as the proper distance or physical cost diverges.

In SHS, the asymptotic scalar potential satisfies:
\begin{equation}
    V(\varphi) \to -V_0, \quad \text{as } \varphi \to \infty,
\end{equation}
which describes an approach to an AdS boundary. The fact that $\Delta_\text{eff}^{(E)}$ diverges in this limit signals the emergence of a nonperturbative instability or tower of light modes, in line with the AdSDC. In this sense, the GSDC and AdSDC jointly capture the swampland constraints on trajectories that flow toward infinite energetic cost in the presence of negative vacuum energy.

\subsection{The Supersymmetric Horizon Conjecture}
\label{subsec6.3:SHC}

\paragraph{Motivation and Setup.}
In the framework developed throughout this work, large canonical field excursions ($ \Delta\varphi \to \infty $) are realized without large displacements of geometric moduli. This is achieved via exponential parametrizations such as:
\begin{equation}
\tau(\varphi) = \tau_{0} + \frac{1}{\xi} e^{-\sqrt{\tfrac{2}{3}} \varphi}, \label{eq:tau_confined}
\end{equation}
which ensures that the Kähler modulus $\tau$ remains bounded as $\varphi$ diverges. The effective theory remains compactified and under control, with moduli confined to a finite domain $ \tau \in [\tau_{0},\,\tau_{0} + \xi^{-1}] $. This separation between field and moduli space leads to a novel UV completion mechanism tied to supersymmetry restoration.

\paragraph{Frame Function and EFT Breakdown.}
Recall that the frame function $ \Omega(\tau) $ appears in the logarithmically corrected Kähler potential (see Section~\ref{subsec4.3:Conformal_equivalence}), defined by:
\begin{equation}
\Omega(\tau) = \xi (\tau - \tau_{0}).
\end{equation}
As $ \tau \to \tau_{0} $, the frame function vanishes, $ \Omega(\tau) \to 0 $, signaling the boundary of the EFT. Crucially, this is not a curvature singularity in the Einstein frame, but a genuine divergence in the Jordan (superconformal) frame. The Weyl rescaling between the two metrics,
\begin{equation}
g_{\mu\nu}^{(\text{E})} = \Omega^{-1}(\tau)\, g_{\mu\nu}^{(\text{J})},
\end{equation}
becomes singular as $ \Omega(\tau) \to 0 $, and so does the curvature. In particular, the Jordan-frame Ricci scalar behaves as\footnote{Here, $ \Box \Omega \equiv g^{\mu\nu}_{(\text{J})} \nabla_\mu \nabla_\nu \Omega $ is the covariant Laplacian in the Jordan frame.}
\begin{equation}
R^{(\text{J})} \sim \Omega^{-1} \Box \Omega,
\end{equation}
which diverges near the boundary. This reflects the fact that the gravitational sector becomes strongly coupled and geodesically incomplete as one approaches $ \Omega \to 0 $. The divergence originates from Weyl rescaling terms in the Einstein-frame action, such as $ (\partial_\mu \ln \Omega)^2 \sim (\partial_\mu \Omega)^2 / \Omega^2 $, which dominate in this limit. We interpret this divergence not as a pathology, but as a geometric signal of a UV transition—potentially into a higher-dimensional or strongly coupled supersymmetric phase.

\paragraph{Asymptotic Supersymmetry and Field Distance.}
As the canonical field displacement $ \Delta \varphi \to \infty $, supersymmetry is restored. The scalar potential asymptotes to
\begin{equation}
V(\varphi) = -\frac{3 W_0^2}{\mathcal{K}^3(\tau(\varphi))},
\end{equation}
corresponding to a supersymmetric AdS vacuum with vanishing F-terms. Meanwhile, the physical displacement in moduli space remains finite:
\begin{equation}
\Delta_{\text{phys}} = \int_{\tau_{0}}^{\tau_{0}+1/\xi} \frac{d\tau}{\tau} = \log\left(1 + \frac{1}{\tau_{0}\xi}\right).
\end{equation}
Thus, even infinite excursions in $\varphi$ map to a compact, finite region in geometric moduli space. The scaling parameter $\xi$, inherited from the conformal frame structure, controls the size of this domain and thereby sets the location of the supersymmetric horizon.

\paragraph{The Conjecture.} We therefore propose:
\begin{quote}
\textit{In gravitational EFTs with stabilized moduli and controlled large-field dynamics, asymptotic canonical displacements can approach a finite boundary in moduli space where supersymmetry re-emerges. This ‘supersymmetric horizon’ suggests that consistent EFTs begin and end with supersymmetry—dynamically broken along the way, yet restored at both ends of cosmological evolution.}
\end{quote}

Unlike the SDC, where towers of light states emerge gradually and the cutoff collapses at infinite field distance, the SHC posits a sudden transition at finite $\tau$. This boundary marks the termination of EFT validity via a vanishing frame function:
\begin{equation}
\Omega(\tau) = \xi(\tau - \tau_{0}) \to 0,
\end{equation}
interpreted as a supersymmetric UV wall. Crucially, the position of this wall is governed by $\xi$, which appears as an integration constant in the canonical field redefinition (cf.~Appendix~\ref{sec:Appendix_CanonicalField}). This suggests that high-energy input—such as flux configurations or quantum corrections—may dynamically determine the maximum IR field range, embedding trans-Planckian dynamics within a compact geometric framework.

\paragraph{Gravitino Dynamics and GDC Compatibility.}
The SHC naturally explains why the gravitino mass remains finite during large-field evolution:
\begin{equation}
m_{3/2} \sim \frac{1}{\mathcal{V}} \sim \tau^{-3/2}.
\end{equation}
Since $\tau$ remains bounded, the cutoff scale $ \Lambda_{\text{EFT}} \sim m_{3/2}^{2/3} $ stays above inflationary energies. This avoids the tower collapse predicted by the GDC, maintaining EFT validity throughout. The SHC thus provides a \emph{geometric origin} for why $m_{3/2}$ remains finite even in large-field inflationary models (cf.\ Section~\ref{subsec6.1:GDC_SHS}).

\paragraph{Metastable dS and UV Endpoint.}
The EFTs developed here allow for metastable de Sitter vacua supported by supersymmetry-breaking ingredients—such as nilpotent sectors—that can mimic the uplift mechanisms found in KKLT or LVS-type constructions. However, these dS vacua reside deep within the interior of moduli space, far from the asymptotic regime where supersymmetry is restored. The true UV endpoint of the theory, characterized by vanishing F-terms and a supersymmetric AdS configuration, lies at parametrically large field displacement (e.g., $\Delta \varphi \to \infty$ or $\tau \to \tau_{0}$). While this limit is formally inaccessible in finite cosmological time, it provides a geometric anchor for the EFT and ensures that SUSY breaking is temporary from a UV perspective. Thus, the universe can remain in a long-lived, metastable de Sitter phase—with finite SUSY-breaking scale and controlled vacuum energy—while being consistently embedded in a supersymmetric, UV-complete theory.

\paragraph{UV-IR Consistency and Dynamical Cost.}
The SHC furnishes a refined UV-IR consistency principle: trans-Planckian excursions in the canonical field $ \varphi $ are permitted, yet geometric moduli remain confined, and the frame function enforces a \emph{hard boundary} in field space. This boundary corresponds not to an infinite geodesic distance, but to a divergence in the \emph{dynamical cost}—mirroring the GSDC~\cite{Debusschere:2024rmi,Mohseni:2024njl}. In this view, the EFT ceases to be valid as the scalar reaches a boundary defined by $ \Omega(\tau) \to 0 $, where the Jordan-frame curvature diverges, and supersymmetry is restored.

This behavior invites a holographic reinterpretation. The canonical scalar $ \varphi $ acts as an emergent radial coordinate in a dual RG flow, reminiscent of the AdS/CFT correspondence~\cite{Maldacena:1997re,Witten:1998qj,Akhmedov:1998vf,deBoer:1999tgo,Skenderis:2002wp}. Under this analogy, the energy scale $\mu$ is related to the scalar field via:
\begin{equation}
\mu(\varphi) \sim e^{\alpha \varphi}, \qquad \text{with} \quad \alpha = \sqrt{\tfrac{2}{3}},
\end{equation}
such that evolution toward large $\varphi$ corresponds to flow toward the UV. The associated beta function governing the evolution of the modulus $\tau$ is:
\begin{equation}
\beta(\tau) \equiv \frac{d \tau}{d \log \mu}
= \frac{d\tau}{d\varphi} \cdot \frac{d\varphi}{d\log\mu}
\sim-\frac{1}{\xi} e^{ -\alpha \varphi },
\end{equation}
which vanishes exponentially as $\varphi \to \infty$. This signals an asymptotically safe fixed point: the RG flow slows, moduli stabilize, and the system flows toward a supersymmetric vacuum.

Furthermore, the frame function $ \Omega(\tau) = \xi (\tau - \tau_{0}) \sim e^{- \alpha \varphi} $ may be interpreted as a c-function controlling degrees of freedom along the RG flow. Its monotonicity is manifest:
\begin{equation}
\frac{d\Omega}{d\varphi} = - \alpha \Omega < 0,
\end{equation}
in agreement with holographic c-theorems~\cite{Freedman:1999gp}. The vanishing of $ \Omega $ at the SUSY horizon thus marks the UV boundary of the EFT, beyond which the theory enters a scale-invariant, supersymmetric regime with frozen moduli.

Consequently, this correspondence unifies several swampland constraints: the SDC becomes a statement about RG trajectory geometry; the GSDC arises from divergent cost rather than distance; and the SHC acquires a dual interpretation as a holographic RG endpoint, geometrically realized via a vanishing frame function.

\begin{figure}[htb]
\centering
\begin{tikzpicture}[
  every node/.style={font=\footnotesize},
  phaseShared/.style={
    rectangle, draw=black, thick, rounded corners=3.5pt,
    minimum width=4.4cm, minimum height=1.5cm,
    align=center, fill=gray!10, drop shadow
  },
  phaseSDC/.style={
    rectangle, draw=red!60!black, thick, rounded corners=3.5pt,
    minimum width=4.4cm, minimum height=1.5cm,
    align=center, fill=red!5, drop shadow
  },
  phaseSHS/.style={
    rectangle, draw=blue!60!black, thick, rounded corners=3.5pt,
    minimum width=4.4cm, minimum height=1.5cm,
    align=center, fill=blue!5, drop shadow
  },
  arrow/.style={->, thick, draw=black},
  endmarker/.style={font=\Large, text=black},
  uvmarker/.style={->, thick, draw=blue!60!black!80!white, line width=0.7pt}
]

\node[phaseShared] (inflation) at (0,5.3) {
  \textbf{Large-Field Inflation}\\
  $\varphi \gg 1$, \quad $\dot{\varphi}^2 \ll V(\varphi)$
};

\node[phaseSDC] (sdc1) at (-4,3.2) {
  \textbf{Moduli Runaway}\\
  $\tau \to \infty$, \quad $\mathcal{V} \to \infty$
};
\node[phaseSDC, below=0.5cm of sdc1] (sdc2) {
  \textbf{Light Tower Emerges} \\
$M_{\text{tower}} \sim m_{3/2}^\delta$, \quad
$m_{3/2} \to 0$
};
\node[phaseSDC, below=0.5cm of sdc2] (sdc3) {
  \textbf{Cutoff Collapse}\\
  $\Lambda_{\text{EFT}} \sim M_{\text{tower}} \to 0$
};

\node[phaseSHS] (shs1) at (4,3.2) {
  \textbf{Moduli Stabilization}\\
  $\tau \sim \tau_{0} + e^{-\Delta \varphi} \to$ finite\\ $\mathcal{V} \to \mathcal{V}_0 \sim$ finite
};
\node[phaseSHS, below=0.5cm of shs1] (shs2) {
  \textbf{Metastable dS Phase}\\
  $V > 0$, \quad $m_{3/2} \sim 1/\mathcal{V}_{0}$
};
\node[phaseSHS, below=0.5cm of shs2] (shs3) {
  \textbf{Supersymmetric Horizon}\\
  $\Omega(\tau) \to 0$, \quad $V<0$,
\\
 $M_{\text{tower}} \sim m_{3/2}^{2/3}$ \,(GDC)
};

\draw[arrow, bend left=15, red!50] (inflation) to (sdc1);
\draw[arrow, red!50] (sdc1) -- (sdc2);
\draw[arrow, red!50] (sdc2) -- (sdc3);
\draw[arrow, bend right=15, blue!50] (inflation) to (shs1);
\draw[arrow, blue!50] (shs1) -- (shs2);
\draw[arrow, blue!50] (shs2) -- (shs3);

\node[endmarker] at (sdc3.south) [yshift=-0.5cm] {\textcolor{red}{\ding{55}}} (sdc3.south) to node[ font=\scriptsize\itshape, xshift=2pt] {EFT breakdown \quad (pathological)} 
  ++(-0.5,-1.0);

\draw[->, thick, draw=blue!40] 
  (shs3.south) to node[ font=\scriptsize\itshape, xshift=2pt] {\hspace{+0.8cm} EFT boundary \quad UV regime (controlled)} 
  ++(0,-1.0);

\end{tikzpicture}
\caption{ Two contrasting large-field trajectories. On the left, the Swampland Distance Conjecture path leads to runaway moduli, emergent towers of light states, and eventual EFT breakdown. On the right, the Supersymmetric Horizon Stabilization scenario confines moduli evolution and stabilizes the gravitino mass along a controlled trajectory. Provided additional moduli directions are stabilized and no hidden infinite-distance limits are accessed, the SHS path can maintain EFT control and terminate at a supersymmetric boundary with finite volume and bounded scalar geometry.}
\label{fig:shs-sdc-unified-tower}
\end{figure}
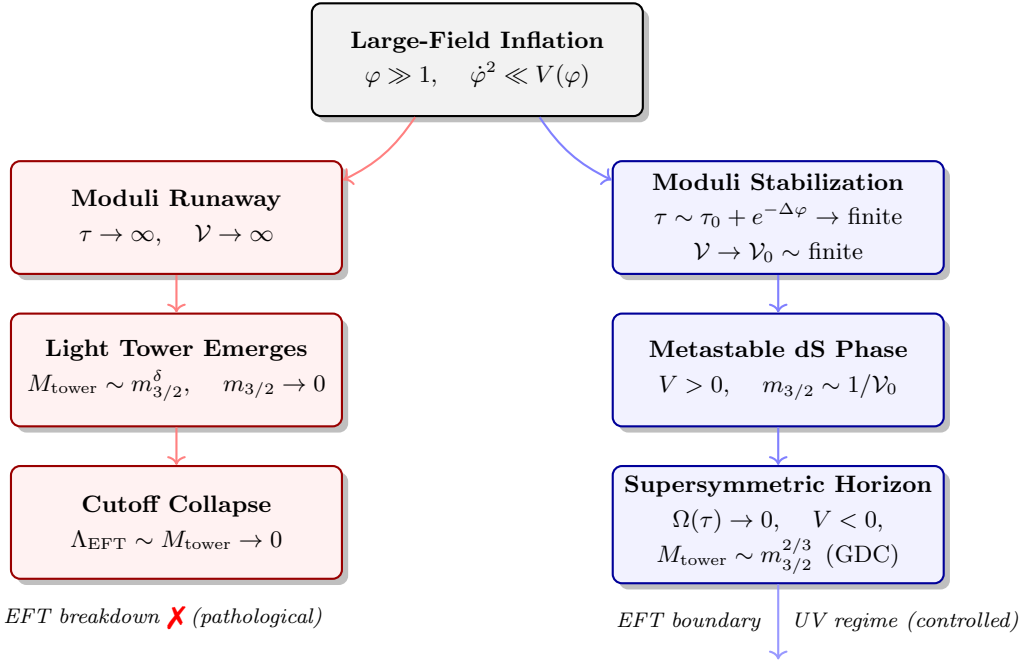

\paragraph{de Sitter Critical Points.}
Crucially, the SHC does not rule out the existence of dS vacua. It refines the perspective by asserting that any such vacua must:
\begin{itemize}
  \item be \emph{metastable},
  \item reside in a finite, compact region of moduli space,
  \item and be separated from the SUSY-restoring boundary by an infinite $\varphi$ excursion.
\end{itemize}
This viewpoint aligns with the dS Swampland Conjecture~\cite{Obied:2018sgi,Garg:2018reu,Ooguri:2018wrx} (see e.g.~\cite{Conlon:2018eyr}) and the AdS Distance Conjecture~\cite{Lust:2019zwm}, both of which challenge the viability of stable non-SUSY vacua at infinite field distance. The SHC offers a geometric reason why: the boundary of moduli space is supersymmetric, and any non-SUSY phase must reside strictly inside.

\medskip
\noindent\textbf{Summary.}
The Supersymmetric Horizon Conjecture proposes a new organizing principle for consistent EFTs derived from string theory. It supports trans-Planckian dynamics without sacrificing moduli stability, reconciling inflation, SUSY breaking, and swampland principles under a unifying geometric constraint. This framework may offer a sharpened alternative to the traditional SDC and motivate new dualities between large-field EFTs and compact UV-complete phases. 

\section{Inflation and Primordial Black Holes}\label{sec7:INF_PBH}

This section explores the interplay between inflationary dynamics and post-inflationary structure formation, focusing in particular on the role of higher-order corrections and their potential to seed primordial black holes.

First, let us show that the string corrections encoded in the small parameter $\varepsilon$ have a negligible effect on inflationary dynamics. Specifically, we consider the fully corrected scalar potential:
\begin{equation}
V(\varphi) = \left(V_0 + V_{\text{inf}}(\varphi)\right) A(\varphi) - V_0,
\qquad
A(\varphi) = \frac{\mathcal{V}_0^{2/3}\, e^{\varepsilon\sqrt{\tfrac{8}{3}} \varphi}}{\mathcal{V}_0^{2/3} + 4\varepsilon\, e^{-\sqrt{\tfrac{8}{3}} \varphi}},
\quad
V_0 = \text{constant}.
\end{equation}

\paragraph{Perturbative expansion.}
For $|\varepsilon| \ll 1$, we expand the correction factor in the inflationary regime—where $\varphi$ is finite and the potential remains positive—as
\begin{equation}
A(\varphi) = 1 + \varepsilon\, \alpha(\varphi) + \mathcal{O}(\varepsilon^2),
\end{equation}
where
\begin{equation}
\alpha(\varphi) = \sqrt{\tfrac{8}{3}}\,\varphi - \tfrac{4}{\mathcal{V}_0^{2/3}}\, e^{-\sqrt{\tfrac{8}{3}} \varphi} + \cdots
\end{equation}
is an $\mathcal{O}(1)$ function across the slow-roll inflationary phase. Substituting this into the potential, we find:
\begin{equation}\label{V_inf_small_eps}
V(\varphi) = V_{\text{inf}}(\varphi) + \varepsilon \left[V_0 + V_{\text{inf}}(\varphi)\right] \alpha(\varphi) + \mathcal{O}(\varepsilon^2).
\end{equation}

Thus, to first order in $\varepsilon$, the corrected potential differs from $V_{\text{inf}}(\varphi)$ by a small multiplicative shift. Since $V_{\text{inf}} \sim V_0$ during inflation, the correction term $\varepsilon \left[V_0 + V_{\text{inf}}(\varphi)\right] \alpha(\varphi)$ is uniformly small and leaves inflationary predictions unaffected at leading order. The shape of the inflationary potential is preserved, and the background dynamics continue to be governed by $V_{\text{inf}}(\varphi)$ up to suppressed $\mathcal{O}(\varepsilon)$ effects.\footnote{We emphasize that this perturbative treatment is valid only in the finite-field inflationary regime. In the asymptotic limit $\varphi \to \infty$, the correction factor $A(\varphi)$ vanishes exponentially, and the potential approaches an AdS minimum—requiring the full non-perturbative structure beyond the expansion above.}

\subsection{Robustness of Inflationary Observables for Small \texorpdfstring{$\boldsymbol\varepsilon$}{ε}}
\label{subsec6.1:InflationaryRobustness}
To assess the impact of corrections on inflationary observables, we explicitly expand the potential and its derivatives to first order in $\varepsilon$. Using~\eqref{V_inf_small_eps}, the derivatives become
\begin{equation}
    \begin{aligned}
V' &= V'_{\text{inf}} + \varepsilon\left[ V'_{\text{inf}}\, \alpha + (V_0 + V_{\text{inf}})\, \alpha' \right] + \mathcal{O}(\varepsilon^2), \\
V'' &= V''_{\text{inf}} + \varepsilon\left[ V''_{\text{inf}}\, \alpha + 2 V'_{\text{inf}}\, \alpha' + (V_0 + V_{\text{inf}})\, \alpha'' \right] + \mathcal{O}(\varepsilon^2).
\end{aligned}
\end{equation}

We then compute the slow-roll parameters:
\begin{equation}
    \begin{aligned}
\epsilon &= \frac{1}{2} \left( \frac{V'}{V} \right)^2 
= \frac{1}{2} \left( \frac{V'_{\text{inf}}}{V_{\text{inf}}} \right)^2 + \varepsilon\, \delta\epsilon(\varphi) + \mathcal{O}(\varepsilon^2), \\
\eta &= \frac{V''}{V}
= \frac{V''_{\text{inf}}}{V_{\text{inf}}} + \varepsilon\, \delta\eta(\varphi) + \mathcal{O}(\varepsilon^2),
\end{aligned}
\end{equation}
where $\delta\epsilon(\varphi)$ and $\delta\eta(\varphi)$ are analytic $\mathcal{O}(1)$ functions depending on $\alpha$, $\alpha'$, $\alpha''$, and the ratio $V_0/V_{\text{inf}}$.

The spectral index and tensor-to-scalar ratio follow as
\begin{equation}
    \begin{aligned}
n_s &= 1 - 6\epsilon + 2\eta 
= 1 - 3\left( \frac{V'_{\text{inf}}}{V_{\text{inf}}} \right)^2 + 2 \frac{V''_{\text{inf}}}{V_{\text{inf}}} + \varepsilon\, \delta n_s(\varphi) + \mathcal{O}(\varepsilon^2), \\
r &= 16\epsilon = 8\left( \frac{V'_{\text{inf}}}{V_{\text{inf}}} \right)^2 + \varepsilon\, \delta r(\varphi) + \mathcal{O}(\varepsilon^2).
\end{aligned}
\end{equation}

The number of e-folds is similarly robust:
\begin{equation}
N = \int_{\varphi_{\rm end}}^{\varphi_*} \frac{V}{V'}\, d\varphi
= \int_{\varphi_{\rm end}}^{\varphi_*} \frac{V_{\text{inf}}}{V'_{\text{inf}}}\, d\varphi + \mathcal{O}(\varepsilon),
\end{equation}
where corrections again appear at $\mathcal{O}(\varepsilon)$ due to the multiplicative structure of the deformation.

\paragraph{Numerical comparisons.}
To validate our analytic expectations, we numerically compute key inflationary observables for several benchmark potentials, comparing the uncorrected (“pure”) case to the fully corrected version incorporating $\varepsilon = -10^{-5}$. In all cases, we define the inflationary potential as
\begin{equation}
    V_{\rm inf}(\varphi) = V_0\,s(\varphi),
\end{equation}
where $s(\varphi)$ specifies the scalar profile for each model:
\begin{equation}
\begin{aligned}
\textbf{Starobinsky:} \quad &s(\varphi) = \left(1 - e^{-\sqrt{\tfrac{2}{3}}\,\varphi}\right)^2 \quad \text{\cite{Starobinsky:1980te}}, \\[4pt]
\textbf{Fibre Inflation:} \quad &s(\varphi) = 3 - 4\,e^{-\varphi/\sqrt{3}} + e^{-4\varphi/\sqrt{3}} + R\left(e^{2\varphi/\sqrt{3}} - 1\right), \quad R = 10^{-6} \quad \text{\cite{Cicoli:2008gp,Cicoli:2020bao}}, \\[4pt]
\textbf{T-model}~(\alpha = 1): \quad &s(\varphi) = \tanh^2\!\left(\frac{\varphi}{\sqrt{6}}\right)\quad \text{\cite{Kallosh:2013hoa,Carrasco:2015rva,Achucarro:2017ing,Linde:2018hmx,Pallis:2022qkw}}, \\[4pt]
\textbf{KKLTI:} \quad &s(\varphi) = \frac{\varphi^2}{\varphi^2 + \tfrac{1}{4}} \quad \text{\cite{Kachru:2003sx,Kallosh:2019eeu}}.
\end{aligned}
\end{equation}

\vspace{1em}

\noindent Table~\ref{tab:inflationary_comparison} summarizes the numerical results for each model. For both the pure and fully corrected potentials, we report the pivot-scale value $\varphi_*$, end-of-inflation field value $\varphi_{\rm end}$, scalar spectral index $n_s$, tensor-to-scalar ratio $r$, and total number of $e$-folds $N$. The differences between pure and corrected values are small, with all observables remaining well within observational bounds.

\begin{table}[htb]
  \centering
  \small
  \begin{tabular}{@{} l 
                   c 
                   cc 
                   cc 
                   cc 
                   cc @{}}
    \toprule
    \textbf{Model} 
      & $\varphi_{*}$ 
      & \multicolumn{2}{c}{$\varphi_{\rm end}$} 
      & \multicolumn{2}{c}{$n_s$} 
      & \multicolumn{2}{c}{$r$} 
      & \multicolumn{2}{c}{$N$} \\
    \cmidrule(lr){2-2}
    \cmidrule(lr){3-4}\cmidrule(lr){5-6}\cmidrule(lr){7-8}\cmidrule(lr){9-10}
      &  
      & Pure & Full 
      & Pure & Full 
      & Pure & Full 
      & Pure & Full \\
    \midrule
    Starobinsky
      & 5.45315
      & 0.940178 & 0.940176
      & 0.967827 & 0.967828
      & 0.002964 & 0.002974
      & 60.00    & 59.95    \\

    Fibre
      & 6.01916
      & 0.917620 & 0.917625
      & 0.970364 & 0.970363
      & 0.005107 & 0.005115
      & 60.00    & 59.97    \\

    T-model
      & 6.23891
      & 1.208390 & 1.208370
      & 0.966885 & 0.966886
      & 0.003210 & 0.003220
      & 60.00    & 59.95    \\

    KKLTI
      & 3.27426
      & 0.590521 & 0.590525
      & 0.974687 & 0.974687
      & 0.001550 & 0.001557
      & 60.00    & 59.92    \\
    \bottomrule
  \end{tabular}
  \caption{Comparison of inflationary observables for the pure potential $V_{\rm inf}$ and the fully corrected potential with $\varepsilon = -10^{-5}$. Each column shows the pivot-scale field value $\varphi_*$, end-of-inflation field $\varphi_{\rm end}$, scalar spectral index $n_s$, tensor-to-scalar ratio $r$, and number of $e$-folds $N$.}
  \label{tab:inflationary_comparison}
\end{table}

\vspace{1em}

To ensure proper CMB normalization, we also adjust the overall scale $V_0$ in each model to match the observed scalar power spectrum:
\begin{equation}
    A_s = \frac{V(\varphi_*)}{24\pi^2\,\epsilon(\varphi_*)}.
\end{equation}
The corresponding values of $V_0$ are presented in Table~\ref{tab:V0_normalizations}. In all cases, the fractional shift in $V_0$ induced by the correction is of order $10^{-3}$, confirming that the inflationary energy scale remains nearly unchanged.

\begin{table}[htb]
  \centering
  \small
  \begin{tabular}{@{} l cc @{}}
    \toprule
    \textbf{Model} 
      & $V_0^{\rm (pure)}$ 
      & $V_0^{\rm (full)}$ \\
    \midrule
    Starobinsky
      & $9.433 \times 10^{-11}$ & $9.462 \times 10^{-11}$ \\
    Fibre
      & $5.518 \times 10^{-11}$ & $5.527 \times 10^{-11}$ \\
    T-model ($\alpha = 1$)
      & $1.023 \times 10^{-10}$ & $1.026 \times 10^{-10}$ \\
    KKLTI
      & $4.931 \times 10^{-11}$ & $4.954 \times 10^{-11}$ \\
    \bottomrule
  \end{tabular}
  \caption{CMB-normalized energy scales $V_0$ for the pure and corrected potentials. All values are fixed to reproduce the observed scalar amplitude $A_s$. The fractional corrections are of order $10^{-3}$.}
  \label{tab:V0_normalizations}
\end{table}

\medskip

In summary, for $|\varepsilon| \lesssim 10^{-5}$, the impact of corrections on inflationary observables such as $n_s$, $r$, and $N$ remains negligible. This confirms the robustness of the SHS framework against small string-theoretic deformations in the scalar potential.

\medskip

Nevertheless, while these corrections leave inflationary predictions essentially unaffected, they can have pronounced effects in the post-inflationary era. In particular, small localized features in the potential—consistent with SHS dynamics—can transiently violate slow-roll conditions, amplifying curvature perturbations at specific scales. As we will explore in the next section, this opens a viable channel for generating primordial black holes (PBHs) with enhanced mass spectra even when $\varepsilon$ remains perturbatively small.

\subsection{Primordial Black Hole Production}\label{subsec7.2:PBH_production}

Cosmological models often extend beyond standard slow-roll inflation by considering scenarios in which features in the inflationary potential (e.g., bumps, dips, or steps) induce transient departures from quasi–de Sitter evolution \cite{Mishra:2019pzq,Bhatt:2022mmn, Geller:2022nkr, Green:2024bam, Ozsoy:2023ryl, Escriva:2022duf}. Such deviations can amplify specific modes in the primordial power spectrum, potentially leading to the formation of \emph{Primordial Black Holes} when overdensities re-enter the cosmological horizon at later times. Within the framework of \emph{finite moduli displacement}, these localized modifications to the potential can be implemented through small deformations in the functions $\mathcal{K}(T,\overline{T})$, $f(T,\overline{T})$, or $g(T)$, while ensuring that the overall volume remains bounded.

A hallmark of the SHS-based approach is that, once the canonical field $\varphi$ crosses the threshold $\tau = \tau_{0} + e^{-\Delta\varphi}/\xi \simeq \tau_{0}$, the theory transitions to a SUSY-restoring limit. Although the inflationary dynamics remain largely unaltered, in PBH-generating scenarios this SHS mechanism may also trigger a post-inflationary phase in which higher-order supergravity corrections or higher-dimensional operators become dynamically significant.

\vspace{3mm}

We now outline two mechanisms for embedding PBH-generating features within the SHS framework:

\begin{enumerate}
    \item \textbf{Localized Corrections via Nilpotent Couplings:}  
    The nilpotent couplings $f(T,\overline{T})$ and $g(T)$ (cf.\ \S\ref{sec5:SHS}) can introduce localized bumps or dips in the scalar potential. This is achieved by introducing small perturbative functions $\delta f(T)$ or $\delta g(T)$ that become non-negligible within a narrow region of $\tau$. By tuning these functions, one can transiently alter the slow-roll parameters and selectively enhance or suppress fluctuations at specific scales, which may later seed PBH formation upon horizon re-entry.  

    \item \textbf{Localized Corrections via Compactification Volume:}  
    Slight localized deformations in $\mathcal{K}(T,\overline{T}) \sim \mathcal{V}$ can similarly produce features in the scalar potential. Around a finite value $\tau = \tau_*$, such features can temporarily enhance the amplitude of short-wavelength modes. The interplay between Kähler geometry (via $\mathcal{K}_{T\overline{T}}$) and the nilpotent sector modifies the slow-roll parameters $(\epsilon, \eta)$, thereby creating favorable conditions for PBH formation.
\end{enumerate}

Once PBHs are generated through either mechanism, they may act as viable dark matter candidates or be probed via observational signatures such as gravitational waves, microlensing, or Hawking radiation, depending on their mass spectrum. Notably, the bounded nature of $\tau$ and the approximate constancy of the compactification volume help maintain consistency with swampland constraints, which are often in tension with large-field inflationary models. Additional observational signatures—such as non-Gaussian tails in the power spectrum—may further validate or constrain these localized bump/dip scenarios.

To illustrate the relevance of these mechanisms, consider the standard phenomenological approach in which a Gaussian bump or dip is manually introduced into the inflationary potential:
\begin{equation}\label{Gaussian_PBH_Method}
    V_{\text{inf}}(\varphi) \to V_{\text{inf}}(\varphi)\left[ 1 \pm G(\varphi) \right],
\end{equation}
where $G(\varphi)$ is a Gaussian profile centered at some $\varphi_0$. Such modifications are typically added by hand to demonstrate PBH formation at a localized scale. In comparison, the approximate SHS potential~\eqref{V_inf_small_eps} can be naturally reorganized as:
\begin{equation}
    V(\varphi) = V_{\text{inf}}(\varphi) \left[ 1 + \varepsilon \alpha(\varphi) \right] + \varepsilon V_0 \alpha(\varphi) + \mathcal{O}(\varepsilon^2).
\end{equation}
Thus, the first-order string corrections involving $\alpha(\varphi)$ can effectively reproduce a Gaussian-like feature, with $\varepsilon \alpha(\varphi)$ effectively playing the role of $G(\varphi)$. In this way, either the nilpotent couplings or volume deformations can naturally realize the bump/dip behavior without the need for phenomenological insertion.

Alternatively, an equivalent effect to the procedure in~\eqref{Gaussian_PBH_Method} can be realized by introducing a localized correction to the scalar coupling $s(\varphi)$ that appears in the nilpotent bilinear term $S \bar{S}$ within the Kähler potential~\eqref{Full_Sugra_Kahler_SHS}. Specifically, we consider the deformation:
\begin{equation}\label{SUGRA_PNH_Procedure}
s(\varphi) \to s(\varphi)\left[ 1 \pm G(\varphi) \right],
\end{equation}
which leads to the following modification in the scalar potential:
\begin{equation}
V(\varphi) \to V(\varphi) \pm  V_{\text{inf}}(\varphi) G(\varphi)\left[ 1 + \varepsilon \alpha(\varphi) \right] .
\end{equation}
In this formulation, the Gaussian profile $G(\varphi)$ couples naturally through the string-theoretic correction $\varepsilon \alpha(\varphi)$, allowing the bump or dip structure to emerge from the underlying geometric data. This provides a more fundamental alternative to phenomenological insertions, embedding PBH-generating features directly into the moduli-dependent effective theory. We will explore the phenomenological implications of this mechanism through numerical analysis and demonstrate that it can substantially enhance the PBH mass spectrum, potentially by up to four orders of magnitude.

\subsection{Power Spectrum Amplification and PBH Abundance}
\paragraph{Primordial power spectrum.}The primordial power spectrum, which seeds the formation of PBHs, is computed by numerically solving the Mukhanov–Sasaki (MS) equation on a time-evolving inflationary background. This equation governs the evolution of scalar perturbations in single-field inflation and is derived from the second-order action for gauge-invariant curvature perturbations.

In canonical form, the MS equation reads:
\begin{equation}
    \label{eq:ms_canonical}
    v^{\prime\prime} - \nabla^{2}v - \frac{z^{\prime\prime}}{z}v = 0,
\end{equation}
where $v(\tau, \mathbf{x})$ is the Mukhanov–Sasaki variable, related to the comoving curvature perturbation via $v(\tau, \mathbf{x})  = z\, \zeta(\tau, \mathbf{x})$, and $z = a M_p \sqrt{2\epsilon_H}$. The dynamics of scalar fluctuations are governed by the effective mass term $z^{\prime\prime}/z$, which encodes the time dependence of the background and is sensitive to the inflationary model. Upon Fourier transformation, the equation decouples into independent equations for each comoving mode $k$:
\begin{equation}
    \label{eq:ms_fouriermodes}
    v^{\prime\prime}_k + \left(k^{2} - \frac{z^{\prime\prime}}{z}\right)v_k = 0,
\end{equation}
with the effective mass term given by \cite{Bhatt:2022mmn}:
\begin{equation}
    \label{eq:effective_mass_term}
    \frac{z^{\prime\prime}}{z} = a^{2}\left[\frac{5}{2}\frac{\dot{\varphi}^{2}}{M_p^{2}} + 2\frac{\dot{\varphi}}{H}\frac{\ddot{\varphi}}{M_p^{2}} + 2H^{2} + \frac{1}{2}\frac{\dot{\varphi}^{4}}{H^{2}M_p^{4}} - V^{\prime\prime}\left(\varphi\right)\right].
\end{equation}
The second-order differential equation in Eq.\eqref{eq:ms_fouriermodes} is solved numerically for a wide range of modes k that re-enter the horizon during the radiation-dominated era and are relevant for PBH formation. 

The background evolution is obtained by solving the classical equations of motion for the inflaton field $\varphi$, its velocity $\dot{\varphi}$, and the Hubble parameter $H$:
\begin{equation}\label{eq:background_dynamics}
        H^{2} = \frac{1}{3M_p^{2}}\left(\frac{1}{2}\dot{\varphi}^{2} + V(\varphi)\right), \quad
        \dot{H} = -\frac{1}{2}\dot{\varphi^{2}}, \quad
        \ddot{\varphi} = - 3H\dot{\varphi}-V^{\prime}(\varphi).
\end{equation}
For numerical purposes, we introduce dimensionless variables:
\begin{equation}
    T = t\,M_p\quad \hat{a} = a\,M_p, \quad
    x = \frac{\varphi}{M_p},\quad y = \frac{\dot{\varphi}}{M_p^{2}},\quad z = \frac{H}{M_p}.
\end{equation}
Background dynamics in Eq.\eqref{eq:background_dynamics} are then recast in terms of dimensionless variables and numerically evolved using either the \texttt{odeint} or \texttt{solve\_ivp} solvers from the \textit{SciPy} library, yielding the following dimensionless dynamical system:
\begin{equation}
        \frac{\textrm{d}\hat{a}}{\textrm{d}T} = \hat{a}z, \quad 
        \frac{\textrm{d}x}{\textrm{d}T} = y, \quad
        \frac{\textrm{d}y}{\textrm{d}T} = - 3zy-\frac{\textrm{d}V(x)}{\textrm{d}x}, \quad
        \frac{\textrm{d}z}{\textrm{d}T} = -\frac{1}{2}y^{2}.
\end{equation}
The background integration continues until the Hubble slow-roll parameter $\epsilon_H = \frac{1}{2} (\dot{\varphi}/H)^2$ reaches unity, signaling the end of inflation.

Once the background evolution is established, the MS equation \eqref{eq:ms_fouriermodes} is solved for a range of comoving modes $k$. The initial conditions are set in the subhorizon regime $(k \gg aH)$ using the Bunch–Davies vacuum:
\begin{equation}
    v_k \rightarrow \frac{1}{\sqrt{2k}}\mathrm{e}^{-ik\tau}.
\end{equation}
 The solution evolves forward in time until the mode exits the horizon $(k \ll aH)$, at which point the curvature perturbation freezes out. The scalar power spectrum is then extracted from:
\begin{equation}
    \mathcal{P}_\mathcal{\zeta}(k) = \frac{k^3}{2\pi^2} \frac{\left|v_k\right|^{2}}{z^{2}} \bigg|_{k \ll aH}.
\end{equation}
In practice, the MS system is evolved in physical time alongside the background equations, using initial data consistent with the vacuum state, and monitored until freeze-out, defined by the condition $ k/(aH) \lesssim 10^{-2}$. The full system of background and perturbation variables is integrated simultaneously. Scalar and tensor fluctuations both evolve, although only scalar modes contribute to PBH formation.

To ensure computational efficiency, we distribute a logarithmic sampling of $k$-modes across MPI processes using \texttt{mpi4py}. Each process evolves a subset of modes independently, and the final power spectrum values are collected for post-processing.

To link the spectrum with cosmological observables, we identify the pivot scale $k_{*} = 0.05\, \mathrm{Mpc}^{-1}$ and extract key inflationary parameters:
\begin{subequations}
    \begin{align}
        &\epsilon_{H} = \frac{1}{2}\frac{y^{2}}{z^{2}},\quad \eta_{H} = -\frac{1}{yz}\frac{\textrm{d}y}{\textrm{d}{T}}\\
        &n_\mathrm{s} = 1 + 2\eta_{H} - 4\epsilon_{H},\quad r = 16\epsilon_{H},\quad A_{s} = \frac{z^{2}}{8\pi^{2}\epsilon_{H}}
    \end{align}
\end{subequations}
evaluated at the time when $k_{*} = a H$.

\begin{figure}[htb]
\centering
\includegraphics[width=\textwidth, height=\textwidth, keepaspectratio]{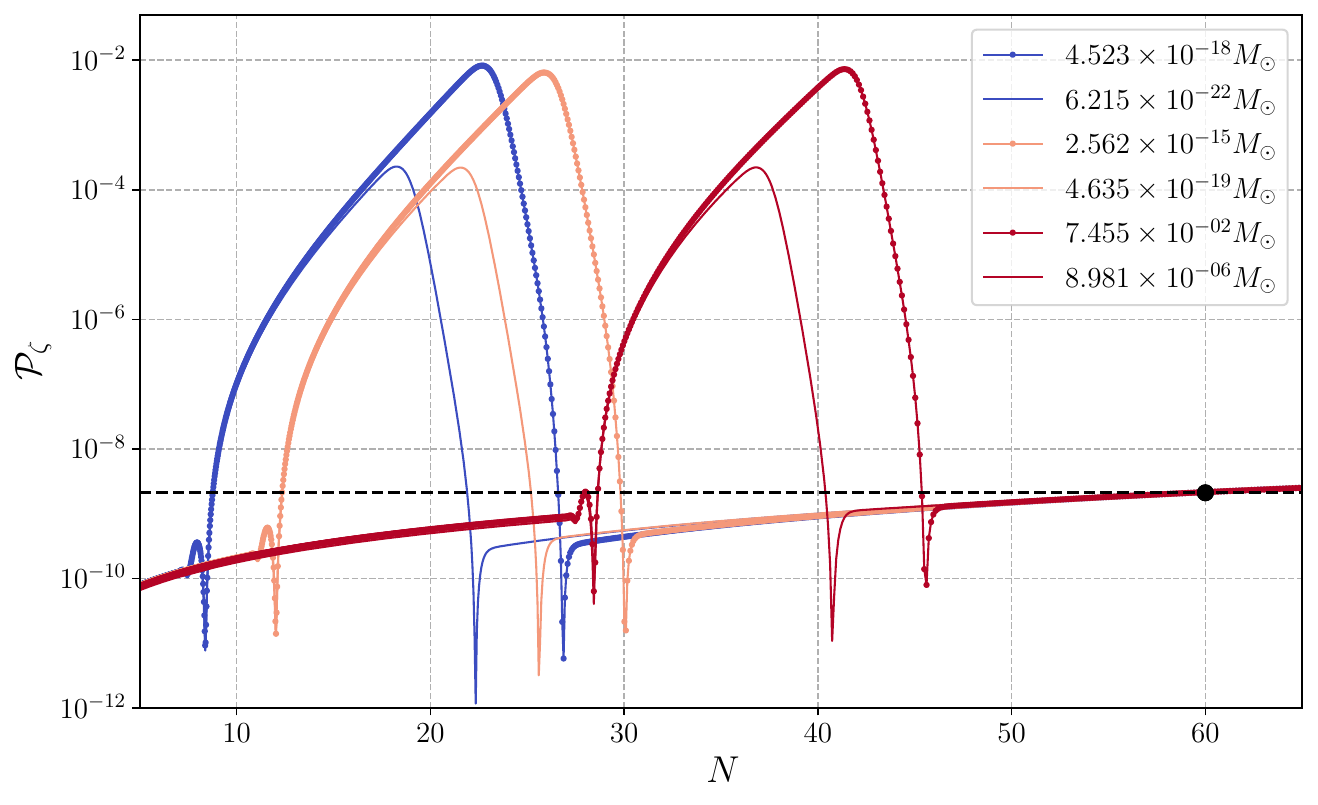
}
\caption{\label{fig:pswrb}%
Amplification of the curvature power spectrum for the pure potential $V_{\mathrm{inf}}$ (solid lines) versus the fully corrected potential in Eq.~\eqref{eq:Scalar_Pot}, with $\varepsilon = -10^{-5}$ and $W_0 = 1$, under the setup defined by Eq.~\eqref{Full_Sugra_Kahler_SHS} and the procedure in Eq.~\eqref{SUGRA_PNH_Procedure}. The resulting PBH masses, indicated in the legend, are extracted from the peak scales of the amplified spectra. The blue, orange, and red curves correspond to the initial conditions labeled \textit{Small}, \textit{Medium}, and \textit{Large} in Table~\ref{tab:initial_values}, with the associated inflationary observables summarized in Table~\ref{tab:obs_results}.}
\end{figure}
\paragraph{Power spectrum amplification.}
Following the framework outlined above, we computed the primordial power spectrum for two potential configurations. The first is the pure inflationary potential $V_{\mathrm{inf}} = V_0\,s(\varphi)$ defined in Eq.\eqref{eq:modelPotential}. The second is the fully corrected potential of Eq.\eqref{eq:Scalar_Pot}, evaluated exactly to capture the full SHS dynamics.

To introduce a localized feature, we next applied the procedure in Eq.~\eqref{SUGRA_PNH_Procedure} (with the plus sign), adding a Gaussian term of the form
\begin{equation}
    G(\varphi) = A\, \exp\left(-\frac{1}{2}\frac{(\varphi - \varphi_0)^{2}}{\sigma^{2}}\right)
\end{equation}
which acts as a bump in the potential. The initial conditions used are summarized in Table~\ref{tab:initial_values}, with the corresponding observable predictions given in Table~\ref{tab:obs_results}.

\begin{table}[htb]
    \small
    \centering
    \begin{tabular}{l ccccccc}
        \toprule
        & \textbf{Model} & $A (\times 10^{-3})$ & $\sigma (\times 10^{-2})$ & $\varphi_0$ & $\varphi_i$ & $V_0 (\times 10^{-10})$ & $\mathcal{V}_{0}^{2/3}(\times 10^{3})$ \\
        \midrule
        \multirow{2}{*}{\textit{Small}} &Pure   & $1.876$ & $1.993$ & $2.0045$ & $3.3424780$ & $0.65876$ & None \\
        & Full     & $1.876$ & $1.993$ & $2.0045$ & $3.2813400$ & None                     & $3.40935$ \\
        \hline
        \multirow{2}{*}{\textit{Medium}} &Pure     & $1.17035$ & $1.59$ & $2.187451$ & $3.3420300$ & $0.6595$ & None \\
        & Full  & $1.17035$ & $1.59$ & $2.187451$ & $3.2411673$ & None                    & $3.30020$ \\
        \hline
        \multirow{2}{*}{\textit{Large}} &Pure         & $0.3502$ & $0.8818$ & $2.7114$ & $3.3621045$ & $0.62994$ & None \\
        & Full     & $0.3502$ & $0.8818$ & $2.7114$ & $3.3002120$ & None                     & $3.46078$ \\
        \bottomrule
    \end{tabular}

    \caption{Initial conditions used for the pure and fully corrected KKLTI model configurations. The parameter $\varphi_i$ denotes the field value at which the numerical integration begins.}
\label{tab:initial_values}
\end{table}

\begin{table}[htb]
    \small
    \centering
    \begin{tabular}{l ccccccc}
        \toprule
        & \textbf{Model} & $N$ & $\varphi_{*}$ & $\varphi_{\rm end}$ & $n_s$ & $r$ & $A_s (\times 10^{-9})$ \\
        \midrule
        \multirow{2}{*}{\textit{Small}} & Pure         & 60.0007 & 3.113 & $0.587$ & 0.9695 & 0.00207 & $2.1000$ \\
                               & Full     & 60.0004 & 3.039 & $0.587$ & 0.9665 & 0.00237 & $2.1002$ \\
        \hline
        \multirow{2}{*}{\textit{Medium}} & Pure         & 60.0002 & 3.113 & $0.587$ & 0.9695 & 0.00207 & $2.1001$ \\
                                & Full     & 60.0000 & 2.989 & $0.590$ & 0.9643 & 0.00261 & $2.1000$ \\
        \hline
        \multirow{2}{*}{\textit{Large}} & Pure         & 60.0009 & 3.137 & $0.589$ & 0.9704 & 0.00198 & $2.1003$ \\
                            & Full     & 60.0000 & 3.062 & $0.590$ & 0.9675 & 0.00227 & $2.1003$ \\
        \bottomrule
    \end{tabular}
    \caption{Observational results for pure and fully corrected models.}
    \label{tab:obs_results}
\end{table}

PBH formation requires the power spectrum to be amplified by approximately seven orders of magnitude relative to its value at the pivot scale. Figure \ref{fig:pswrb} shows that the fully corrected potential configuration gives rise to larger amplification up to two orders of magnitude on the power spectrum compared to the one with the pure potential configuration. Moreover, the resulting PBH masses extracted from the spectrum peaks show a mass enhancement of up to four orders of magnitude, potentially lifting them above the evaporation threshold.

\begin{figure}[htb]
\centering
\includegraphics[width=\textwidth, height=\textwidth, keepaspectratio]{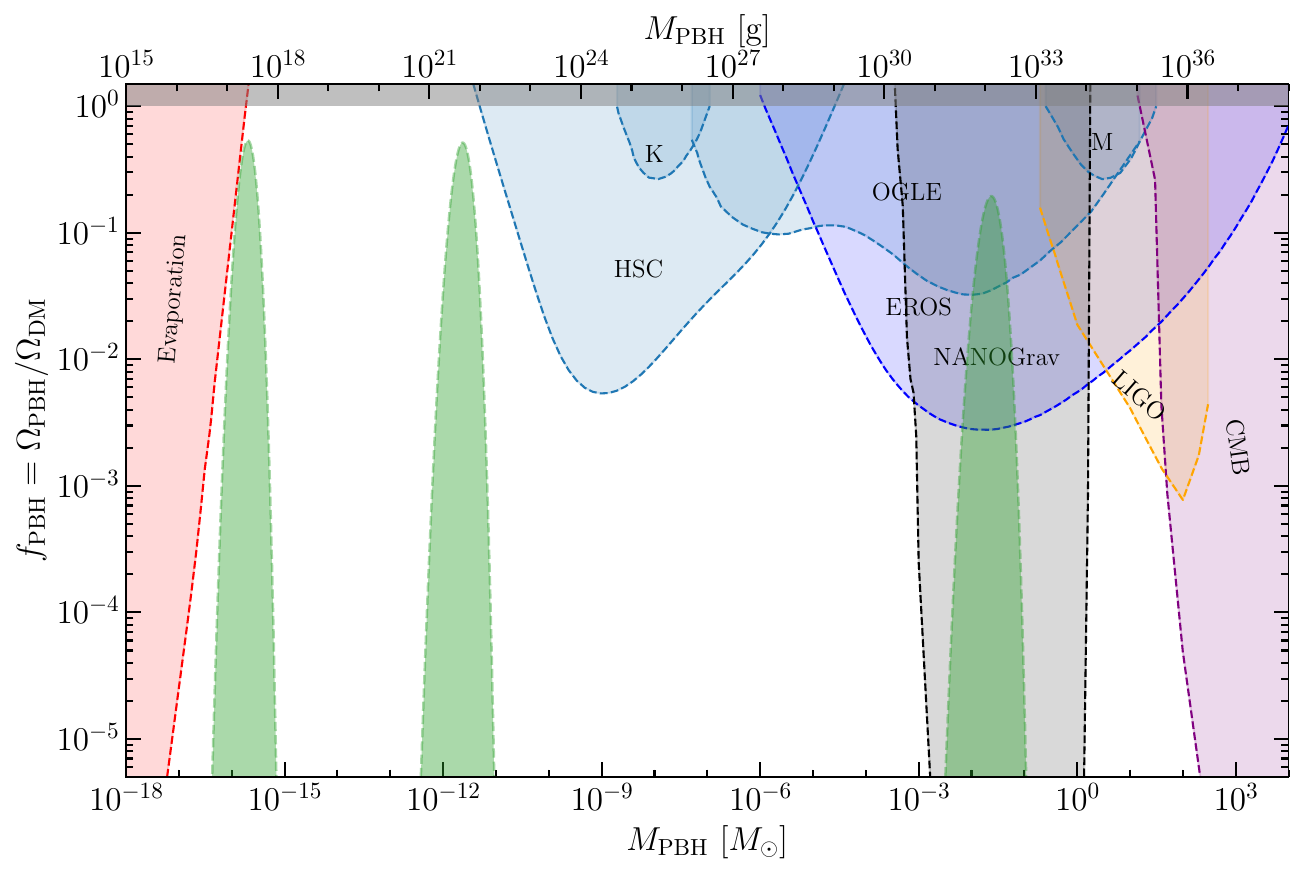}
\caption{\label{fig:f_abundance_plot}Fractional PBH abundances $f_{\mathrm{PBH}}(M_{\mathrm{PBH}})$ as a function of mass, computed for three different parameter sets—\textit{Small}, \textit{Medium}, and \textit{Large}—corresponding to the fully corrected potential configurations. Shaded green regions indicate the resulting mass spectra, based on initial conditions largely drawn from Table~\ref{tab:initial_values}. Current observational constraints are overlaid: evaporation limits \cite{evap1, evap2, evap3, evap4, evap5, evap6, evap7, evap8, evap9, evap10, evap11, evap12} (red), microlensing constraints \cite{ml1, ml2, ml3, ml4, ml5, ml6} (light and dark blue), CMB constraints \cite{cmbbound} (purple), gravitational wave limits from LIGO \cite{ligo1, ligo2} (orange), and PTA constraints from NANOGrav \cite{nanograv} (grey). The plot was generated using the open-source \textit{PBHbounds}~\cite{pbhbounds}.}
\end{figure}
\paragraph{PBH abundance.}
To compute the PBH abundance, we follow the procedure outlined in~\cite{Mishra:2019pzq,Bhatt:2022mmn}, relating the curvature power spectrum to the PBH mass via the horizon mass at reentry:
\begin{equation}
\frac{M_{\mathrm{PBH}}}{M_\odot} = 1.13 \times 10^{15} \left(\frac{\gamma}{0.2}\right) \left(\frac{g_*}{106.75}\right)^{-1/6} \left(\frac{k_{\mathrm{PBH}}}{k_*}\right)^{-2},
\end{equation}
where $\gamma \simeq 0.2$ is the collapse efficiency and $g_*$ is the effective number of relativistic degrees of freedom, assuming $N_{\mathrm{eff}} = 3.046$.

The corresponding fractional abundance is defined via the mass function:
\begin{equation}
\label{eq:mass_abundance}
f_{\mathrm{PBH}}\left(M_{\mathrm{PBH}}\right)=1.68 \times 10^8\left(\frac{\gamma}{0.2}\right)^{1 / 2}\left(\frac{g_*}{106.75}\right)^{-1 / 4}\left(\frac{M_{\mathrm{PBH}}}{M_{\odot}}\right)^{-1 / 2} \beta\left(M_{\mathrm{PBH}}\right),
\end{equation}
where $\beta(M)$ denotes the mass fraction at formation. In the Press–Schechter formalism~\cite{psformalism}, this is given by
\begin{equation}
\label{eq:mass_fraction}
\beta\left(M_{\mathrm{PBH}}\right) = \gamma \int_{\delta_{\mathrm{th}}}^1 \frac{d\delta}{\sqrt{2\pi} \sigma_{M_{\mathrm{PBH}}}} \exp\left(-\frac{\delta^2}{2\sigma_{M_{\mathrm{PBH}}}^2}\right) \approx \gamma \frac{\sigma_{M_{\mathrm{PBH}}}}{\sqrt{2\pi} \delta_{\mathrm{th}}} \exp\left(-\frac{\delta_{\mathrm{th}}^2}{2\sigma_{M_{\mathrm{PBH}}}^2}\right),
\end{equation}
with $\delta_{\mathrm{th}}$ the critical threshold for collapse, and $\sigma_{M{\mathrm{PBH}}}$ the variance of the density contrast:
\begin{equation}
\label{eq:variance_density_contrast}
\sigma_{M_{\mathrm{PBH}}}^2 = \int \frac{dk}{k} P_\delta(k) W^2(k, k_{\mathrm{PBH}}), \qquad
W(k, k_{\mathrm{PBH}}) = \exp\left(-\frac{k^2}{2k_{\mathrm{PBH}}^2}\right).
\end{equation}
Here, $W(k, k_{\mathrm{PBH}})$ is a Gaussian window function that smooths the density contrast over the comoving Hubble scale at reentry $R = 1/k_{\mathrm{PBH}}$.

During the radiation-dominated era, the density contrast power spectrum $\mathcal{P}_\delta$ is related to the curvature power spectrum $\mathcal{P}_\zeta$ via~\cite{Green:2004wb}:
\begin{equation}
\label{eq:densitycontrast_pspec}
P_\delta(k) = \frac{16}{81} \left(\frac{k}{k_{\mathrm{PBH}}}\right)^4 P_\zeta(k).
\end{equation}
Substituting the numerically computed $\mathcal{P}_\zeta(k)$ into Eq.~\eqref{eq:densitycontrast_pspec}, we evaluate $\sigma_{M_{\mathrm{PBH}}}$, determine $\beta(M)$ via Eq.~\eqref{eq:mass_fraction}, and thereby obtain the fractional PBH abundance $f_{\mathrm{PBH}}(M)$. We performed three separate calculations of the fractional PBH abundance corresponding to the \textit{Small}, \textit{Medium}, and \textit{Large} mass spectrum configurations, following the procedure outlined above. The resulting distributions are shown in Figure~\ref{fig:f_abundance_plot}. We adopt a density contrast threshold of $\delta_{\mathrm{th}} = 0.61$, consistent with the theoretical bound $0.3 < \delta_{\mathrm{th}} < 0.66$~\cite{dth1, dth2, dth3, dth4, dth5, dth6, dth7, dth8}. Initial values for $A$, $\sigma$, $V_0$, and $\mathcal{V}_0$ were taken from Table\ref{tab:initial_values}, with minor adjustments to $\varphi_0$ and $\varphi_i$ to mitigate numerical instabilities. For reproducibility, the $(\varphi_0, \varphi_i)$ pairs used are: $(2.0000055,\ 3.25861)$ for \textit{Small}, $(2.187451,\ 3.24117)$ for \textit{Medium}, and $(2.67,\ 3.2804250)$ for \textit{Large}, with deviations from Table~\ref{tab:initial_values} not exceeding $0.04,M_p$.

These results highlight a key implication of the SHS-induced corrections: the amplified curvature power spectrum systematically shifts the PBH mass distribution toward larger values, significantly improving the survivability of PBHs that would otherwise evaporate. In particular, the \textit{Small} mass configuration—which lies below the evaporation threshold in the pure potential case—is pushed into a regime where Hawking evaporation is avoided. This opens the possibility for such PBHs to persist until the present epoch and contribute to the dark matter abundance, underscoring the broader cosmological relevance of the SHS mechanism.

\section{Conclusion}\label{sec8:Concl}

This work developed the \emph{Supersymmetric Horizon Stabilization} mechanism as a technically controlled and conceptually unified framework for embedding metastable de Sitter vacua, trans-Planckian inflation, and asymptotic supersymmetry within a unified effective $\mathcal{N}=1$ supergravity framework. Built from minimal ingredients—a logarithmic Kähler potential corrected by a nilpotent superfield—the SHS setup yields a single scalar potential that smoothly connects distinct cosmological epochs without staged uplifting, auxiliary brane sectors, or hierarchically separated stabilization schemes. This offers not only technical simplicity but a new organizing principle for supergravity models rooted in string-theoretic corrections, standing in contrast to conventional scenarios such as KKLT or LVS, which assemble vacua through multi-step, scale-separated ingredients.

At the core of SHS lies a scalar potential (Eq.~\eqref{V_log}) that governs the dynamics of the light scalar degrees of freedom and encapsulates SUSY breaking across moduli space, derived from Kähler potentials of the form
\begin{equation}
    K = -3 \log\left[\mathcal{K}(T,\overline{T}) + f(T,\overline{T})\,S\overline{S} + g(T)S + \overline{g(T)}\,\overline{S}\right],
\end{equation}
paired with a constant superpotential $W = W_0$. The coupling functions $f(T, \overline{T})$ and $g(T)$ control the scalar interactions of the nilpotent sector and shape the pattern of SUSY breaking across the moduli space. After imposing nilpotency and eliminating auxiliary fields, this structure yields a rich scalar dynamics with only a handful of tunable functions.

To isolate physically viable models, we imposed four robust constraints: ghost-free dynamics (via the no-ghost conditions~\eqref{NG1}–\eqref{NG2}), metastability of de Sitter vacua, smooth transitions to Minkowski space, and asymptotic supersymmetry restoration, i.e., $D_T W \to 0$ as $\varphi \to \infty$. These criteria were translated into analytic inequalities involving the nilpotent couplings and the moduli geometry (Eq.~\eqref{branch1}–\eqref{branch2}), leading to a complete classification of viable EFT branches. Representative examples, shown in Figures~\ref{fig:1} and \ref{fig:2}, exhibit rich scalar-field trajectories with continuous transitions between dS, Minkowski, and supersymmetric AdS phases.

A key analytic result of this work is the emergence of exponential moduli profiles,
\begin{equation}
    \tau(\varphi) = \tau_0 + \frac{1}{\xi} e^{-\Delta \varphi},
\end{equation}
where $\varphi$ is the canonical inflaton and $\tau = \text{Re}(T)$ encodes physical volume. This form arises from both conformal frame transformations and the balancing of flux and non-perturbative terms in type IIB compactifications. As shown in Appendix~\ref{sec:Appendix_CanonicalField}, the exponential relation is not an ansatz, but a consequence of the Weyl-rescaled kinetic structure. This ensures that even deep into the trans-Planckian regime, $\tau$ remains bounded and geodesic distances are finite. As a result, SHS evades the Gravitino Distance Conjecture by preventing $m_{3/2} \to 0$, and complements refined Swampland Distance Conjectures by allowing infinite canonical field excursions without triggering light towers or decompactification.

This geometric and dynamical consistency motivates the \emph{Supersymmetric Horizon Conjecture}: the proposal that consistent gravitational EFTs may asymptotically flow toward supersymmetric endpoints—not in the ultraviolet, but as well-defined infrared boundaries. Supersymmetry, in this picture, reemerges not as a fading imprint of high-energy physics, but as an infrared attractor governing the scalar-field trajectory. Asymptotic canonical field limits, traditionally viewed as signals of breakdown, are here recast as dynamically stable fixed points—a shift with potentially broad implications for cosmology and EFT design.

We also remark on the implications of the Trans-Planckian Censorship Conjecture (TCC)~\cite{Bedroya:2019snp}, which limits the lifetime of inflationary or quasi-de Sitter phases to $ T \leq \frac{1}{H} \log\left( \frac{M_p}{H} \right)$, in order to prevent sub-Planckian modes from exiting the Hubble horizon. In our setup, the metastable de Sitter-like phase arises as part of a continuous scalar-field evolution governed by a single EFT, and this evolution concludes in a supersymmetric AdS vacuum that satisfies both the Generalized Distance and AdS Swampland Conjectures. While the TCC may introduce additional constraints on the maximal duration or energy scale of the intermediate phase, a detailed analysis lies beyond the scope of this work. We acknowledge that a literal interpretation of the TCC may be in mild tension with our setup, and view this as a meaningful direction for future investigation—including possible refinements or loopholes motivated by the presence of supersymmetric asymptotics. (See also~\cite{Cribiori:2025oek,Bedroya:2025fwh,Basile:2025zjc,Vafa:2025nst,Casas:2024oak,Bedroya:2024zta,Kitazawa:2024deb,Eichhorn:2024rkc,Brandenberger:2024vgt,Apers:2024ffe,Montero:2024qml,Hebecker:2023qke,Brandenberger:2023ver,vandeHeisteeg:2023uxj,Tsukahara:2023xzy,Blamart:2023ixr,Andriot:2022brg,Bedroya:2022tbh,Calderon-Infante:2022nxb,Andriot:2022xjh,Rudelius:2022gbz,Cunillera:2022hdb,Schneider:2022gkg,Dine:2021gxg,Rudelius:2021azq,Guleryuz:2021zik,Brandenberger:2021pzy,Andriot:2020lea} for related discussions.)

On the phenomenological side, we have shown in Section~\ref{subsec7.2:PBH_production} that the SHS framework permits controlled, localized features in the scalar potential that can transiently amplify curvature perturbations and trigger the formation of primordial black holes. As illustrated in Fig.~\ref{fig:pswrb}, the resulting PBHs can attain masses several orders of magnitude larger than those predicted by conventional inflationary scenarios, providing a novel observational window into trans-Planckian dynamics governed by string-theoretic corrections. Importantly, this mechanism emerges naturally from the underlying moduli geometry, without requiring ad hoc modifications to the potential. The same framework robustly supports standard slow-roll inflation, maintaining compatibility with current cosmological data while offering new opportunities to probe high-scale physics beyond the reach of CMB observables.

The theoretical consistency of SHS is further supported by its geometric origins. In Appendix~\ref{sec:Appendix_Superconformal}, we reconstruct the Jordan-frame Lagrangian from superconformal supergravity, demonstrating how logarithmic Kähler potentials and nilpotent sectors emerge from symmetry principles. In Appendix~\ref{sec:Appendix_GeometricOrigin}, we trace the scalar–gravity coupling $\Omega(\tau) R$ to the dimensional reduction of the 10D NS–NS sector of type IIB string theory,
\begin{equation}
    S_{10} = \frac{1}{2\kappa_{10}^2} \int d^{10}X \sqrt{-G_{10}}\, e^{-2\phi}  R_{10}  \quad \to \quad \frac{1}{2\kappa_4^2} \int d^4x \sqrt{-g_4}\, e^{-2\phi_4(x)} R_4,
\end{equation}
which leads, upon compactification, to a Jordan-frame structure with $\Omega \propto e^{-2\phi_4} \sim \tau - \tau_{0}$. While our formulation remains at the level of four-dimensional effective theory, its ingredients and structure align with known features of string compactifications, including $\alpha'^3$ and loop corrections, warped volume dependence, and nilpotent brane dynamics.

Looking forward, several promising directions warrant deeper investigation. A key avenue is the explicit realization of SHS within compactified string constructions—such as flux-stabilized Calabi–Yau orientifolds with nilpotent branes—to assess UV consistency, anomaly cancellation, and moduli stabilization in concrete settings. Another is the extension of the framework to include additional moduli, visible-sector matter, or axion-like fields, enabling connections to reheating, low-energy phenomenology, and potential mediation of supersymmetry breaking to observable sectors. We have also initiated a study of non-perturbative vacuum decay using both Euclidean and Hamiltonian approaches, and it would be valuable to extend this to include multi-field dynamics, gravitational backreaction, or holographic interpretations of metastability.

In parallel, it would be worthwhile to deepen the study of PBH phenomenology within this setup—particularly the mapping of PBH mass spectra, merger rates, and associated gravitational-wave signals arising from localized SHS-induced features. Such signatures could be probed by upcoming experiments like LISA~\cite{LISA:2017pwj}, SKA~\cite{Janssen:2014dka}, ET~\cite{ET:2019dnz} and microlensing surveys~\cite{Niikura:2017zjd,Niikura:2019kqi,Blaineau:2022nhy}, offering a rare observational handle on high-scale string-inspired inflation.

Finally, the SUSY-restoring asymptotic boundary of SHS geometries invites further exploration in the context of holography—particularly regarding possible dual descriptions in terms of supersymmetric RG flows or IR limits of emergent conformal field theories.

In conclusion, the SHS mechanism challenges the conventional notion that metastable de Sitter vacua must be engineered through staged uplifting or fine-tuned hierarchies. It offers a structurally minimal, dynamically coherent, and observationally testable framework in which inflation, vacuum stability, and asymptotic supersymmetry emerge from a single effective potential. More broadly, it opens a new perspective on gravitational EFTs—one in which supersymmetry is not merely hidden or broken, but serves as a unifying boundary condition—at both the origin and the asymptotic edge of cosmic evolution—shaped not by conjectural uplifts or fine-tuned hierarchies, but by geometry, symmetry, and analytic control.


\section*{Acknowledgments}
The authors thank Miguel Montero for his insightful comments and suggestions during the early stages of this work. The contribution of O.G. was supported in part by the Istanbul Technical University Research Fund (grant number TGA-2024-45577). O.G. also acknowledges the scientific exchange and collaboration fostered by COST Action CA21106 COSMIC WISPers (European Cooperation in Science and Technology). The numerical calculations reported in this paper were partially performed at TUBITAK ULAKBIM, High Performance and Grid Computing Center (TRUBA resources).

The authors are also grateful to Mehmet Özkan and A. Savaş Arapoğlu for the broader academic guidance and support they have provided throughout the years, which helped cultivate the environment in which this collaboration could take root.


\appendix
\section{Geometric Origin of the Canonical Field and the Role of Loop Corrections}
\label{sec:Appendix_GeometricOrigin}

This appendix presents a self-contained derivation of the Jordan-frame supergravity Lagrangian from its superconformal origin, illustrating how nilpotent constraints, non-minimal gravitational couplings, and string-inspired corrections combine into a coherent scalar–gravity system with potential relevance for cosmological model building. 

We begin by embedding the theory in the superconformal framework, where the action is constructed using a conformal compensator multiplet and projective coordinates. Through a field-dependent gauge-fixing procedure, we recover the Jordan frame, characterized by a non-minimal coupling between the scalar field and gravity via the frame function $\Omega(\tau)$.

A central outcome of this construction is the identification of both the scalar potential and the kinetic structure in the Jordan frame, followed by their transformation to the Einstein frame. Crucially, the kinetic sector is governed not solely by the Jordan-frame prefactor $K_{\tau\tau}$, but by the full Einstein-frame metric $G_{\tau\tau}$, which incorporates contributions from the Weyl rescaling. This distinction becomes essential once $\alpha'^3$ and string loop corrections are included, as they deform the field-space geometry in a nontrivial, yet controllable, way.

In particular, the exponential behavior of the canonically normalized field $\varphi$, which underlies many inflationary mechanisms, is not an imposed ansatz but a consequence of the full Einstein-frame kinetic metric. Specifically, the mapping
\begin{equation}
    \Omega(\tau) = \xi (\tau - \tau_{0}) =  e^{\pm \sqrt{\tfrac{2}{3}} \varphi}
\end{equation}
arises naturally once the Weyl contribution dominates over the Jordan-frame term—a regime reliably attained in the presence of small higher-derivative or loop-induced corrections.

This reflects a deeper geometric origin: the flattening of the potential and the emergence of exponential hierarchies follow directly from the interplay of nilpotency, Kähler invariance, and the conformal gauge-fixing of the compensator. The consistency of the potential across Jordan and Einstein frames further emphasizes the physical meaningfulness of the superconformal embedding.

\paragraph{String-Frame Origin of the Jordan Coupling.}
To close the circle, we briefly trace the non-minimal coupling $\Omega(\tau) R$ back to its geometric origin in ten-dimensional string theory. In string theory, this coupling arises from dimensional reduction of the 10D NS–NS sector in string frame:
\begin{equation}
    S_{10} = \frac{1}{2 \kappa_{10}^2} \int d^{10}X \sqrt{-G_{10}}\, e^{-2\phi} \left[
    R_{10} + 4(\partial \phi)^2 - \frac{1}{2} |H_3|^2
    \right].
\end{equation}
Assuming a product metric 
$
ds^2_{10} = g_{\mu\nu}(x)\, dx^\mu dx^\nu + g_{mn}(y)\, dy^m dy^n
$
and vanishing three-form $H_3 = 0$, the internal volume is:
\begin{equation}
    \mathcal{V} = \int d^6 y \sqrt{g_6}.
\end{equation}
Compactifying to four dimensions gives:
\begin{equation}
    S_4 = \frac{\mathcal{V}}{2\kappa_{10}^2} \int d^4x \sqrt{-g_4}\, e^{-2\phi(x)} R_4
    = \frac{1}{2\kappa_4^2} \int d^4x \sqrt{-g_4}\, e^{-2\phi_4(x)} R_4,
\end{equation}
where $e^{-2\phi_4} \equiv \mathcal{V} e^{-2\phi}$ and $\kappa_4^2 = \kappa_{10}^2/\mathcal{V}$. This leads precisely to the Jordan-frame structure $\Omega(\tau) R$, with $\Omega \propto e^{-2\phi_4} \sim \tau - \tau_{0}$, as used in Eq.~\eqref{eq:Omega_Jordan}; see \cite{Giddings:2001yu} for a detailed derivation.

\medskip

Altogether, the results in this appendix demonstrate that the superconformal embedding and its associated gauge-fixing are not merely technical conveniences, but rather suggest that the superconformal embedding captures important geometric and physical features of the scalar–gravity system. The canonical field $\varphi$, the exponential potential structure, and the effective dynamics can all be understood within this unified framework—bridging 4D supergravity, higher-dimensional string theory, and cosmological phenomenology in a conceptually unified manner.

\subsection{Superconformal Embedding and Jordan Frame Lagrangian}\label{sec:Appendix_Superconformal}

We present the derivation of the Jordan-frame supergravity Lagrangian for a model involving a nilpotent superfield $S$ and a chiral superfield $T$. The construction begins from a superconformal embedding and proceeds through gauge fixing, leading to a Lagrangian of the form
\begin{equation}
    \mathcal{L}_\textrm{J} = \sqrt{-g_\textrm{J}} \left[
\frac{1}{2} \Omega(\tau)\, R_\textrm{J}
- \frac{1}{4} \Omega(\tau) K_{\tau\tau}(\tau)\, g_\textrm{J}^{\mu\nu} \partial_\mu \tau \partial_\nu \tau
- V_\textrm{J}(\tau)
\right],
\end{equation}
with $T = \bar{T} = \tau \in \mathbb{R}$.

\paragraph{Supergravity Setup.}
The Einstein-frame Kähler potential is given by:
\begin{equation}
    K(T,\bar{T},S,\bar{S}) = -3 \log\left[\mathcal{K}(T,\bar{T}) + f(T,\bar{T})\, S\bar{S} + g(T) S + \bar{g}(\bar{T}) \bar{S} \right],
\end{equation}
and the superpotential is constant:
\begin{equation}
    W = W_0.
\end{equation}
The field $S$ is nilpotent, satisfying $S^2 = 0$. The function $\mathcal{K}(T,\bar{T})$ contains the kinetic information of the scalar field $T$.

\paragraph{Superconformal Embedding.}
We begin in the superconformal (or conformal compensator) formulation to derive our single-field action from the underlying supergravity. This framework includes a chiral compensator multiplet $X^0$ and physical multiplets $X^1$ and $X^2$, governed by:
\begin{equation}
    \mathcal{N}(X,\bar{X}) \quad \text{(the $D$-term density),} \qquad \mathcal{W}(X) \quad \text{(the $F$-term superpotential).}
\end{equation}
The bosonic action in superconformal gauge reads:
\begin{equation}
    \mathcal{L} = [\mathcal{N}(X,\bar{X})]_D + [\mathcal{W}(X)]_F \longrightarrow \int \! d^4x\, \sqrt{-g_\textrm{J}}\, \left[
-\frac{1}{6}\, \mathcal{N}\, R_\textrm{J} - \mathcal{N}_{I\bar{J}} \mathcal{D}_\mu X^I \mathcal{D}^\mu \bar{X}^{\bar{J}} - V_{\text{sc}}
\right],
\end{equation}
with $I,J \in \{0,1,2\}$ and scalar potential:
\begin{equation}
    V_{\text{sc}} = \mathcal{N}^{I\bar{J}} \mathcal{W}_I \overline{\mathcal{W}_J}
\end{equation}
We define projective coordinates:
\begin{equation}
    T = \frac{X^1}{X^0}, \qquad S = \frac{X^2}{X^0},
\end{equation}
and note the derivatives:
\begin{align}
\frac{\partial T}{\partial X^1} &= \frac{1}{X^0}, & \frac{\partial T}{\partial X^0} &= -\frac{T}{X^0}, \\
\frac{\partial S}{\partial X^2} &= \frac{1}{X^0}, & \frac{\partial S}{\partial X^0} &= -\frac{S}{X^0}.
\end{align}
The Kähler potential in the superconformal frame is:
\begin{equation}
\mathcal{N}(X,\bar{X}) = -3 |X^0|^2 \exp\left[-\tfrac{1}{3} K(T,\bar{T},S,\bar{S})\right].
\end{equation}
In terms of the component functions, we obtain:
\begin{equation}
    \mathcal{N}(X,\bar{X}) = -3 |X^0|^2 \left[
\mathcal{K}(T, \bar{T}) + f(T,\bar{T}) |S|^2 + g(T) S + \bar{g}(\bar{T}) \bar{S}
\right].
\end{equation}
This superconformal Kähler potential has Weyl weight 2 and satisfies the scaling relations:
\begin{equation}
    \mathcal{N}(X, \bar{X})=X^I \mathcal{N}_I=\bar{X}^I \mathcal{N}_{\bar{I}}=X^I \bar{X}^{\bar{J}} \mathcal{N}_{I \bar{J}} .
\end{equation}
Since the superfield $ S $ is nilpotent, all terms involving $ S $ or $ \bar{S} $ vanish on-shell in the bosonic sector. In particular, all terms linear or higher order in $ S $, $ \bar{S} $ drop out. The superconformal Kähler metric then simplifies to:
\begin{equation}
    \mathcal{N}_{I\bar{J}} = -3
\begin{pmatrix}
\mathcal{K} - T \mathcal{K}_T - \bar{T} \mathcal{K}_{\bar{T}} + T\bar{T} \mathcal{K}_{T\bar{T}} & \mathcal{K}_{\bar{T}} - T \mathcal{K}_{T\bar{T}} & \bar{g} \\
\mathcal{K}_T - \bar{T} \mathcal{K}_{T\bar{T}} & \mathcal{K}_{T\bar{T}} & 0 \\
g & 0 & f
\end{pmatrix}.
\end{equation}
The superconformal superpotential is given by:
\begin{equation}
    \mathcal{W}(X) = (X^0)^3 W_0,
\end{equation}
which satisfies the homogeneity condition of degree three, corresponding to Weyl weight 3 in the superconformal framework. The derivatives of $ \mathcal{W} $ satisfy:
\begin{equation}
\mathcal{W}_I \equiv \frac{\partial \mathcal{W}}{\partial X^I}, \qquad X^I \mathcal{W}_I = 3\, \mathcal{W}.
\end{equation}

\paragraph{Gauge Fixing and Conformal Compensator.}
To recover a Jordan-frame Lagrangian with non-minimal coupling to gravity, we fix the conformal compensator $ X^0 $ to be a field-dependent function rather than a constant. While a common choice for transitioning to Einstein frame is $ X^0 = \sqrt{3} $, in our case we adopt the gauge:
\begin{equation}
    X^0 = \sqrt{\frac{\Omega(\tau)}{\mathcal{K}(T, \bar{T})}} \quad \longrightarrow \quad X^0 = \sqrt{\frac{\Omega(\tau)}{\mathcal{K}(\tau)}},
\end{equation}
which ensures that the gravitational term in the Lagrangian takes the canonical Jordan-frame form:
\begin{equation}\label{Jordan_frame_coupling}
    \mathcal{L} \supset \tfrac{1}{2} \Omega(\tau)\, R_\textrm{J}.
\end{equation}
Here, we have restricted to the real slice $ T = \bar{T} = \tau \in \mathbb{R} $, as appropriate for the bosonic sector.

\paragraph{Choice of Frame Function.}
We choose the frame function:
\begin{equation}
    \Omega(T) = \xi (T - \tau_{0}),
\end{equation}
with real constants $\xi$ and $\tau_{0}$, and define:
\begin{equation}
    \mathcal{K}(T, \bar{T}) = \mathcal{V}_0^{2/3} + \varepsilon [\Omega(T) + \Omega(\bar{T})]^2,
\end{equation}
so that along the real direction $T = \bar{T} = \tau$, the frame function simplifies to:
\begin{equation}\label{Real_Kahler_pot}
    \mathcal{K}(\tau) = \mathcal{V}_0^{2/3} + 4\varepsilon \xi^2 (\tau - \tau_{0})^2.
\end{equation}

\paragraph{Vanishing of the $U(1)$ Gauge Field in the Bosonic Sector.}

In the superconformal formulation of supergravity, the chiral multiplets $ X^I $ are charged under the local $ U(1)_R $ symmetry. As a result, their covariant derivative includes a gauge field $ A_\mu $, defined as:
\begin{equation}
\mathcal{D}_\mu X^I = \left( \partial_\mu - \mathrm{i} A_\mu \right) X^I.
\end{equation}
The kinetic term in the Lagrangian takes the gauge-invariant form:
\begin{equation}
\mathcal{L}_{\text{kin}} = -\mathcal{N}_{I\bar{J}} \mathcal{D}_\mu X^I \mathcal{D}^\mu \bar{X}^{\bar{J}}.
\end{equation}

However, in the purely bosonic sector, the gauge field $ A_\mu $ is not an independent degree of freedom—it is an auxiliary field whose on-shell value is determined by the Kähler potential. Specifically, one finds:
\begin{equation}
A_\mu = \frac{\mathrm{i}}{2\mathcal{N}} \left( \mathcal{N}_{\bar{I}} \partial_\mu \bar{X}^{\bar{I}} - \mathcal{N}_I \partial_\mu X^I \right).
\end{equation}

Now, we consider the nilpotent superfield $ S $ and restrict to the real direction $ T = \bar{T} $. In this limit, the Kähler potential $ K(T,\bar{T},S,\bar{S}) $ becomes a real function of $ \tau $, so $ \partial_T K = \partial_{\bar{T}} K $, and the function $ \Phi = e^{-K/3} $ is also real, with $ \partial_T \Phi = \partial_{\bar{T}} \Phi $. Using this, the gauge field simplifies to:
\begin{equation}
    A_\mu  = 0,
\end{equation}
because the two terms cancel exactly due to the real nature of the scalar sector. Therefore, along the real trajectory where $ T = \bar{T} = \tau \in \mathbb{R} $ and $ S = 0 $, the covariant derivative reduces to an ordinary derivative:
\begin{equation}
    \mathcal{D}_\mu X^I \longrightarrow \partial_\mu X^I,
\end{equation}
and the kinetic term becomes:
\begin{equation}
    \mathcal{L}_{\text{kin}} = -\mathcal{N}_{I\bar{J}} \partial_\mu X^I \partial^\mu \bar{X}^{\bar{J}}.
\end{equation}

\paragraph{Jordan-Frame Kinetic Term.}
After superconformal embedding and gauge fixing, we now derive the scalar kinetic term in the Jordan frame. With $S = 0$, only $I, J \in \{0,1\}$ contribute. Using the gauge-fixing condition:
\begin{equation}\label{Gauge_Fixed_X0}
    X^0 = \sqrt{G}, \quad G \equiv \frac{\Omega(\tau)}{\mathcal{K}(T,\bar{T})},
\end{equation}
the spacetime derivatives are:
\begin{equation}
\begin{aligned}
    \partial_\mu X^1 &= X^0 \partial_\mu T + T \partial_\mu X^0, \\
    \partial_\mu X^0 &= \frac{1}{2\sqrt{G}} \left( G_T \partial_\mu T + G_{\bar{T}} \partial_\mu \bar{T} \right).
\end{aligned}
\end{equation}
Substituting into the kinetic term yields:
\begin{equation}
\begin{aligned}
\mathcal{L}_{\text{kin}} =
& -3 \mathcal{K}_{T\bar{T}} G \partial_\mu T \partial^\mu \bar{T} \\
& -\tfrac{3}{2} \mathcal{K}_T \left(G_T \partial_\mu T \partial^\mu T + G_{\bar{T}} \partial_\mu T \partial^\mu \bar{T}\right) \\
& -\tfrac{3}{2} \mathcal{K}_{\bar{T}} \left(G_T \partial_\mu \bar{T} \partial^\mu T + G_{\bar{T}} \partial_\mu \bar{T} \partial^\mu \bar{T}\right) \\
& -\tfrac{3}{4} \mathcal{K} \cdot \frac{1}{G} \left[
G_T^2 \partial_\mu T \partial^\mu T + 2 G_T G_{\bar{T}} \partial_\mu T \partial^\mu \bar{T} + G_{\bar{T}}^2 \partial_\mu \bar{T} \partial^\mu \bar{T}
\right]. \label{eq:kinetic-expanded}
\end{aligned}
\end{equation}
Following this and combining all the terms with the gauge-fixing~\eqref{Gauge_Fixed_X0}, the kinetic Lagrangian simplifies to:
\begin{equation}
\mathcal{L}_{\text{kin}} = \frac{\Omega(\tau)}{\mathcal{K}^2} \left[
-3 \mathcal{K}_{T\bar{T}} \mathcal{K} \partial_\mu T \partial^\mu \bar{T}
+ \tfrac{3}{4} \mathcal{K}_T^2 \partial_\mu T \partial^\mu T
+ \tfrac{3}{2} \mathcal{K}_T \mathcal{K}_{\bar{T}} \partial_\mu T \partial^\mu \bar{T}
+ \tfrac{3}{4} \mathcal{K}_{\bar{T}}^2 \partial_\mu \bar{T} \partial^\mu \bar{T}
\right].
\end{equation}

\paragraph{Kinetic Lagrangian along the Real Direction \texorpdfstring{$T = \bar{T}$}{T = T̄}.}
Specializing to the real direction $T = \bar{T} = \tau \in \mathbb{R}$, we use:
\begin{equation}
    K_{T\bar{T}} = \frac{3}{\mathcal{K}^2} (\mathcal{K}_T \mathcal{K}_{\bar{T}} - \mathcal{K} \mathcal{K}_{T\bar{T}}),
\end{equation}
and hence:
\begin{equation}
    \mathcal{L}_{\text{kin}}\big|_{T = \bar{T}} = -\Omega(\tau) K_{T\bar{T}}\big|_{T = \bar{T}} \partial_\mu T \partial^\mu T.
\end{equation}
Given the Kähler potential in the real direction with~\eqref{Real_Kahler_pot}, we compute:
\begin{equation}
    K_{\tau\tau} = 4 K_{T\bar{T}}\big|_{T = \bar{T}} = \frac{24 \varepsilon \xi^2}{\mathcal{K}^2} \left[4 \varepsilon \xi^2 (\tau - \tau_{0})^2 - \mathcal{V}_0^{2/3}\right].
\end{equation}
Thus, the final form of the Jordan-frame kinetic term is:
\begin{equation}
\mathcal{L}_{\text{kin}} = -\frac{\Omega(\tau)}{4} K_{\tau\tau} \partial_\mu \tau \partial^\mu \tau.
\end{equation}

\paragraph{Superconformal Equivalence of the Einstein Frame Scalar Potential.}
We derive the scalar potential from the superconformal formulation, focusing on the inverse Kähler metric and constant superpotential. Note that, considering the non-minimal coupling term to gravity with~\eqref{Jordan_frame_coupling}, the produced scalar potential becomes a Jordan frame potential and given by:
\begin{equation} \label{eq:Vsc_general}
V_{\text{sc}}^{\textrm{J}} = \mathcal{N}^{I\bar{J}} \mathcal{W}_I \overline{\mathcal{W}_J},
\end{equation}
where indices run over $I, J \in \{0,1,2\}$. For a constant superpotential
\begin{equation} \label{eq:W_constant}
\mathcal{W}(X) = (X^0)^3 W_0,
\end{equation}
we compute the derivatives:
\begin{equation} \label{eq:W_derivatives}
\mathcal{W}_0 = 3 (X^0)^2 W_0, \qquad \mathcal{W}_1 = \mathcal{W}_2 = 0.
\end{equation}
Thus, the scalar potential reduces to:
\begin{equation} \label{eq:Vsc_N00}
V_{\text{sc}}^{\textrm{J}} = 9\, |X^0|^4\, |W_0|^2 \cdot \mathcal{N}^{0\bar{0}}
\end{equation}

We extract the $(0\bar{0})$ component of the inverse Kähler metric $\mathcal{N}^{I\bar{J}}$, which takes the form:
\begin{equation} \label{eq:N_inverse}
\mathcal{N}^{I\bar{J}} = \frac{1}{\Delta}
\begin{pmatrix}
\mathcal{K}_{T\bar{T}} f & f(\mathcal{K}_{T\bar{T}} T - \mathcal{K}_{\bar{T}}) & \mathcal{K}_{T\bar{T}} \bar{g} \\
 f(-\mathcal{K}_T + \mathcal{K}_{T\bar{T}} \bar{T}) & \mathcal{K} f - \mathcal{K}_T T f + \mathcal{K}_{T\bar{T}} T \bar{T} f - \mathcal{K}_{\bar{T}} \bar{T} f - g \bar{g} & \bar{g}(\mathcal{K}_T - \mathcal{K}_{T\bar{T}} \bar{T}) \\
 \mathcal{K}_{T\bar{T}} g & g(-\mathcal{K}_{T\bar{T}} T + \mathcal{K}_{\bar{T}}) & \mathcal{K} \mathcal{K}_{T\bar{T}} - \mathcal{K}_T \mathcal{K}_{\bar{T}}
\end{pmatrix},
\end{equation}
with determinant factor:
\begin{equation} \label{eq:Delta}
\Delta = \frac{1}{3\left(-\mathcal{K} \mathcal{K}_{T\bar{T}} f + \mathcal{K}_T \mathcal{K}_{\bar{T}} f + \mathcal{K}_{T\bar{T}} g \bar{g} \right)}.
\end{equation}
The relevant component is therefore:
\begin{equation} \label{eq:N00_component}
\mathcal{N}^{0\bar{0}} = \frac{\mathcal{K}_{T\bar{T}} f}{3 \left(- \mathcal{K} \mathcal{K}_{T\bar{T}} f + \mathcal{K}_T \mathcal{K}_{\bar{T}} f + \mathcal{K}_{T\bar{T}} g \bar{g} \right)}.
\end{equation}
Substituting into Eq.~\eqref{eq:Vsc_N00} using the gauge fixing~\eqref{Gauge_Fixed_X0}, we obtain:
\begin{equation} \label{eq:Vsc_final_expanded}
V_{\text{sc}}^{\textrm{J}} = \frac{3\, \Omega^2(\tau)\, \mathcal{K}_{T\bar{T}} f\, |W_0|^2}{\mathcal{K}^2 \left( - \mathcal{K} \mathcal{K}_{T\bar{T}} f + \mathcal{K}_T \mathcal{K}_{\bar{T}} f + \mathcal{K}_{T\bar{T}} g \bar{g} \right)}.
\end{equation}
It is convenient to express the result in rational form by factoring out the common numerator:
\begin{equation} \label{eq:Vsc_rational_form}
V_{\text{sc}}^{\textrm{J}} =
\frac{3\,\Omega^2(\tau)\, |W_0|^2}{\mathcal{K}^3}
\left[
      \frac{f\,|\mathcal{K}_T|^2 + |g|^2\,\mathcal{K}_{T\overline{T}}}
           {f\,|\mathcal{K}_T|^2 + |g|^2\,\mathcal{K}_{T\overline{T}} - f\,\mathcal{K}_{T\overline{T}}\,\mathcal{K}}
      - 1
    \right].
\end{equation}

Here, one can switch to the Einstein frame after a conformal transformation as denoted in section (\ref{subsec4.3:Conformal_equivalence}), where the Einstein frame potential becomes
\begin{equation} \label{eq:Vsc_Einstein}
V_{\text{sc}}^{\textrm{E}} = \frac{V_{\text{sc}}^{\text{J}}}{\Omega^2(\tau)} = \frac{3 \, |W_0|^2}{\mathcal{K}^3}
\left[
      \frac{f\,|\mathcal{K}_T|^2 + |g|^2\,\mathcal{K}_{T\overline{T}}}
           {f\,|\mathcal{K}_T|^2 + |g|^2\,\mathcal{K}_{T\overline{T}} - f\,\mathcal{K}_{T\overline{T}}\,\mathcal{K}}
      - 1
    \right],
\end{equation}
as follows from the conformal rescaling relation~\eqref{Einstein_Frame_to_Jordan_Pot}, and the Jordan-frame origin of the potential via non-minimal coupling in~\eqref{Jordan_frame_coupling}. As a result, the superconformal scalar potential, when expressed in Einstein frame, takes precisely the same form as the supergravity potential~\eqref{V_log} derived from the logarithmic Kähler potential~\eqref{kahler_potential}. The overall factor of $\Omega^2(\tau)$, arising from the gauge-fixing of the conformal compensator, thus plays a central role in mediating between the Jordan and Einstein frames, while preserving the underlying structure of the scalar potential. 

This matching confirms the internal consistency of the superconformal embedding and gauge-fixing procedure: upon transition to Einstein frame, the scalar potential derived via superconformal methods precisely reproduces the standard supergravity result~\eqref{V_log}. In addition, it clarifies the origin of the non-minimal coupling term, which plays a critical role in enabling the canonical field diffeomorphism presented in equation~\eqref{canonical_diffeomorphism}.

\subsection{Canonical Field Redefinition and the Role of $\alpha'^3$ and Loop Corrections}
\label{sec:Appendix_CanonicalField}

In this section, we derive the field redefinition between the original modulus $\tau$ and the canonically normalized scalar field $\varphi$, accounting for both $\alpha'^3$ and loop corrections via the corrected Jordan-frame kinetic structure.

\paragraph{Small‑$\boldsymbol{x}$ Expansion and Definition of $\boldsymbol{K_0}$.}

We define the shifted modulus:
\begin{equation}
    x \equiv \tau - \tau_{0},
    \qquad
    \mathcal{K}(\tau) = \mathcal{V}_0^{2/3} + 4 \varepsilon \xi^2 x^2.
\end{equation}
In the regime $ \mathcal{V}_0^{2/3} \gg 4 |\varepsilon| \xi^2 x^2 $, the kinetic prefactor expands as:
\begin{equation}
    K_{\tau\tau}(x) \approx -\frac{24 \varepsilon \xi^2}{\mathcal{V}_0^{2/3}} \equiv K_0 > 0.
\end{equation}
Naively identifying this with the Einstein-frame metric yields:
\begin{equation}
    \frac{d\varphi}{d\tau} = \sqrt{ \frac{1}{2} \Omega(\tau) K_0 }
    \quad \Rightarrow \quad
    \varphi(\tau) = \frac{\sqrt{2 \xi K_0}}{3} \tau^{3/2} + \text{const},
\end{equation}
suggesting a power-law dependence inconsistent with the exponential field redefinition used in the main text. The discrepancy arises because the Einstein-frame kinetic term receives a nontrivial contribution from the Weyl rescaling of the gravitational sector.

\paragraph{Einstein‑Frame Kinetic Metric.}

Starting from the Jordan-frame action:
\begin{equation}
    \mathcal{L}_\textrm{J} = \sqrt{-g_\textrm{J}} \left[
    \frac{1}{2} \Omega(\tau)\, R_\textrm{J}
    - \frac{1}{4} \Omega(\tau) K_{\tau\tau} g_J^{\mu\nu} \partial_\mu \tau \partial_\nu \tau
    - V_\textrm{J}(\tau)
    \right],
\end{equation}
with $ \Omega(\tau) = \xi x $, we apply the Weyl rescaling:
\begin{equation}
    g_{\textrm{J}\,\mu\nu} = \Omega^{-1}(\tau)\, g_{\textrm{E}\,\mu\nu}.
\end{equation}
This yields the Einstein-frame kinetic term:
\begin{equation}
    \mathcal{L}_{\text{kin},E} \supset -\frac12 \sqrt{-g_\textrm{E}} \left[
    \tfrac12 K_{\tau\tau}(\tau) +
    \tfrac32 \left( \partial_\tau \ln \Omega \right)^2
    \right] g_\textrm{E}^{\mu\nu} \partial_\mu \tau \partial_\nu \tau,
\end{equation}
with kinetic metric:
\begin{equation} \label{eq:Gtautau_app}
    G_{\tau\tau}(\tau) = \frac{1}{2} K_{\tau\tau}(\tau) + \frac{3}{2} \left( \partial_\tau \ln \Omega \right)^2.
\end{equation}

\paragraph{Weyl‑Dominated Regime and Parametric Estimate.}

In the small-$x$ regime where $\Omega = \xi x$, we find:
\begin{equation}
    G_{\tau\tau}(x) = \frac{K_0}{2} + \frac{3}{2 x^2}.
\end{equation}
The second term dominates when $ x^2 \ll \frac{3}{K_0} \sim \frac{\mathcal{V}_0^{2/3}}{8 |\varepsilon| \xi^2} $. For realistic parameters, e.g., $ \mathcal{V}_0^{2/3} \sim 10^3 $, this condition is easily satisfied for moderate values of $\xi$ and small $|\varepsilon|$, making the Weyl term the leading contribution to the field-space metric. In this regime:
\begin{equation}
    G_{\tau\tau}(x) \approx \frac{3}{2 x^2} = \frac{3}{2 (\tau - \tau_{0})^2}.
\end{equation}

\paragraph{Canonical Field and Exponential Mapping.}
\label{app:fieldredef}

Using the dominant kinetic term, the canonically normalized field satisfies:
\begin{equation}
    \frac{d\varphi}{d\tau} = \sqrt{G_{\tau\tau}} \approx \sqrt{ \frac{3}{2} } \cdot \frac{1}{\tau - \tau_{0}}.
\end{equation}
Integrating:
\begin{equation}
    \varphi = \pm \sqrt{ \frac{3}{2} } \ln(\tau - \tau_{0}) + \varphi_0.
\end{equation}
Matching this to the frame function,
\begin{equation}
    \Omega(\tau) = \xi(\tau - \tau_{0}) = e^{\pm \sqrt{ \tfrac{2}{3} } \varphi},
\end{equation}
requires setting the integration constant to $ \varphi_0 = - \sqrt{\tfrac{3}{2}} \log \xi $, i.e.,
\begin{equation}
    \tau = \tau_{0} + \frac{1}{\xi} e^{- \sqrt{ \tfrac{2}{3} } \varphi}.
\end{equation}
This suggests that the integration constant may be physically determined by the slope of the frame function $\xi$, determining the geometric field range and the location of the SUSY-restoring boundary. For further discussion of this point in the EFT context, see Section~\ref{subsec6.3:SHC}.



\bibliography{references}

\end{document}